\documentclass[fleqn,10pt]{wlscirep}
\usepackage[utf8]{inputenc}
\usepackage[T1]{fontenc}

\usepackage{times}
\usepackage{epsfig}
\usepackage{transparent}
\usepackage{graphicx}
\usepackage[export]{adjustbox}
\usepackage{array}
\usepackage{amsmath}
\newcommand{\inlineeqnum}{\refstepcounter{equation}~~\mbox{(\theequation)}}
\usepackage{amsthm}
\usepackage{amssymb}
\usepackage{mathtools}
\usepackage{tikz}
\usepackage{multicol}
\usepackage{caption}
\usepackage{subcaption}
\usepackage{pgfplots}
\pgfplotsset{compat=newest}
\usepackage{tcolorbox}
\usepackage{dblfloatfix}
\usepackage[parfill]{parskip}
\usepackage{float}
\usepackage[vlined,ruled,linesnumbered,noresetcount,algo2e]{algorithm2e}

\SetCommentSty{mycommfont}
\usepackage{algorithm}
\usepackage[noend]{algpseudocode}
\usepackage{import}
\usetikzlibrary{quotes,arrows.meta}
\usetikzlibrary{positioning}

\def\edgecolor{rgb:blue,4;red,1;green,4;black,3}
\newcommand{\midarrow}{\tikz \draw[-Stealth,line width =0.8mm,draw=\edgecolor] (-0.3,0) -- ++(0.3,0);}

\usepackage{Ball}
\usepackage{Box}
\usepackage{RightBandedBox}

\usetikzlibrary{positioning}
\usetikzlibrary{3d} 

\def\ConvColor{rgb:yellow,5;red,2.5;white,5}

\def\PoolColor{rgb:red,1;black,0.3}
\def\UnpoolColor{rgb:blue,2;green,1;black,0.3}

\def\SoftmaxColor{rgb:magenta,5;black,7}

\newtheorem{theorem}{Theorem}

\usepackage{color}
\usepackage[normalem]{ulem}
\usepackage{xurl} 
\usepackage{authblk}

\usepackage[belowskip=-5pt,aboveskip=10pt]{caption}
\title{An End-to-End Computer Vision Methodology for Quantitative Metallography}

\author[1,*]{Matan Rusanovsky}
\author[2]{Ofer Beeri}
\author[1,3,*]{Gal Oren}

\affil[1]{Scientific Computing Center, Nuclear Research Center – Negev, Be'er-Sheva, Israel}
\affil[2]{Department of Materials, Nuclear Research Center – Negev, Be'er-Sheva, Israel}
\affil[3]{Department of Computer Science, Technion – Israel Institute of Technology, Haifa, Israel}

\affil[*]{matanr@nrcn.org.il, galoren@cs.technion.ac.il}

\begin{abstract}
Metallography is crucial for a proper assessment of material properties. It mainly involves investigating the spatial distribution of grains and the occurrence and characteristics of inclusions or precipitates.
This work presents a holistic few-shot artificial intelligence model for Quantitative Metallography, including Anomaly Detection, that automatically quantifies the degree of the anomaly of impurities in alloys. 
We suggest the following examination process: 
(1) Deep semantic segmentation is performed on the inclusions (based on a suitable metallographic dataset of alloys and corresponding tags of inclusions), producing inclusions masks that are saved into a separated dataset. (2) Deep image inpainting is performed to fill the removed inclusions parts, resulting in 'clean' metallographic images, which contain the background of grains. (3) Grains' boundaries are marked using deep semantic segmentation (based on another metallographic dataset of alloys), producing boundaries that are ready for further inspection on the distribution of grains' size. (4) Deep anomaly detection and pattern recognition is performed on the inclusions masks to determine spatial, shape, and area anomaly detection of the inclusions. 
Finally, the end-to-end model recommends an expert on areas of interest for further examination. The physical result can re-tune the model according to the specific material at hand. 
Although the techniques presented here were developed for metallography analysis, most of them can be generalized to a broader set of microscopy problems that require automation. 
All source-codes as well as the datasets that were created for this work, are publicly available at \url{https://github.com/Scientific-Computing-Lab-NRCN/MLography}.

\end{abstract}
\begin{document}

\flushbottom
\maketitle
%
%
\thispagestyle{empty}

\begin{figure}
    \centering
\tikzset{every picture/.style={line width=0.75pt}} 
\resizebox{0.9\textwidth}{!}{
\begin{tikzpicture}[x=0.75pt,y=0.75pt,yscale=-1,xscale=1]

\draw (542.53,541.76) node  {\includegraphics[width=137.49pt,height=91.66pt]{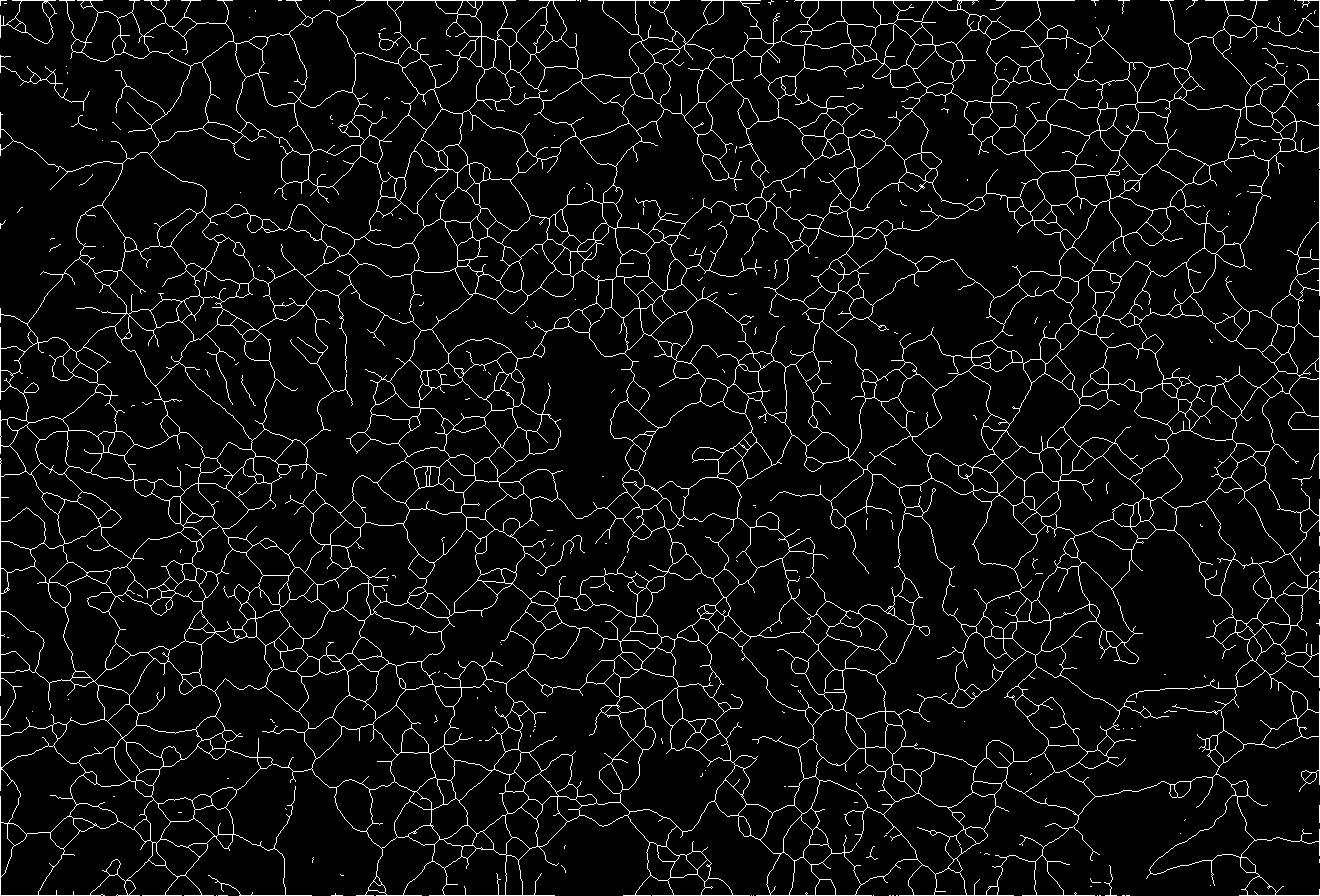}};
\draw (225.81,781.53) node  {\includegraphics[width=293.22pt,height=195.48pt]{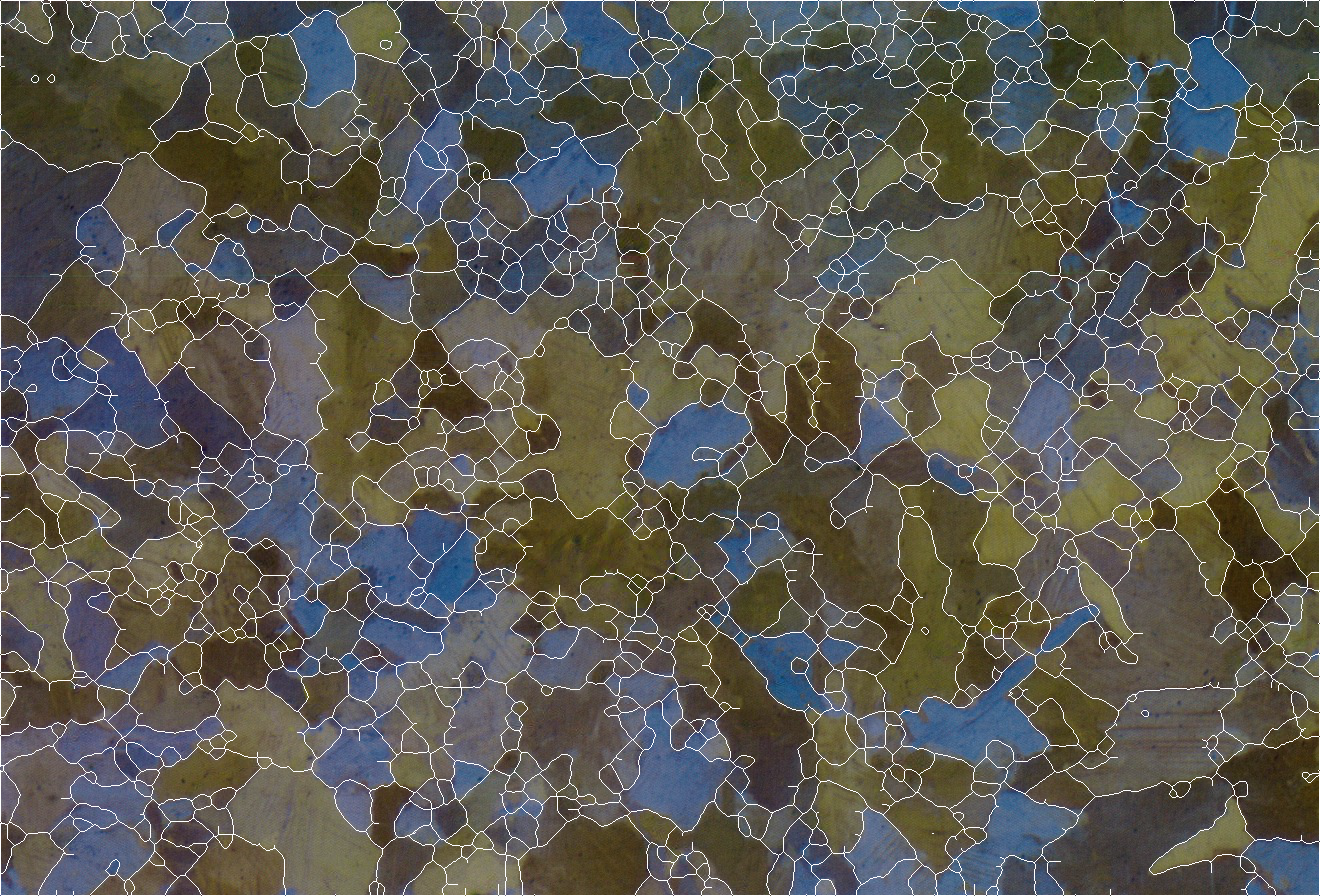}};
\draw (542.82,712.98) node  {\includegraphics[width=137.49pt,height=91.66pt]{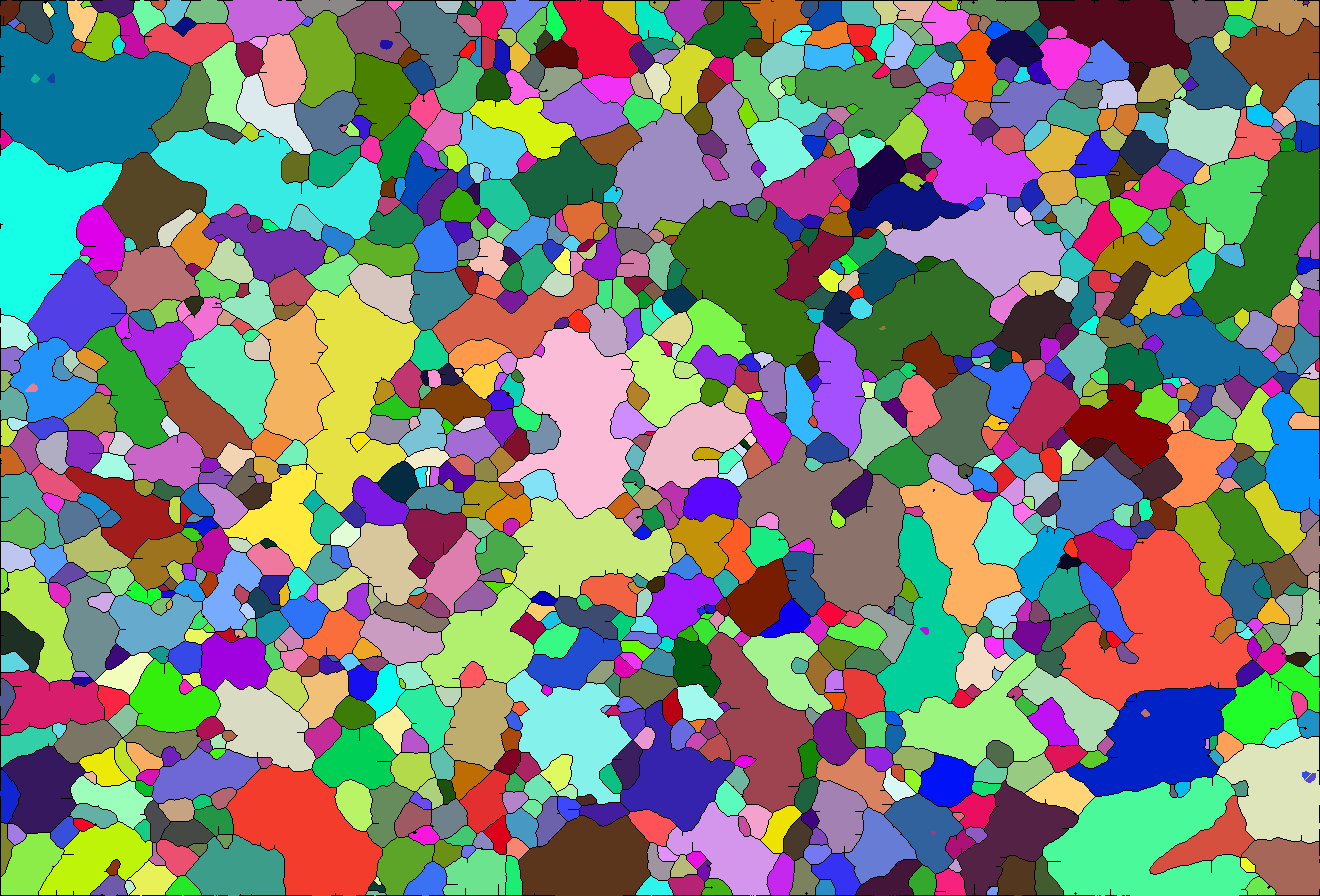}};
\draw (120.92,541.7) node  {\includegraphics[width=137.49pt,height=91.66pt]{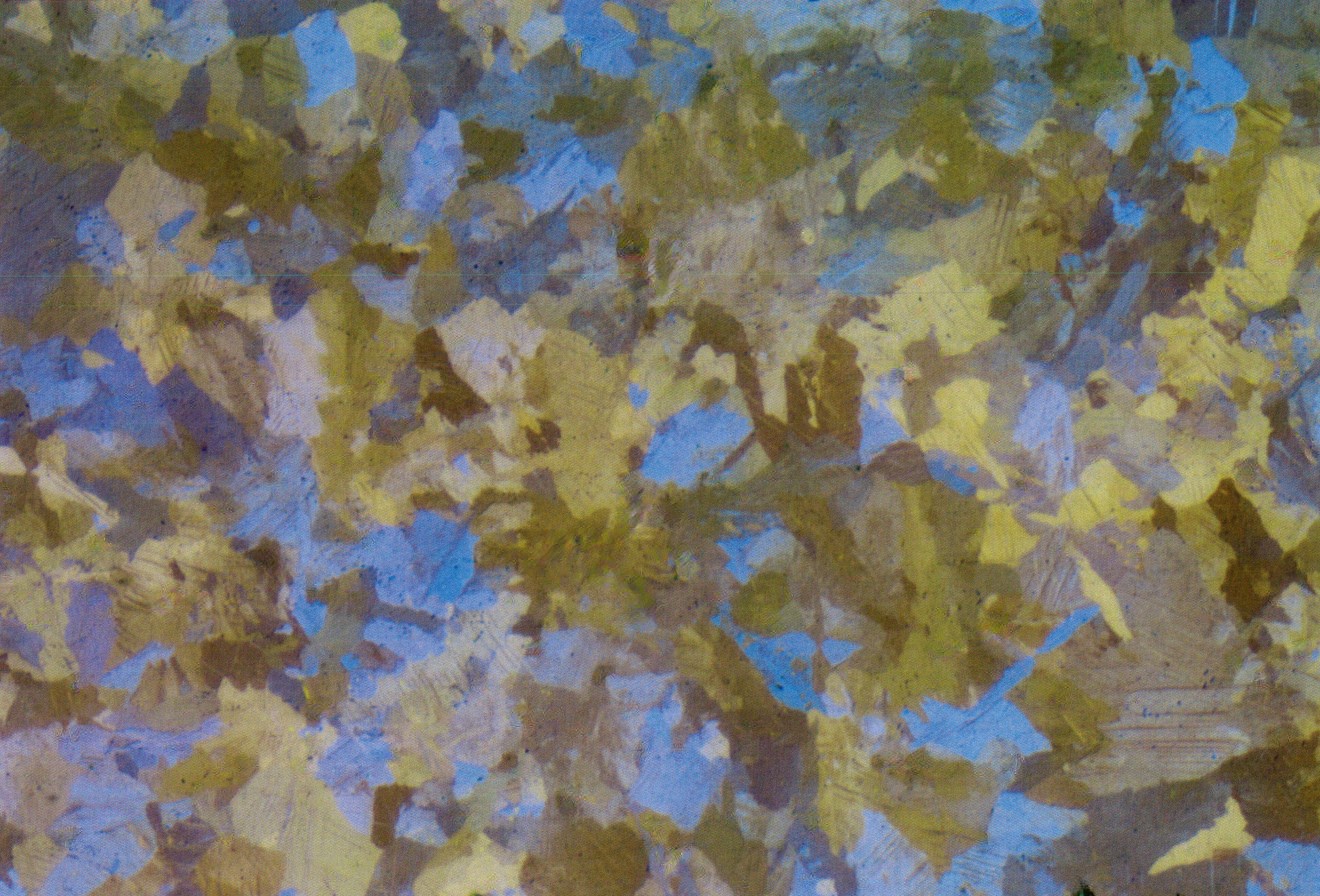}};
\draw (330.74,371.37) node  {\includegraphics[width=137.49pt,height=91.66pt]{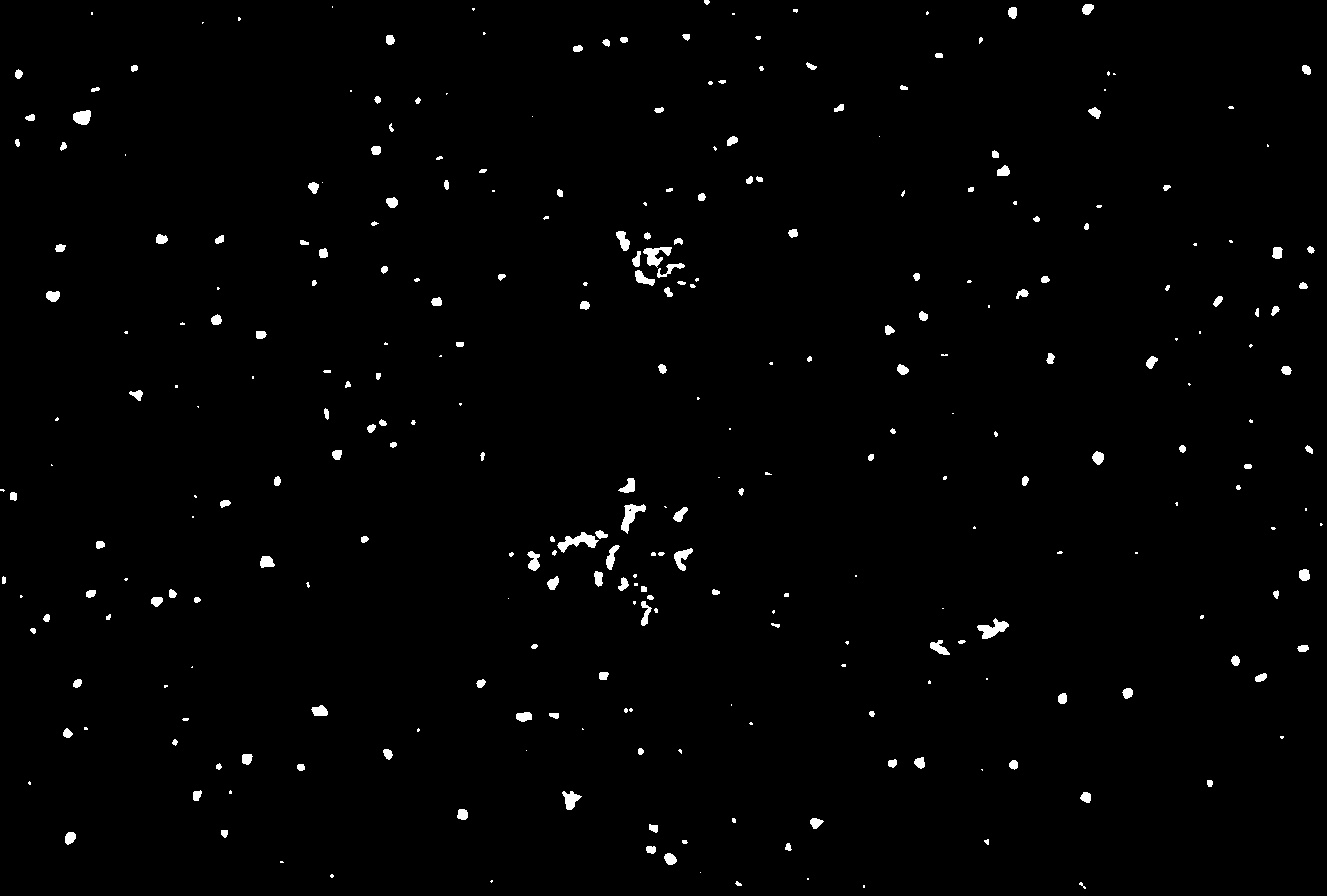}};
\draw (226.8,150.38) node  {\includegraphics[width=293.3pt,height=195.53pt]{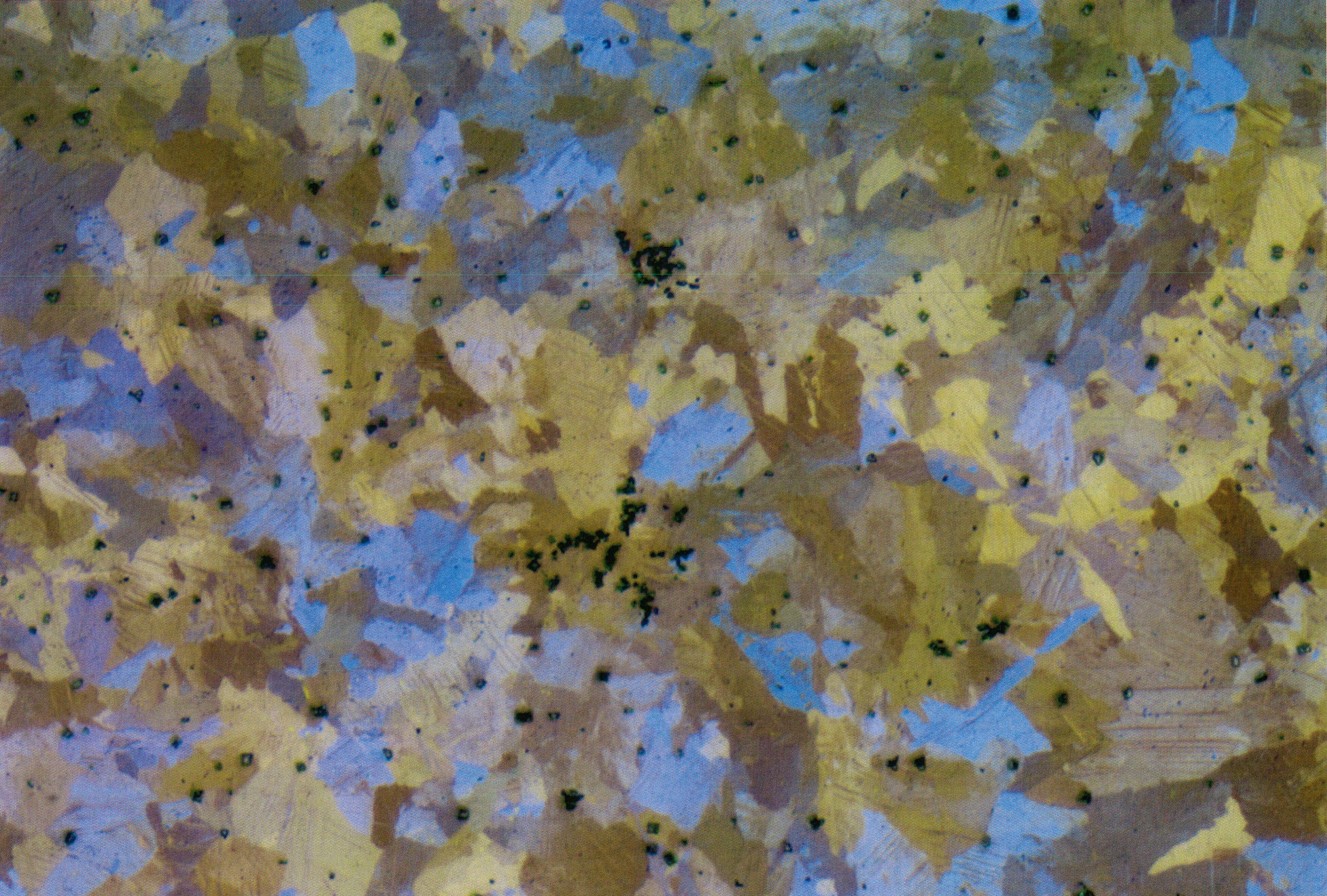}};
\draw (543.46,219.01) node  {\includegraphics[width=137.49pt,height=91.66pt]{Segmentation/big/25.jpg}};
\draw   (207.71,602.86) -- (207.81,611.44) .. controls (207.89,618.54) and (213.71,624.23) .. (220.8,624.15) -- (233.18,624.01) .. controls (240.28,623.93) and (245.97,618.11) .. (245.89,611.02) -- (245.89,611.02) -- (251.54,610.95) -- (243.37,603.23) -- (235.37,611.13) -- (241.02,611.07) -- (241.02,611.07) .. controls (241.07,615.48) and (237.53,619.1) .. (233.12,619.14) -- (220.75,619.28) .. controls (216.34,619.33) and (212.72,615.8) .. (212.68,611.39) -- (212.58,602.81) -- cycle ;
\draw (541.55,370.93) node  {\includegraphics[width=137.49pt,height=91.66pt]{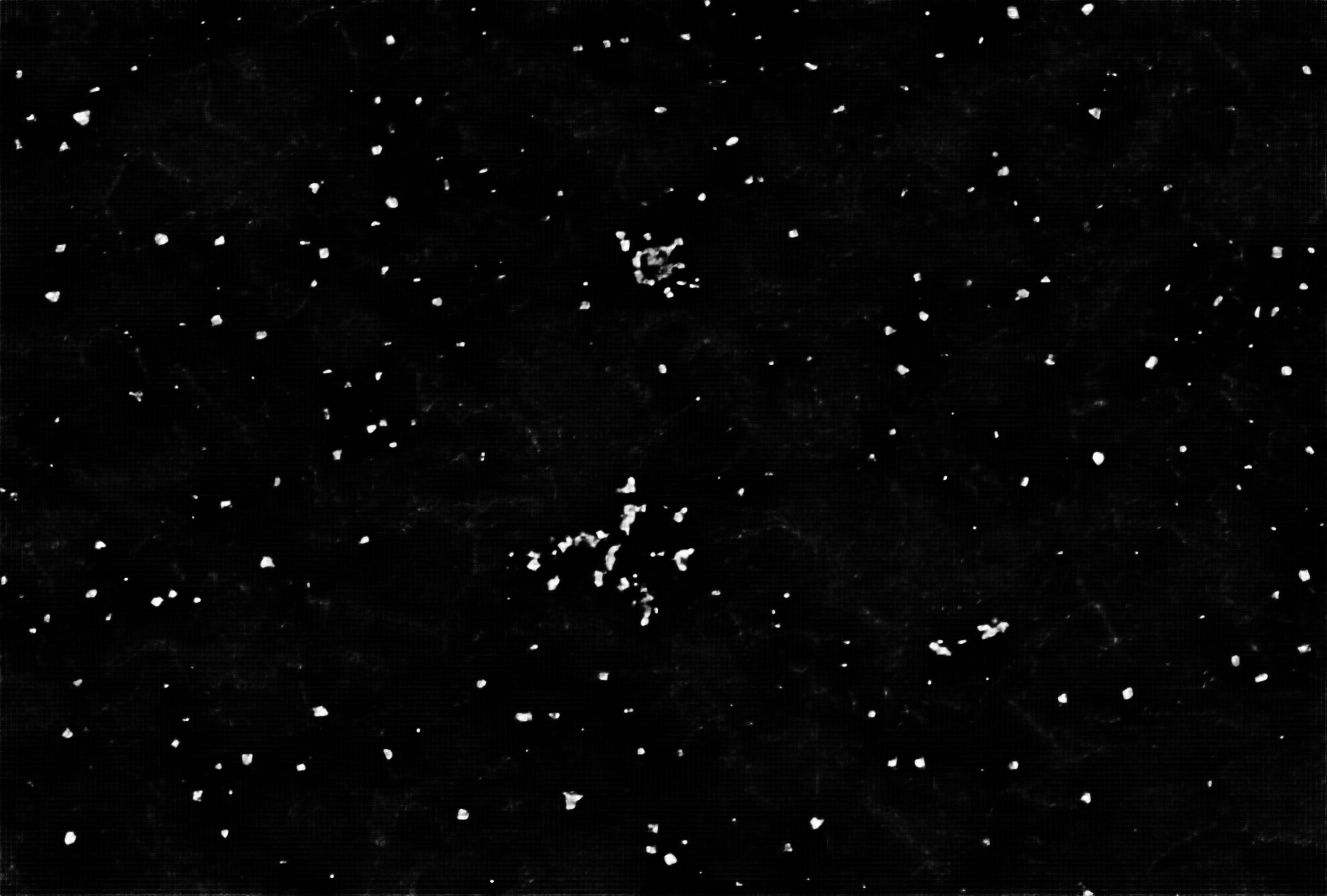}};
\draw   (634.19,598) -- (642.77,598) .. controls (649.87,598) and (655.62,603.76) .. (655.62,610.86) -- (655.62,657.48) .. controls (655.62,664.58) and (649.87,670.33) .. (642.77,670.33) -- (642.77,670.33) -- (642.77,675.99) -- (634.95,667.9) -- (642.77,659.82) -- (642.77,665.47) -- (642.77,665.47) .. controls (647.18,665.47) and (650.76,661.89) .. (650.76,657.48) -- (650.76,610.86) .. controls (650.76,606.45) and (647.18,602.87) .. (642.77,602.87) -- (634.19,602.87) -- cycle ;
\draw (119.58,371.37) node  {\includegraphics[width=137.49pt,height=91.66pt]{Segmentation/without_impurities/25.jpg}};
\draw   (27.92,427.61) -- (19.34,427.61) .. controls (12.24,427.61) and (6.49,433.37) .. (6.49,440.46) -- (6.49,487.09) .. controls (6.49,494.19) and (12.24,499.94) .. (19.34,499.94) -- (19.34,499.94) -- (19.34,505.59) -- (27.16,497.51) -- (19.34,489.42) -- (19.34,495.08) -- (19.34,495.08) .. controls (14.93,495.08) and (11.36,491.5) .. (11.36,487.09) -- (11.36,440.46) .. controls (11.36,436.05) and (14.93,432.48) .. (19.34,432.48) -- (27.92,432.48) -- cycle ;
\draw (330.06,542.01) node  {\includegraphics[width=137.49pt,height=91.66pt]{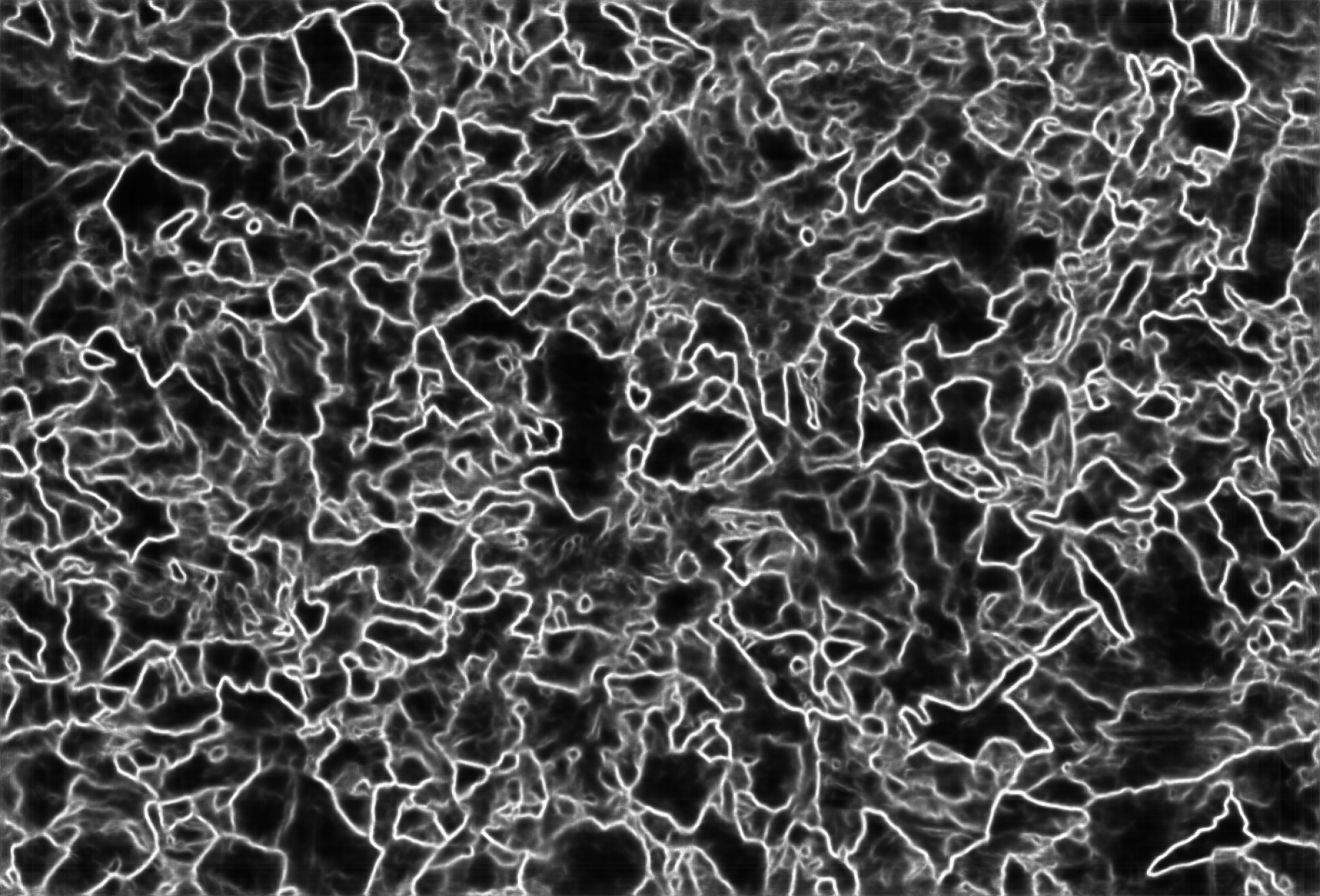}};
\draw   (635.12,275.26) -- (643.7,275.26) .. controls (650.79,275.26) and (656.55,281.01) .. (656.55,288.11) -- (656.55,334.74) .. controls (656.55,341.83) and (650.79,347.59) .. (643.7,347.59) -- (643.7,347.59) -- (643.7,353.24) -- (635.88,345.15) -- (643.7,337.07) -- (643.7,342.72) -- (643.7,342.72) .. controls (648.11,342.72) and (651.68,339.15) .. (651.68,334.74) -- (651.68,288.11) .. controls (651.68,283.7) and (648.11,280.12) .. (643.7,280.12) -- (635.12,280.12) -- cycle ;
\draw   (454.75,432.09) -- (454.65,440.67) .. controls (454.58,447.77) and (448.76,453.46) .. (441.66,453.38) -- (429.29,453.24) .. controls (422.19,453.16) and (416.5,447.34) .. (416.58,440.24) -- (416.58,440.24) -- (410.93,440.18) -- (419.1,432.45) -- (427.09,440.36) -- (421.44,440.3) -- (421.44,440.3) .. controls (421.39,444.71) and (424.93,448.32) .. (429.34,448.37) -- (441.71,448.51) .. controls (446.13,448.56) and (449.74,445.03) .. (449.79,440.61) -- (449.89,432.04) -- cycle ;
\draw  [draw opacity=0][fill={rgb, 255:red, 255; green, 255; blue, 255 }  ,fill opacity=0.27 ] (29.25,480.59) -- (212.58,480.59) -- (212.58,602.81) -- (29.25,602.81) -- cycle ; \draw  [color={rgb, 255:red, 208; green, 2; blue, 27 }  ,draw opacity=1 ] (41.47,480.59) -- (41.47,602.81)(53.7,480.59) -- (53.7,602.81)(65.92,480.59) -- (65.92,602.81)(78.14,480.59) -- (78.14,602.81)(90.36,480.59) -- (90.36,602.81)(102.58,480.59) -- (102.58,602.81)(114.81,480.59) -- (114.81,602.81)(127.03,480.59) -- (127.03,602.81)(139.25,480.59) -- (139.25,602.81)(151.47,480.59) -- (151.47,602.81)(163.69,480.59) -- (163.69,602.81)(175.91,480.59) -- (175.91,602.81)(188.14,480.59) -- (188.14,602.81)(200.36,480.59) -- (200.36,602.81) ; \draw  [color={rgb, 255:red, 208; green, 2; blue, 27 }  ,draw opacity=1 ] (29.25,492.81) -- (212.58,492.81)(29.25,505.04) -- (212.58,505.04)(29.25,517.26) -- (212.58,517.26)(29.25,529.48) -- (212.58,529.48)(29.25,541.7) -- (212.58,541.7)(29.25,553.92) -- (212.58,553.92)(29.25,566.14) -- (212.58,566.14)(29.25,578.37) -- (212.58,578.37)(29.25,590.59) -- (212.58,590.59) ; \draw  [color={rgb, 255:red, 208; green, 2; blue, 27 }  ,draw opacity=1 ] (29.25,480.59) -- (212.58,480.59) -- (212.58,602.81) -- (29.25,602.81) -- cycle ;
\draw  [draw opacity=0][fill={rgb, 255:red, 255; green, 255; blue, 255 }  ,fill opacity=0.27 ] (238.4,480.91) -- (421.72,480.91) -- (421.72,603.12) -- (238.4,603.12) -- cycle ; \draw  [color={rgb, 255:red, 208; green, 2; blue, 27 }  ,draw opacity=1 ] (250.62,480.91) -- (250.62,603.12)(262.84,480.91) -- (262.84,603.12)(275.06,480.91) -- (275.06,603.12)(287.28,480.91) -- (287.28,603.12)(299.51,480.91) -- (299.51,603.12)(311.73,480.91) -- (311.73,603.12)(323.95,480.91) -- (323.95,603.12)(336.17,480.91) -- (336.17,603.12)(348.39,480.91) -- (348.39,603.12)(360.62,480.91) -- (360.62,603.12)(372.84,480.91) -- (372.84,603.12)(385.06,480.91) -- (385.06,603.12)(397.28,480.91) -- (397.28,603.12)(409.5,480.91) -- (409.5,603.12) ; \draw  [color={rgb, 255:red, 208; green, 2; blue, 27 }  ,draw opacity=1 ] (238.4,493.13) -- (421.72,493.13)(238.4,505.35) -- (421.72,505.35)(238.4,517.57) -- (421.72,517.57)(238.4,529.79) -- (421.72,529.79)(238.4,542.01) -- (421.72,542.01)(238.4,554.24) -- (421.72,554.24)(238.4,566.46) -- (421.72,566.46)(238.4,578.68) -- (421.72,578.68)(238.4,590.9) -- (421.72,590.9) ; \draw  [color={rgb, 255:red, 208; green, 2; blue, 27 }  ,draw opacity=1 ] (238.4,480.91) -- (421.72,480.91) -- (421.72,603.12) -- (238.4,603.12) -- cycle ;
\draw  [draw opacity=0][fill={rgb, 255:red, 255; green, 255; blue, 255 }  ,fill opacity=0.27 ] (451.79,157.9) -- (635.12,157.9) -- (635.12,280.12) -- (451.79,280.12) -- cycle ; \draw  [color={rgb, 255:red, 208; green, 2; blue, 27 }  ,draw opacity=1 ] (464.01,157.9) -- (464.01,280.12)(476.24,157.9) -- (476.24,280.12)(488.46,157.9) -- (488.46,280.12)(500.68,157.9) -- (500.68,280.12)(512.9,157.9) -- (512.9,280.12)(525.12,157.9) -- (525.12,280.12)(537.34,157.9) -- (537.34,280.12)(549.57,157.9) -- (549.57,280.12)(561.79,157.9) -- (561.79,280.12)(574.01,157.9) -- (574.01,280.12)(586.23,157.9) -- (586.23,280.12)(598.45,157.9) -- (598.45,280.12)(610.68,157.9) -- (610.68,280.12)(622.9,157.9) -- (622.9,280.12) ; \draw  [color={rgb, 255:red, 208; green, 2; blue, 27 }  ,draw opacity=1 ] (451.79,170.13) -- (635.12,170.13)(451.79,182.35) -- (635.12,182.35)(451.79,194.57) -- (635.12,194.57)(451.79,206.79) -- (635.12,206.79)(451.79,219.01) -- (635.12,219.01)(451.79,231.24) -- (635.12,231.24)(451.79,243.46) -- (635.12,243.46)(451.79,255.68) -- (635.12,255.68)(451.79,267.9) -- (635.12,267.9) ; \draw  [color={rgb, 255:red, 208; green, 2; blue, 27 }  ,draw opacity=1 ] (451.79,157.9) -- (635.12,157.9) -- (635.12,280.12) -- (451.79,280.12) -- cycle ;
\draw   (243.94,432.53) -- (243.84,441.11) .. controls (243.76,448.21) and (237.94,453.9) .. (230.85,453.82) -- (218.47,453.68) .. controls (211.38,453.6) and (205.69,447.78) .. (205.77,440.68) -- (205.77,440.68) -- (200.11,440.62) -- (208.29,432.89) -- (216.28,440.8) -- (210.63,440.74) -- (210.63,440.74) .. controls (210.58,445.15) and (214.12,448.76) .. (218.53,448.81) -- (230.9,448.95) .. controls (235.31,449) and (238.93,445.47) .. (238.98,441.05) -- (239.07,432.48) -- cycle ;
\draw  [draw opacity=0][fill={rgb, 255:red, 255; green, 255; blue, 255 }  ,fill opacity=0.27 ] (449.89,309.82) -- (633.21,309.82) -- (633.21,432.04) -- (449.89,432.04) -- cycle ; \draw  [color={rgb, 255:red, 208; green, 2; blue, 27 }  ,draw opacity=1 ] (462.11,309.82) -- (462.11,432.04)(474.33,309.82) -- (474.33,432.04)(486.55,309.82) -- (486.55,432.04)(498.77,309.82) -- (498.77,432.04)(510.99,309.82) -- (510.99,432.04)(523.22,309.82) -- (523.22,432.04)(535.44,309.82) -- (535.44,432.04)(547.66,309.82) -- (547.66,432.04)(559.88,309.82) -- (559.88,432.04)(572.1,309.82) -- (572.1,432.04)(584.32,309.82) -- (584.32,432.04)(596.55,309.82) -- (596.55,432.04)(608.77,309.82) -- (608.77,432.04)(620.99,309.82) -- (620.99,432.04) ; \draw  [color={rgb, 255:red, 208; green, 2; blue, 27 }  ,draw opacity=1 ] (449.89,322.04) -- (633.21,322.04)(449.89,334.26) -- (633.21,334.26)(449.89,346.49) -- (633.21,346.49)(449.89,358.71) -- (633.21,358.71)(449.89,370.93) -- (633.21,370.93)(449.89,383.15) -- (633.21,383.15)(449.89,395.37) -- (633.21,395.37)(449.89,407.59) -- (633.21,407.59)(449.89,419.82) -- (633.21,419.82) ; \draw  [color={rgb, 255:red, 208; green, 2; blue, 27 }  ,draw opacity=1 ] (449.89,309.82) -- (633.21,309.82) -- (633.21,432.04) -- (449.89,432.04) -- cycle ;
\draw   (416.86,603.18) -- (416.95,611.75) .. controls (417.03,618.85) and (422.85,624.54) .. (429.95,624.46) -- (442.32,624.32) .. controls (449.42,624.24) and (455.11,618.43) .. (455.03,611.33) -- (455.03,611.33) -- (460.68,611.27) -- (452.51,603.54) -- (444.51,611.45) -- (450.17,611.38) -- (450.17,611.38) .. controls (450.21,615.79) and (446.68,619.41) .. (442.27,619.46) -- (429.89,619.6) .. controls (425.48,619.65) and (421.87,616.11) .. (421.82,611.7) -- (421.72,603.12) -- cycle ;
\draw   (423,115.67) -- (436.31,115.67) .. controls (447.19,115.67) and (456.01,124.48) .. (456.01,135.36) -- (456.01,147.3) -- (461.67,147.3) -- (452.31,157.9) -- (442.95,147.3) -- (448.61,147.3) -- (448.61,135.36) .. controls (448.61,128.57) and (443.1,123.07) .. (436.31,123.07) -- (423,123.07) -- cycle ;
\draw   (463.56,773.76) -- (463.56,787.07) .. controls (463.56,797.95) and (454.74,806.77) .. (443.86,806.77) -- (431.93,806.77) -- (431.93,812.42) -- (421.32,803.07) -- (431.93,793.71) -- (431.93,799.36) -- (443.86,799.36) .. controls (450.65,799.36) and (456.16,793.86) .. (456.16,787.07) -- (456.16,773.76) -- cycle ;

\draw (150,283.91) node [anchor=north west][inner sep=0.75pt]   [align=left] {{\fontfamily{ptm}\selectfont {\small Input of metallographic scan}}};
\draw (476.33,435.04) node [anchor=north west][inner sep=0.75pt]   [align=left] {\begin{minipage}[lt]{93.23pt}\setlength\topsep{0pt}
\begin{center}
{\small {\fontfamily{ptm}\selectfont Impurities Segmentation}}\\{\small {\fontfamily{ptm}\selectfont over windows}}
\end{center}

\end{minipage}};
\draw (330.7,418.58) node [anchor=north west][inner sep=0.75pt]   [align=left] {5};
\draw (474.46,282.79) node [anchor=north west][inner sep=0.75pt]   [align=left] {{\fontfamily{ptm}\selectfont {\small Divide to sliding windows}}};
\draw (258.46,435.46) node [anchor=north west][inner sep=0.75pt]   [align=left] {{\fontfamily{ptm}\selectfont {\small Union of sliding windows}}};
\draw (39.79,435.46) node [anchor=north west][inner sep=0.75pt]   [align=left] {\begin{minipage}[lt]{109.82pt}\setlength\topsep{0pt}
\begin{center}
{\fontfamily{ptm}\selectfont {\small Impurities inpainting based }}\\{\fontfamily{ptm}\selectfont {\small on input and impurities mask}}
\end{center}

\end{minipage}};
\draw (48.46,605.46) node [anchor=north west][inner sep=0.75pt]   [align=left] {{\fontfamily{ptm}\selectfont {\small Divide to sliding windows}}};
\draw (281,605.7) node [anchor=north west][inner sep=0.75pt]   [align=left] {\begin{minipage}[lt]{68.26pt}\setlength\topsep{0pt}
\begin{center}
{\small {\fontfamily{ptm}\selectfont GB Segmentation}}\\{\small {\fontfamily{ptm}\selectfont over windows}}
\end{center}

\end{minipage}};
\draw (480,606.12) node [anchor=north west][inner sep=0.75pt]   [align=left] {{\fontfamily{ptm}\selectfont {\small Union of sliding windows}}};
\draw (478.46,777.46) node [anchor=north west][inner sep=0.75pt]   [align=left] {\begin{minipage}[lt]{91.21pt}\setlength\topsep{0pt}
\begin{center}
{\fontfamily{ptm}\selectfont {\small Watershed for boundary}}\\{\fontfamily{ptm}\selectfont {\small  completion}}
\end{center}

\end{minipage}};
\draw (110,913.91) node [anchor=north west][inner sep=0.75pt]   [align=left] {{\fontfamily{ptm}\selectfont {\small Output of final segmented metallographic scan}}};
\end{tikzpicture}
}

    \caption{Proposed segmentation pipeline. This pipeline serves, in part, as the data pre-processing to the anomaly detection pipeline.}
    \label{fig:seg_pipeline}
\end{figure}

\begin{figure}
    \centering
\tikzset{every picture/.style={line width=0.75pt}} 
\resizebox{0.7\textwidth}{!}{        

\begin{tikzpicture}[x=0.75pt,y=0.75pt,yscale=-1,xscale=1]

\draw (178.19,395.27) node  {\includegraphics[width=255.28pt,height=172.09pt]{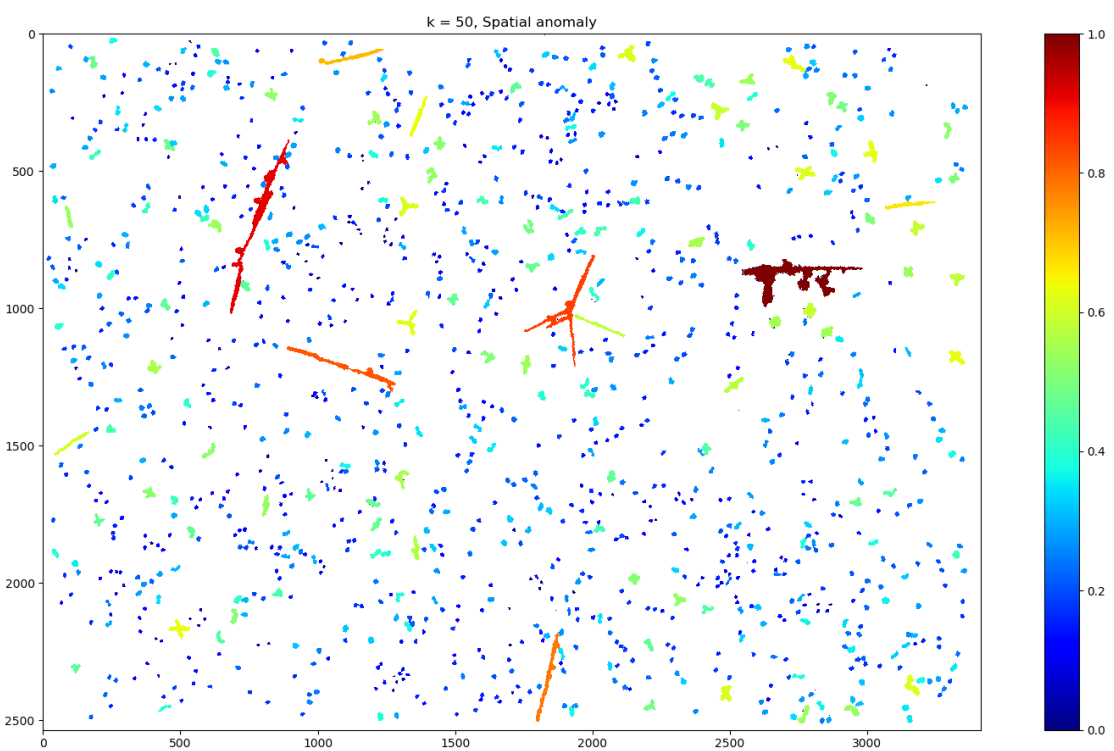}};
\draw (302.29,169.73) node  {\includegraphics[width=213.44pt,height=157.91pt, frame]{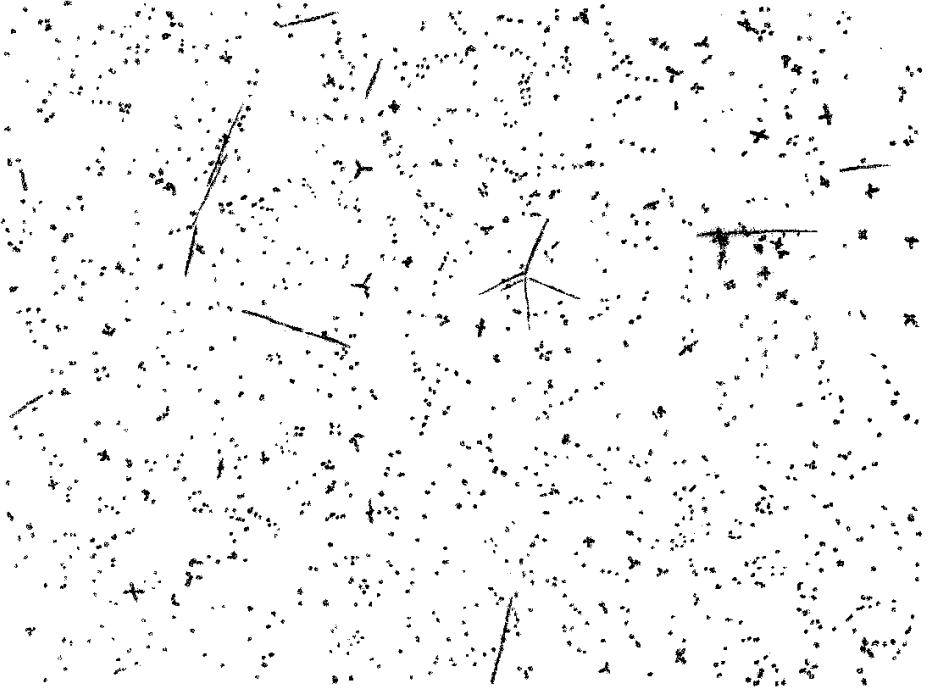}};
\draw (486.68,395.77) node  {\includegraphics[width=257.52pt,height=168.34pt]{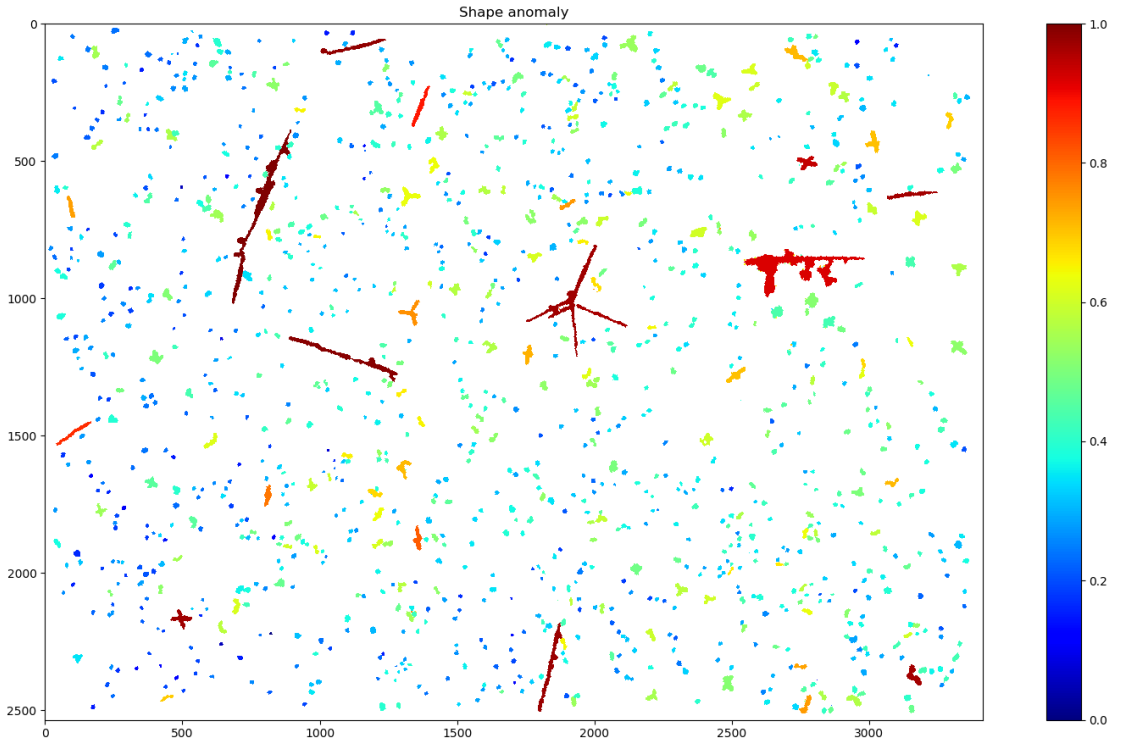}};
\draw (316.57,639.9) node  {\includegraphics[width=258.85pt,height=169.35pt]{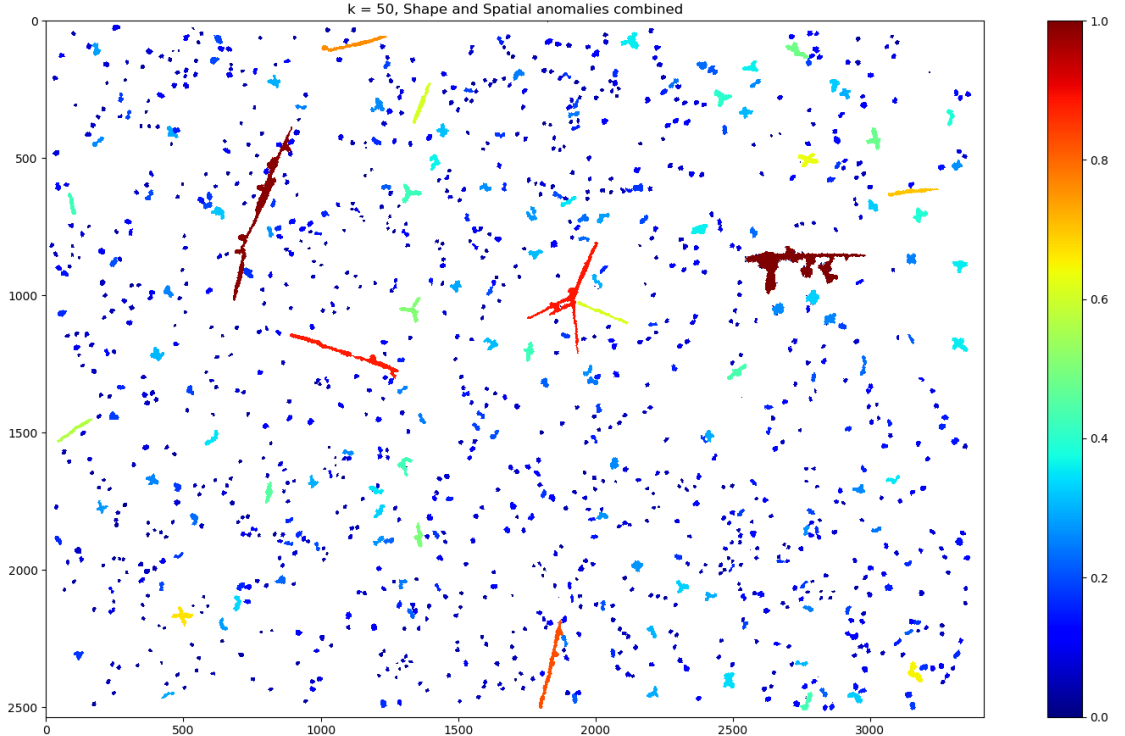}};
\draw (323.38,915.45) node  {\includegraphics[width=264.57pt,height=183.83pt]{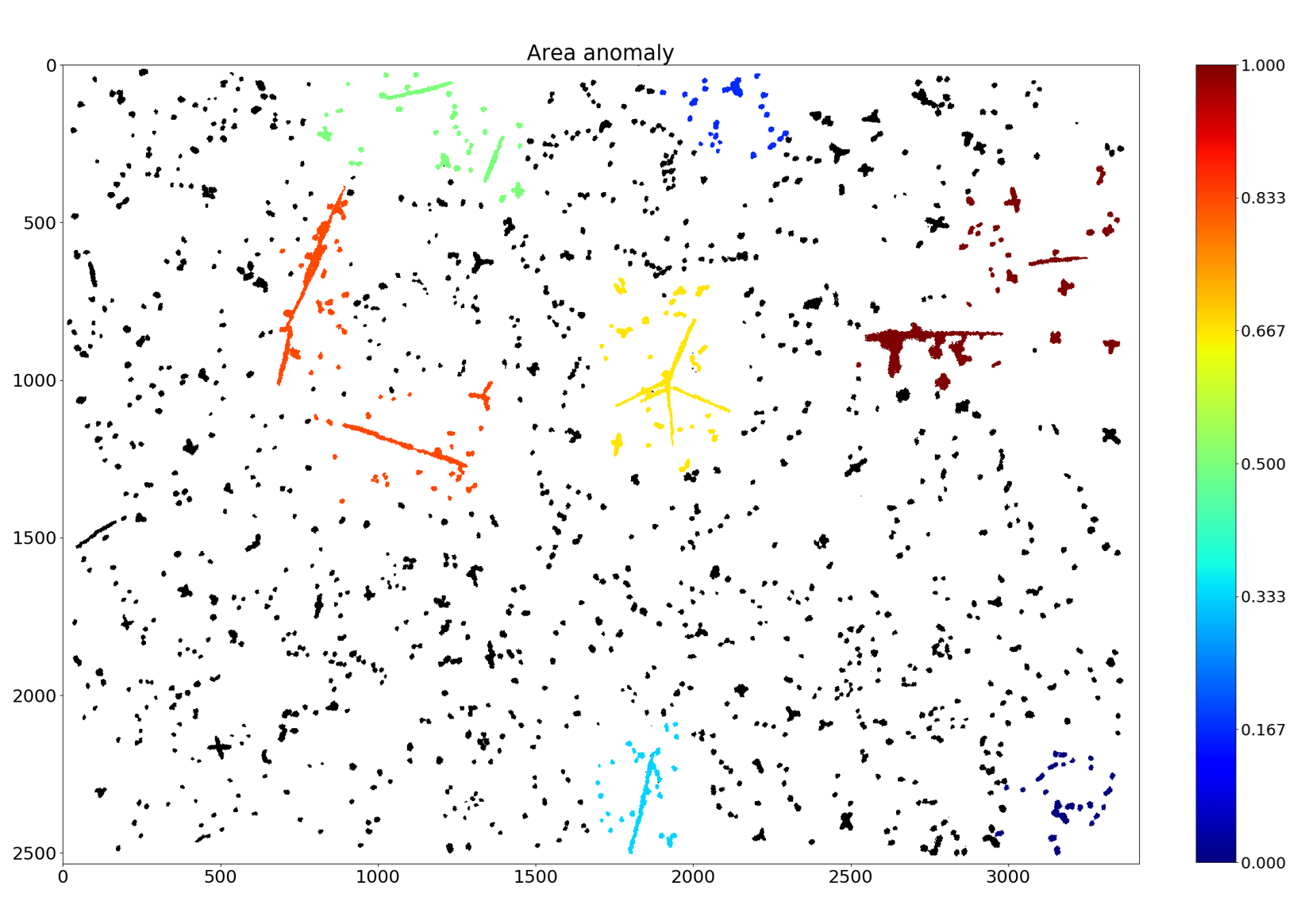}};
\draw   (445,240.67) -- (458.31,240.67) .. controls (469.19,240.67) and (478.01,249.48) .. (478.01,260.36) -- (478.01,272.3) -- (483.67,272.3) -- (474.31,282.9) -- (464.95,272.3) -- (470.61,272.3) -- (470.61,260.36) .. controls (470.61,253.57) and (465.1,248.07) .. (458.31,248.07) -- (445,248.07) -- cycle ;
\draw   (158.67,241.67) -- (145.35,241.67) .. controls (134.48,241.67) and (125.66,250.48) .. (125.66,261.36) -- (125.66,273.3) -- (120,273.3) -- (129.36,283.9) -- (138.72,273.3) -- (133.06,273.3) -- (133.06,261.36) .. controls (133.06,254.57) and (138.56,249.07) .. (145.35,249.07) -- (158.67,249.07) -- cycle ;
\draw   (534.45,532.45) -- (534.45,545.77) .. controls (534.45,556.64) and (525.63,565.46) .. (514.76,565.46) -- (502.82,565.46) -- (502.82,571.12) -- (492.21,561.76) -- (502.82,552.4) -- (502.82,558.06) -- (514.76,558.06) .. controls (521.55,558.06) and (527.05,552.56) .. (527.05,545.77) -- (527.05,532.45) -- cycle ;
\draw   (104.21,530.45) -- (104.21,543.77) .. controls (104.21,554.64) and (113.03,563.46) .. (123.91,563.46) -- (135.84,563.46) -- (135.84,569.12) -- (146.45,559.76) -- (135.84,550.4) -- (135.84,556.06) -- (123.91,556.06) .. controls (117.12,556.06) and (111.62,550.56) .. (111.62,543.77) -- (111.62,530.45) -- cycle ;
\draw  [fill={rgb, 255:red, 255; green, 255; blue, 255 }  ,fill opacity=1 ] (312,767.83) -- (316.95,767.83) -- (316.95,749) -- (325.87,749) -- (325.87,767.83) -- (330.83,767.83) -- (321.41,781.12) -- cycle ;

\draw (252,36) node [anchor=north west][inner sep=0.75pt]  [font=\large] [align=left] {{\fontfamily{ptm}\selectfont Impurities input}};
\draw (60,211) node [anchor=north west][inner sep=0.75pt]  [font=\large] [align=left] {\begin{minipage}[lt]{44.87pt}\setlength\topsep{0pt}
\begin{center}
{\fontfamily{ptm}\selectfont Spatial}\\{\fontfamily{ptm}\selectfont anomaly}
\end{center}

\end{minipage}};
\draw (486,208) node [anchor=north west][inner sep=0.75pt]  [font=\large] [align=left] {\begin{minipage}[lt]{44.87pt}\setlength\topsep{0pt}
\begin{center}
{\fontfamily{ptm}\selectfont Shape}\\{\fontfamily{ptm}\selectfont anomaly}
\end{center}

\end{minipage}};
\draw  [draw opacity=0][fill={rgb, 255:red, 255; green, 255; blue, 255 }  ,fill opacity=1 ]  (270,781) -- (379,781) -- (379,810) -- (270,810) -- cycle  ;
\draw (273,785) node [anchor=north west][inner sep=0.75pt]  [font=\large] [align=left] {{\fontfamily{ptm}\selectfont Area anomaly}};
\draw  [draw opacity=0][fill={rgb, 255:red, 255; green, 255; blue, 255 }  ,fill opacity=1 ]  (216,504) -- (419,504) -- (419,533) -- (216,533) -- cycle  ;
\draw (219,508) node [anchor=north west][inner sep=0.75pt]  [font=\large] [align=left] {\begin{minipage}[lt]{135.63pt}\setlength\topsep{0pt}
\begin{center}
{\fontfamily{ptm}\selectfont Spatial and Shape anomaly}
\end{center}

\end{minipage}};

\end{tikzpicture}

}
    \caption{Proposed anomaly detection pipeline.}
    \label{fig:anomaly_pipeline}
\end{figure}

\section{Introduction}
\vspace{-0.2cm}

\subsection{Material science and Quantitative Metallography}

\begin{figure}
    \centering
    \begin{tikzpicture}
            
            \draw[] (-1.5,4) rectangle (-0.5,3.5) node[pos=.5] {Metal};
            
            \node[inner sep=0pt] (metallography) at (-1,2) {\includegraphics[frame, width=.15\textwidth]{Segmentation/big/25.jpg}};
            
            \node[inner sep=0pt] (impurities) at (-1,-0.5) {\includegraphics[frame,width=.15\textwidth]{Segmentation/big/s25.jpg}};
            
            \node[inner sep=0pt] (inpainted_impurities) at (-4.5,2) {\includegraphics[frame,width=.15\textwidth]{Segmentation/without_impurities/25.jpg}};
            
            \node[inner sep=0pt] (gb) at (-4.5,-0.5) {\includegraphics[frame,width=.15\textwidth]{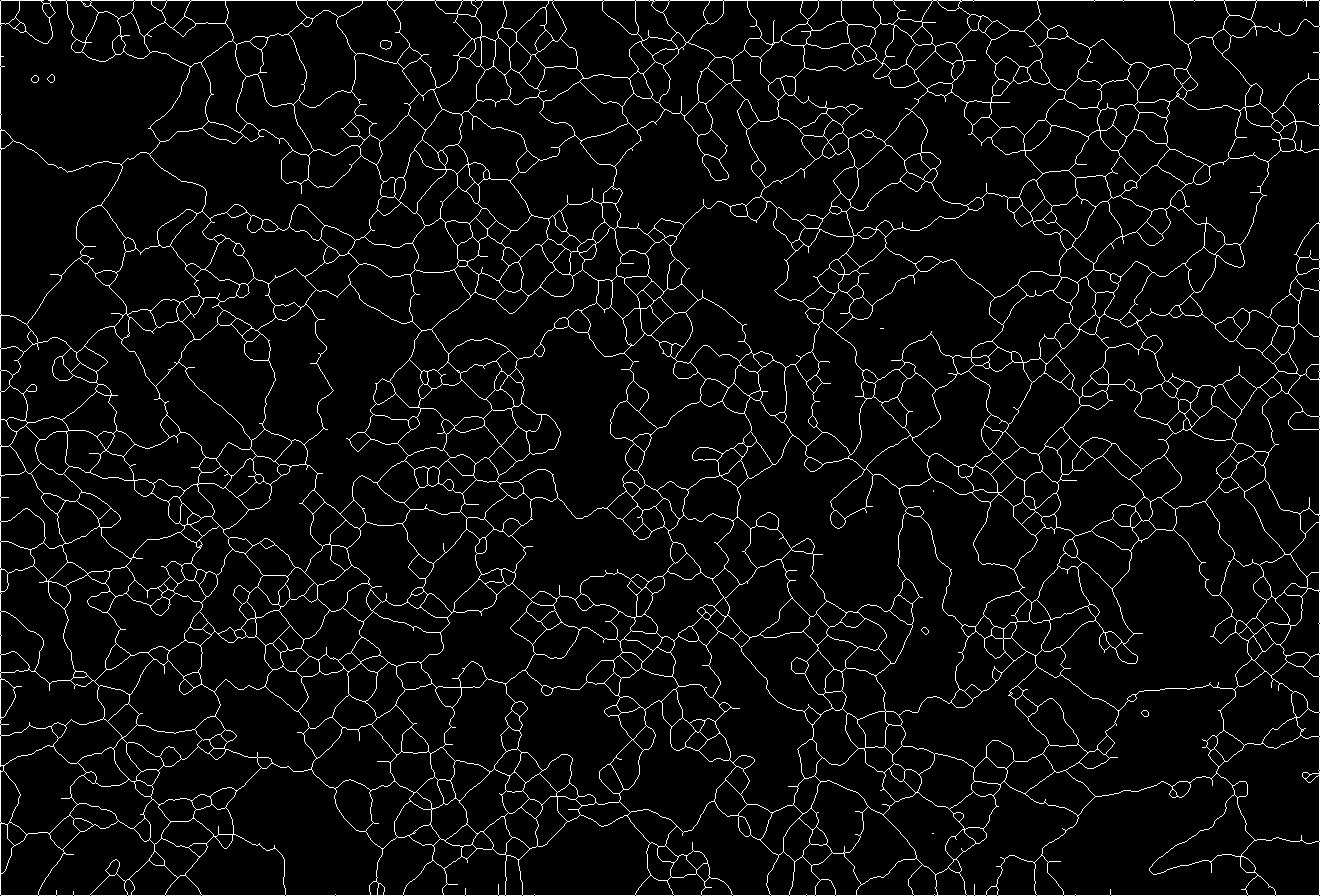}};
            
            \draw[] (-8.8,0) rectangle (-6.5,-1) node[pos=.5, align=left] {Grains'\\ sizes};
            
            \draw[] (-8.8,2.5) rectangle (-6.5,1.5) node[pos=.5, align=left] {Grains below \\ threshold?};
            
            \draw[] (-8.8,4.25) rectangle (-6.5,3.25) node[pos=.5, align=left, color=green] {OK};
            
            \draw[] (-5.65,4.25) rectangle (-3.35,3.25) node[pos=.5, align=left, color=red] {Suspicious};

            \node[inner sep=0pt] (anomaly) at (2.5,2) {\includegraphics[frame,width=.15\textwidth]{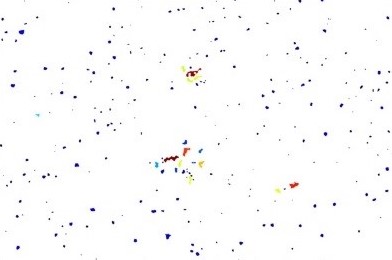}};
            
            \node[inner sep=0pt] (area_anomaly) at (2.5,-0.5) {\includegraphics[frame,width=.15\textwidth]{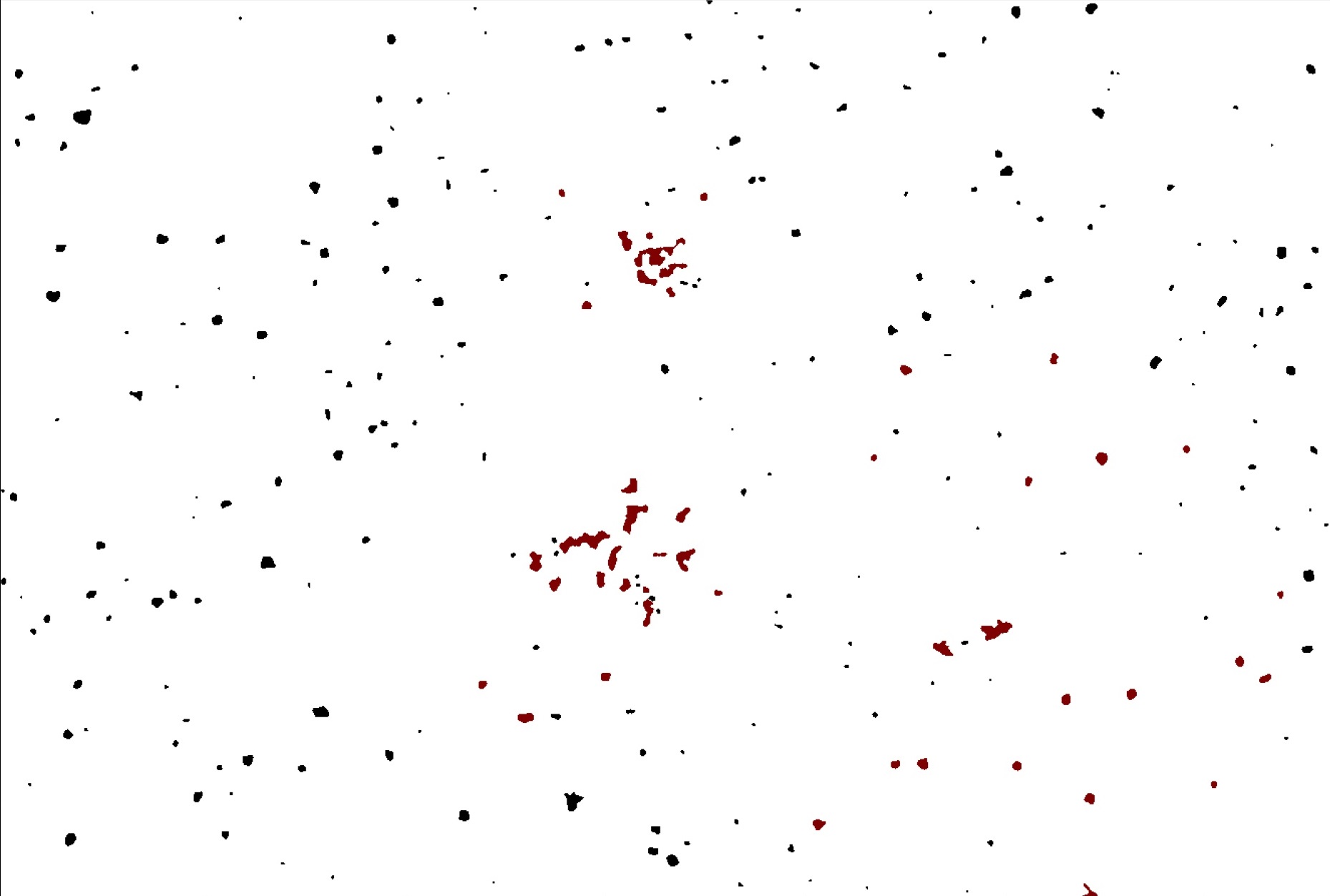}};
            
            \draw[] (4.5,0) rectangle (6.8,-1) node[pos=.5, align=left] {Clusters with \\ anomaly scores};
            
            \draw[] (4.5,2.5) rectangle (6.8,1.5) node[pos=.5, align=left] {Clusters below \\ threshold?};
            
            \draw[] (4.5,4.25) rectangle (6.8,3.25) node[pos=.5, align=left, color=green] {OK};
            
            \draw[] (1.35,4.25) rectangle (3.65,3.25) node[pos=.5, align=left, color=red] {Suspicious};

            \draw[->, anchor=south] (-1,3.5) -- node[right, scale=0.8] {$1$} (metallography.north);
            
            \draw[->, anchor=south] (metallography.south) -- node[right, scale=0.8] {$2$} (impurities.north);
            
            \draw[->, anchor=south] (metallography) -- node[above, scale=0.8] {$3$} (inpainted_impurities);
            
            \draw[->, anchor=south] (impurities) -- node[right=0.1cm, scale=0.8] {$3$} (inpainted_impurities);
            
            \draw[->, anchor=south] (inpainted_impurities) -- node[right, scale=0.8] {$4$} (gb);
            
            \draw[->, anchor=south] (impurities) -- node[right=0.1cm, scale=0.8] {$5$} (anomaly);
            
            \draw[->, anchor=south] (anomaly) -- node[right, scale=0.8] {$6$} (area_anomaly);
            
            \draw[->, anchor=south] (gb) -- node[above, scale=0.8] {$7.1$} (-6.5,-0.5);
            
            \draw[->, anchor=south] (-7.65, 0) -- node[right, scale=0.8] {$8.1$} (-7.65,1.5);
            
            \draw[->, anchor=south] (-7.65, 2.5) -- node[right=0.15cm, scale=0.8] {$9.1$} (-7.65,3.25);
            
            \draw[->, anchor=south] (-7.65, 2.5) -- node[] {} (-5.65,3.25);
            
            \draw[->, anchor=south, bend right=45] (-3.35, 3.75) -- node[below, scale=0.8] {$10.1$} (-1.5,3.75);
            
            \path[every node/.style={font=\sffamily\small}]
            (-1.5,3.75) edge[->, bend right=45] node [below, scale=0.8] {$11.1$} (-6.5,3.75);

            \draw[->, anchor=south] (area_anomaly) -- node[above, scale=0.8] {$7.2$} (4.5,-0.5);
            
            \draw[->, anchor=south] (5.65, 0) -- node[right, scale=0.8] {$8.2$} (5.65,1.5);
            
            \draw[->, anchor=south] (5.65, 2.5) -- node[left=0.15cm, scale=0.8] {$9.2$} (5.65,3.25);
            
            \draw[->, anchor=south] (5.65, 2.5) -- node[] {} (3.65,3.25);
            
            \draw[->, anchor=south, bend right=45] (1.35, 3.75) -- node[below, scale=0.8] {$10.2$} (-0.5,3.75);
            
            \path[every node/.style={font=\sffamily\small}]
            (-0.5,3.75) edge[->, bend left=45] node [below, scale=0.8] {$11.2$} (4.5,3.75);
            
    \end{tikzpicture}
    \caption{The workflow of our proposed end-to-end model for QM on a given input metal. First, in $1$, metallographic imaging is performed. Then, in $2$, semantic segmentation for impurities is carried out. Using the metallographic image and the mask of segmented impurities, the impurities are inpainted in $3$, creating a 'clean' metallographic image. Next, grains' boundaries are semantically segmented over the metallographic image without the impurities in $4$. In $5$, spatial and shape anomaly detection measures are calculated from the segmented impurities, and in $6$, the anomalies are clustered based on area anomaly. In $7.1$, the grains' sizes are calculated, and in $7.2$, each cluster of anomalous impurities receives a numeric value representing its anomaly score. In $8.1$ and $8.2$, statistics based on these values are calculated and compared against pre-set or learned corresponding thresholds. These thresholds determine in $9.1$, $9.2$ whether the input metal is OK (under the threshold) or suspicious (over the threshold). If one of the values exceeds the threshold, a suitable physical test is performed ($10.1$, $10.2$) on the input metal in the advised area from the model. If the physical test establishes no fault in the material, the sample is set as OK and the corresponding threshold is tuned respectively in $11.1$ and $11.2$.}
    \label{fig:system_workflow}
\end{figure}

Material science is the study of material properties, which is based on, for example, understanding how it is influenced by its chemical composition, microstructure, and manufacturing process \cite{sinha2003physical}.
Metallography is the study of the microstructure of metallic alloys, or more generally of any material, in length scales usually ranging from nanometers to millimeters. Investigating the microstructure of a material allows one to discover important properties of that material, such as mechanical properties \cite{wang1995effect, naghizadeh2019effects, armstrong1970influence}, corrosion behavior \cite{ralston2010effect, ralston2011effect, brunner2012impact}, electrical properties \cite{zeng2013grain, ivanov2015grain, andrews1969effect}, etc. Microstructure study is therefore considered to be among the most beneficial and effective fields for understanding materials' properties.

Different techniques \cite{donald1994science} are used to reveal various microstructural features of metals. Most investigations are carried out with incident light microscopy dedicated to metallography (metallographic microscope), which can operate in different modes such as bright field, dark field, polarized light, etc. Another common technique for metallographic investigation is Scanning Electrons Microscopy (SEM). This method is based on electron beam emission from an electron gun which interacts with the sample, then monitoring the different signals resulting from this interaction, mostly backscattered electrons, secondary electrons, and characteristic x-ray.

Metallography is usually performed on a flat sample sizing from a few millimeters to a few centimeters, and its' preparation includes mechanical grinding on silicon-carbide polishing paper and final polishing on clothes soaked with diamond paste. Sometimes this preparation is not sufficient for revealing the features of interest, and some extra preparation is required (such as chemical or electro-chemical etching or controlled oxidation).

Many correlations between microstructure and macroscopic properties have been explained using metallography. A notable example of such correlation is the Hall-Petch strengthening mechanism \cite{liu2003normal, volpp1997grain, naik2020hall} in which there is a negative correlation between the yield strength of a material and the square root of its' average grain size. Another example is the correlation between the nature and concentration of inclusions in some material to its' mechanical endurance under different loads, such as under quasi-static loading \cite{wu2003effects, thornton1971influence, wu2003effects} or fatigue (cyclic loading) \cite{wu2003effects, meurling2001influence}. The ability of a material to hold quasi-static loading is relevant to all aspects of life: From the capability of buildings and bridges to endure loads to the strength of a kitchen knife. Fatigue is an important issue when dealing with moving parts such as shafts and motors \cite{rankine1843causes, braithwaite1854fatigue}, and it has crucial safety aspects, for example, in the case of airplane wings and landing gear \cite{bagnoli2008fatigue, franco2006fatigue}.
Another important macroscopic property is anisotropy. In most cases, isotropic properties (uniform behavior in all directions) are desired, but in some cases, anisotropy (material’s tendency to react differently to stresses applied in different directions) is preferred. An example of such a case is jet engine blades, in which the ability to hold longitudinal (radial) load is more important than in the transverse direction. 

As the material properties are highly affected by its microstructure, and the demand for superior materials' properties constantly increases for edge usage, safety reasons, or economic constraints, the need for tide controlling of the microstructure is increased as well. Hence, metallography is used in almost all stages during the lifetime of a component: From the initial materials development to inspection, production, manufacturing process control, and even failure analysis if required. The principles of metallography help to ensure product reliability.

Features of interest mainly and re-daily studied in metallography are the grains' spatial distribution and the occurrence and characteristics of inclusions or precipitates.

Grains are areas in the sample where the atoms are arranged in a specific crystallographic orientation. Grains are originated from phase transition, e.g., solidification, in which different areas in the material start to transform in different orientations and are growing until filling the entire volume of the material, or from recrystallization, where new grains of the same phase are growing on the expense of old ones. There are various reasons for the spatial distribution of grains. Different growing kinetics in different crystallographic orientations during phase transition and mechanical and thermo-mechanical shaping are among those. The interface between two adjacent grains is called a grain boundary (GB), and often it is the GB’s that are seen in the metallographic image.

Inclusions and precipitates are small material areas that differ from their surrounding matrix either by composition, crystallographic structure, or both. Inclusions are originated from outside the sample, e.g., oxide particles that were added to the melt before solidification, while precipitates are formed from the sample itself, e.g., carbides formation in steels. Henceforth, we will define both of them as inclusions or impurities. Inclusions hold essential information that will be considered in this paper, including the nature of each inclusion (composition, crystallographic structure, size, and shape) as well as more general information such as the surface concentration of inclusions (on a cross-section) and their distribution.

During the development of new material, metallographic characterization is done comprehensively and deeply, while during routine manufacturing, it is usually done only once in a while to ensure the manufacturing stability \cite{astm20113}. In both cases, but more often during routine manufacturing, the metallographic images are analyzed by their similarity to previous samples \cite{standarde112}. This similarity is determined by Quantitative Metallography (QM) parameters such as grain size and the surface concentration of inclusions and also by 'softer' parameters such as shape and distribution of the inclusions, which are usually based on an 'expert opinion' \cite{astm199745}. Based on computational algorithms, such an approach is less preferred than a more objective and quantitative one. Moreover, the expert is not always a qualitative source, as the ability to detect anomalies and analyze the material state quantitatively is a very complex task, and as such, it inherently requires and should heavily rely on computations. 

\begin{figure}
\centering
\begin{minipage}{0.33\textwidth}
\subcaptionbox{Example scan.\label{fig:example_scan}}{\includegraphics[height=6.5cm, frame]{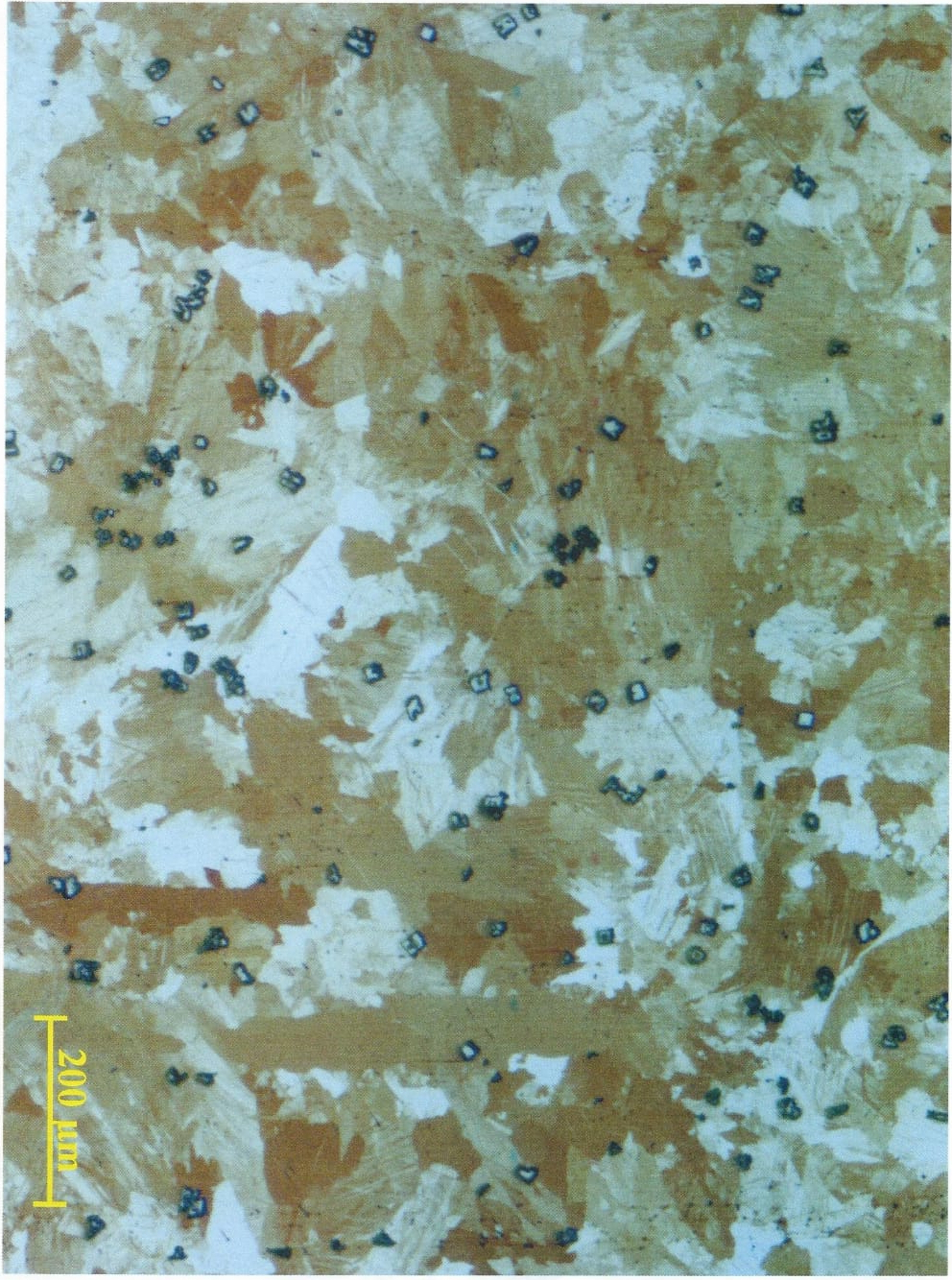}}
\end{minipage}
\begin{minipage}{0.33\textwidth}
\subcaptionbox{Example slide.\label{fig:example_tag}}{\includegraphics[height=6.5cm, frame]{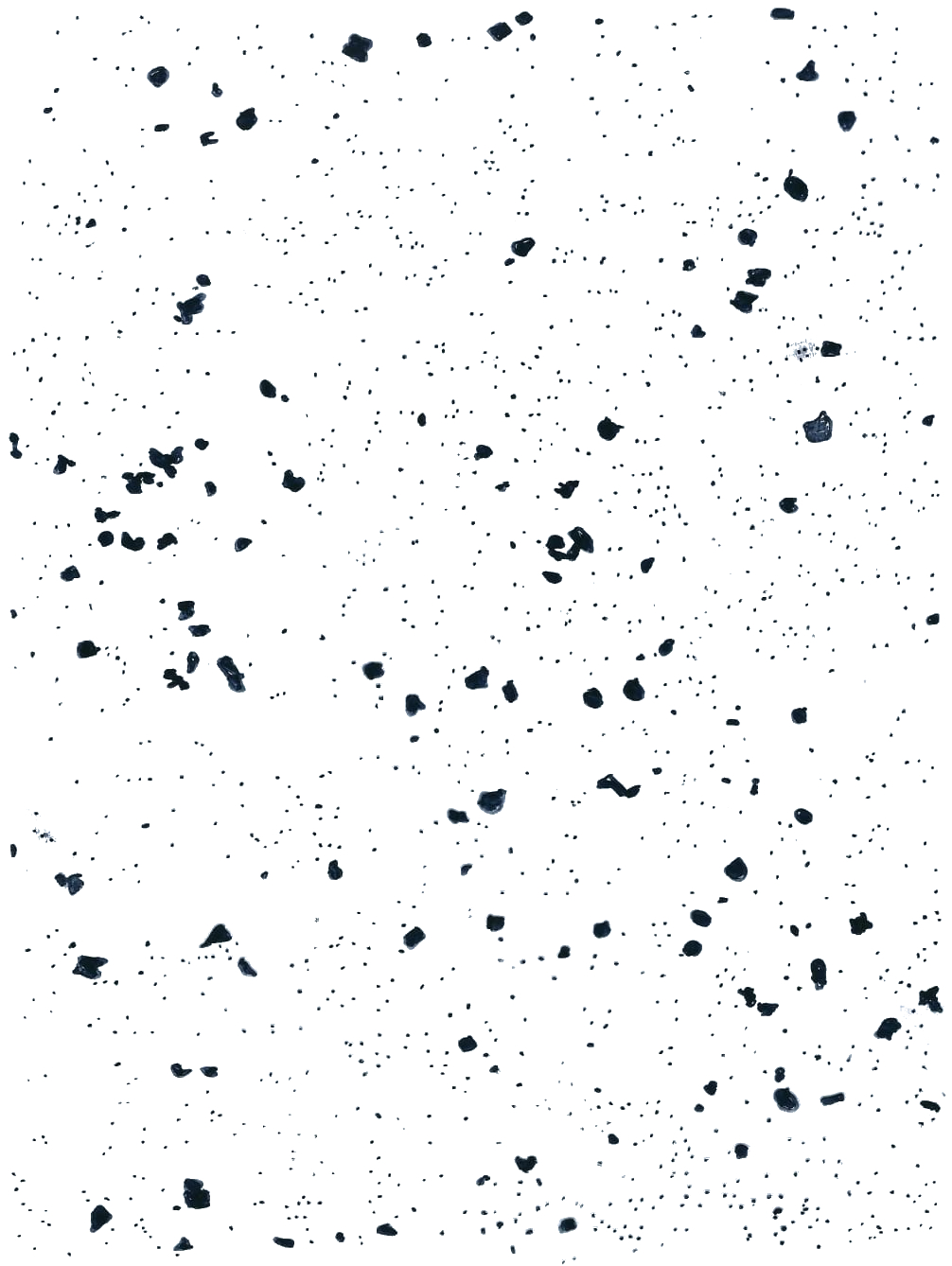}}
\end{minipage}
\begin{minipage}{0.33\textwidth}
\subcaptionbox{Scan with tags.\label{fig:example_scan_tag}}{\includegraphics[height=6.5cm, frame]{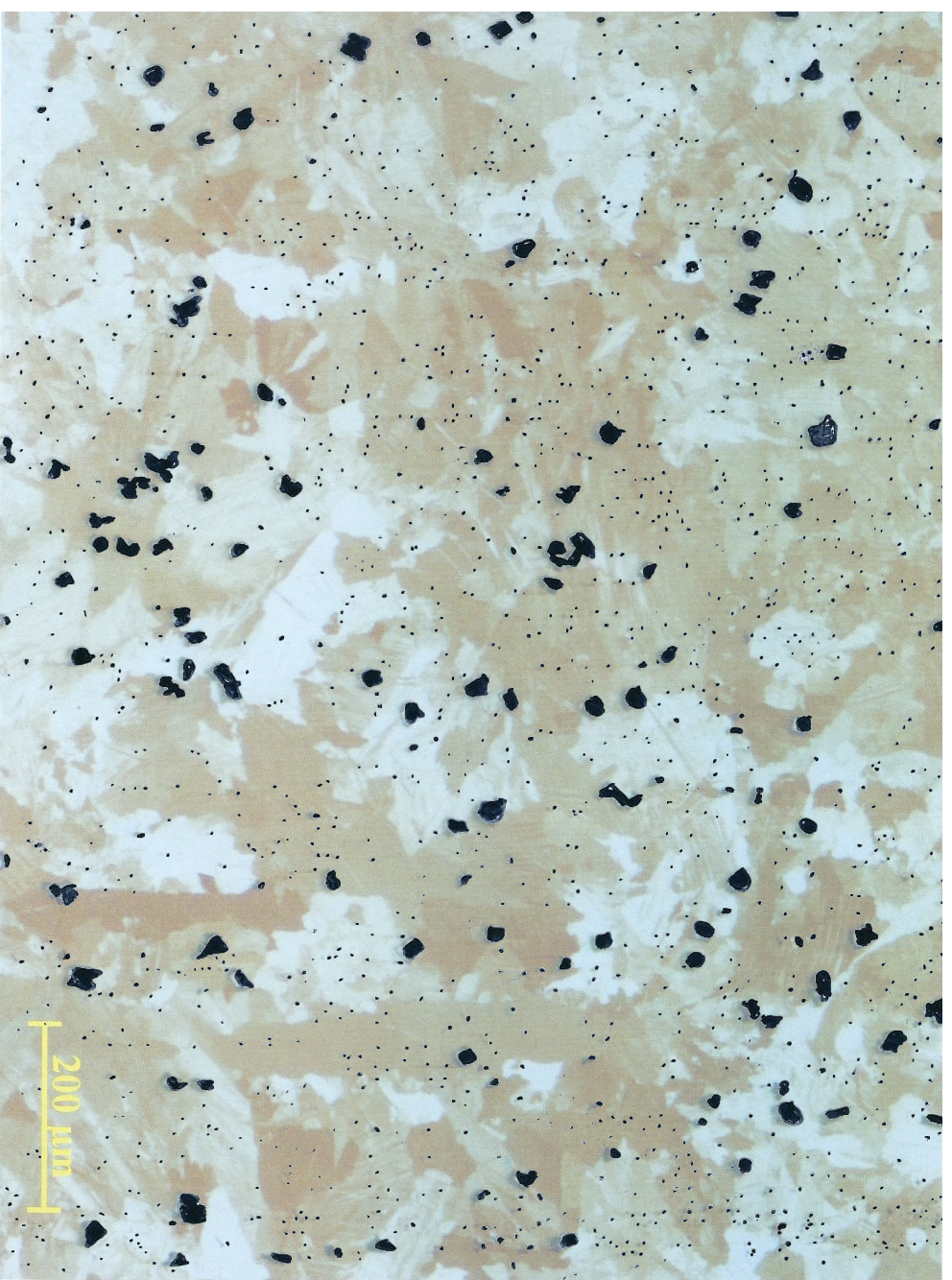}}
\end{minipage}
\caption{Example of metallographic sample with impurities tags.}
\end{figure}

However, for using these algorithms, it is first necessary to identify each relevant feature in the metallographic picture, specifically the inclusions and the GB’s.
In many cases, this is not a trivial task. For example, in Fig. \ref{fig:example_scan} a metallographic image of U-0.1\%Cr sample is shown (this alloy is used as a nuclear fuel, and as such, the importance of repeatable manufacturing is substantial). It can be seen that the contrast between the inclusions and the matrix is not uniform and that the GB’s are not easily noticed. Because of this complexity, classic image processing tools did not yield satisfactory results, and hence the identification of these features is made manually by an expert as an input for the computational analysis. This procedure is very tedious and time-consuming and does not allow the analysis of a high number of pictures for improved statistics.



\subsection{Previous Work}

As metallographic imaging heavily relies on semantic understanding of complex combinations of colors, boundaries, structures, and their corresponding distributions, many image processing and computer vision techniques were introduced over the last decade to extract the needed characteristics automatically. The need for such automation is well motivated from the material science perspective \cite{iskakov2020framework, duval2014image, ramprasad2017machine, dimiduk2018perspectives, decost2020scientific, ge2020deep, kesireddy2015application}, as well as from the perspective of microscopic imaging in general \cite{xing2017deep, moen2019deep}. In relatively-simple cases, classic algorithms for image pre-processing, pixel classification, region extraction, and edge detection, occasionally combined with data mining methods, proved to yield moderate to good results \cite{han2019segmenting, chen2014watershed, zhang2019aluminum, liu2012improved, peregrina2013automatic, marin2017automatic, wu2020study}. However, as the image complexity and resolution increases, and the image domain varies, the error rate in the needed tasks using those methods increases as well, thus demanding human expertise in tuning said algorithms \cite{han2019segmenting}. For example, in cases where segmentation depends on the color histogram or some hyperparameters of other edge detection algorithms, the expert is required to tune the algorithm until reaching the desired result \cite{han2019segmenting}. 

Nevertheless, as many alloys -- even of the same materials and under similar conditions -- exhibit extreme variance in colors, boundaries, structures, and corresponding distributions. Therefore the tasks of classifying the different objects (grains and inclusions) and segmenting those objects correctly (where and how to define the borders) almost completely depends on human expertise, which often cannot precisely determine the accurate classification nor the border edges \cite{american2013e112, iso2012643, astm2015e1382}. Thus, in recent years the introduction of computer vision-based on deep learning (CVDL) techniques \cite{decost2015computer, azimi2018advanced, decost2018high, ma2020deep, kondo2017microstructure} to the QM field is growing, as those techniques proved to minimize the error rate of previously exclusive human tasks, and sometimes to even introduce new capabilities. For example, in the task of image inpainting (inpainting of inclusions) -- by using generative models such as Generative Adversarial Networks (GANs) \cite{ma2020deep}.

The need for such segmentation and knowledge of one sample's properties is crucial, and many features can be extracted that way to understand the state of the material. However, reliable statistics, anomaly detection, and patterns recognition cannot be gained sufficiently in this form, and there is a need to quantify the material's features and metric their behavior over a considerable amount of samples \cite{underwood1985quantitative}, which is far greater than human comprehensibility in the age of Big Data \cite{habeeb2019real}. For example, material scientists often rely on some quantifications in order to provide a quantitative explanation of how anomalous each impurity/whole sample is, such as the distribution of grains' size or the total sum of impurities' area, and comparing those to previously gained results \cite{lehto2014influence, hall1954variation}. While those are essential features, they hardly represent the material status, which should be quantified for each object in terms of spatial, shape, and area distribution in the lone sample and the entire history combined \cite{vander1984metallographic, beeley2001foundry}. 

This methodical and statistical observation of the materials' properties is also crucial in the perspective of Quality Control (QC) and Quality Assurance (QA) in order to meet repeatable manufacturing standards \cite{10.31399/asm.hb.v10.9781627082136}. The goal of quality and reliability assurance in materials' research and development is the continuous improvement of performance, efficiency, and ease-of-use of the formed material \cite{10.31399/asm.hb.v10.9781627082136}. The acquisition and analysis of quantitative observations made during the process of production allow it to be continuously monitored and maintained to meet the desired specifications, or ultimately rapidly diagnosed and remedied if problems arise and the results deviate from the expected ones \cite{10.31399/asm.hb.v10.9781627082136}. As metallography plays a key role in understanding materials and alloys' behavior under production, the need to automatically and reliably diagnose metallography while comparing it to the previous diagnosis and quantify the changes is necessary. Those requirements can only be met using novel data mining and machine learning approaches -- specifically in the field of anomaly detection and pattern recognition -- which can automatically gain the needed insights in a reliable fashion \cite{zhang2020towards}.

\subsection{Contribution}
The benefits of our proposed methodology versus the above-mentioned existing works are rooted, first and foremost, in the ability to bind in a pipeline fashion all of the building blocks of automatic QM while entirely relying on artificial intelligence in general and CVDL in particular. Specifically, we propose a segmentation pipeline in Fig. \ref{fig:seg_pipeline} and an anomaly detection pipeline in Fig. \ref{fig:anomaly_pipeline}. The segmentation pipeline consists of (1) deep semantic segmentation; (2) deep image inpainting, resulting in 'clean' metallographic images; and (3) grains' boundaries deep semantic segmentation. The impurities anomaly detection pipeline consists of (4) spatial anomaly detection; (5) shape anomaly detection; and (6) area anomaly detection.
To the best of our knowledge, this is the first introduction of such a pipeline, including a vast state-of-the-art human expertise labeled QM dataset based on actual metallurgy (see Subsection \ref{data}). This pipeline was established while reinventing each process step technology to adjust the problem and even devise new computer science methods. Among those, we include the weighted variant of \textit{K$^{th}$-Nearest-Neighbor}, a novel training methodology for Auto-Encoders that favors normal examples while discriminating abnormal ones, and the novel \textit{Market-Clustering} anomaly detection algorithm, that are all based on the physical distribution and properties on the material. Moreover, our usage of immensely complicated images, with many different colors, shapes, and textures, while using only a tiny portion of this data for training  (about 1\%), strongly suggests that those findings can be replicated easily for any other material with minimal effort. An additional novel concept is our end-to-end QM model's ability to constantly calibrate itself with respect to new experiments and not solely rely on previous knowledge (Fig. \ref{fig:system_workflow}). This way, we enhance and adjust the pipeline regarding the actual state in the lab or the factory.

\subsection{Dataset Creation}\label{data}

While there are many datasets of electron microscopy, almost none of them is suited for machine learning purposes \cite{morgan2022machine,ede2021deep}, and virtually none of those is suited to the goal of a unified automated approach to the entire metallographic investigation process (Fig. \ref{fig:system_workflow}), including labeling of impurities and GB's over different and complex backgrounds. This state is not a surprise, as the common hypothesis among material science practitioners is that this task is inherently tricky and prone to ambiguation \cite{holm2020overview}: The data itself is not natural, includes many textures, and primarily represents a 2D cross-section of a 3D reality (see Appendix A for more examples). Furthermore, as only experts can label this data, the labeled data is naturally scarce, and there is no real option to reach labeled big data. As a result, our goal in this work was to establish an end-to-end model on \textit{as minimal as possible labeled images}. Iteratively, we added batches of labeled images until reaching the desired performance, all of this while designing the model as a few-shot model. In fact, only 1\% of the data that eventually been evaluated was labeled. Even in comparison to U-net initial dataset \cite{ronneberger2015u} we used half as much labeled data, while our complexity is much greater. We managed to do so by the usage of sliding window over the scans, redesigning U-net (see Section \ref{imps-seg}), and performing physics-oriented post-process (see Section \ref{gb_seg}).

The outcome of this iterative process resulted in the open-sourced \textit{MLography} dataset, which includes extremely detailed metallographic images, with all of the features for QM labeled by experts. Specifically, we used the U-0.1wt\%Cr alloy, which is used as a nuclear fuel. In this alloy, the most abundance inclusion is \textit{Uranium Carbide}, which appears on the 2D metallographic cross-section as dots, spots, or long rods, depending on the impurity concentration in the alloy and the thermal profile during casting and cooling to room temperature \cite{nomine1974physical}. For this work, metallographic images (approximately 1.2mm $\times$ 0.8mm) were used (see example in Fig. \ref{fig:example_scan}). An expert tagged each inclusion and its boundaries on it (see example in Figures \ref{fig:example_tag}, \ref{fig:example_scan_tag}). As these kinds of datasets are rarely public, except for a few exceptions \cite{decost2017uhcsdb}, we have created a novel dataset of manual tags of impurities from 243 metallographic scans for anomaly detection, and we make it publicly available at \cite{MLography_tags_png_copped}. We present our anomaly detection results on a sample image of tagged impurities from a uranium-chromium alloy scan from the dataset in Figures \ref{fig:spatial} -- \ref{fig:area}. Additionally, we have created two datasets for metallographic semantic segmentation, consisting of cropped squares (128 $\times$ 128 pixels) of metallographic scans with corresponding manual tags of impurities as ground truth (261 images in \cite{MLography_impurities_seg}), as well as a dataset of cropped squares with manual tags of grains' boundary as ground truth (320 images in \cite{MLography_gb_seg}). We also provide unlabeled 32 big metallographic images in \cite{MLography_big} and present the output of the segmentation pipeline in \cite{MLography_full_seg}.

\subsection{Paper Organization}

The rest of the paper is organized in accordance to the methodology pipeline: Section \ref{seg-pipeline} presents the segmentation pipeline including impurities’ segmentation (subsection \ref{imps-seg}); generative inpainting over the segmented impurities, creating 'clean' metallographic images (subsection \ref{imps-inpaint}); and grains’ boundary segmentation on the 'clean' images (subsection \ref{gb_seg}). In Section \ref{Segmentation-Pipeline-Evaluation}, we evaluate this entire pipeline as a whole. Next, in Section \ref{anomaly_detection_measures}, we introduce new anomaly detection measures for metallography. Finally, in Section \ref{ex-co}, we suggest a new \textit{physical} evaluation to the presented methodology using both computerized and mechanical measures.

\section{Segmentation pipeline} \label{seg-pipeline}
\subsection{Impurities' Segmentation} \label{imps-seg}

As we stated, the above anomaly detection measures are only applicable to image samples representing a slide containing the tagged impurities. The two main approaches for achieving these images are: Providing manual and pricey expert-made tags; Generating machine-automotive tags from metallographic scans.
The latter might be a much more efficient alternative as it allows the experts to save precious time for other work while completing the task of tagging the impurities in a much faster way. However, it is risky to produce non-accurate impurities tags, as this is a delicate task that involves determining whether each pixel is an impurity or a part of the background.

The above task is called Semantic Segmentation, and it is a well-studied topic in Computer Vision. Due to the emergence of Deep Convolutional Neural Networks (CNN) in recent years, lots of CNN models were proposed \cite{long2015fully}, achieving state-of-the-art performance for segmentation. U-Net \cite{ronneberger2015u} is an example of such a model that is specifically tailored for the segmentation of microscopic biomedical images that resemble our images' characteristics. U-Net's architecture consists of an Encoder that extracts and progresses the underlying abstract information from the images to the Decoder. The latter utilizes the \textit{global} information (low-resolution details such as structures) by up-sampling the latent information from the encoder and the \textit{local} information (high-resolution fine details such as textures) directly from the corresponding layers from the encoder. This 'skip' mechanism allows learning deep semantic information while preserving high-resolution shallow information that might get lost due to down-sampling and up-sampling that CNNs require for the training process.

\begin{figure}
\centering
\begin{minipage}{0.49\textwidth}
\centering
  \subcaptionbox{Training loss. Impurities Segmentation in blue, GB segmentation in orange.\label{fig:seg_loss}}{\includegraphics[height=6cm]{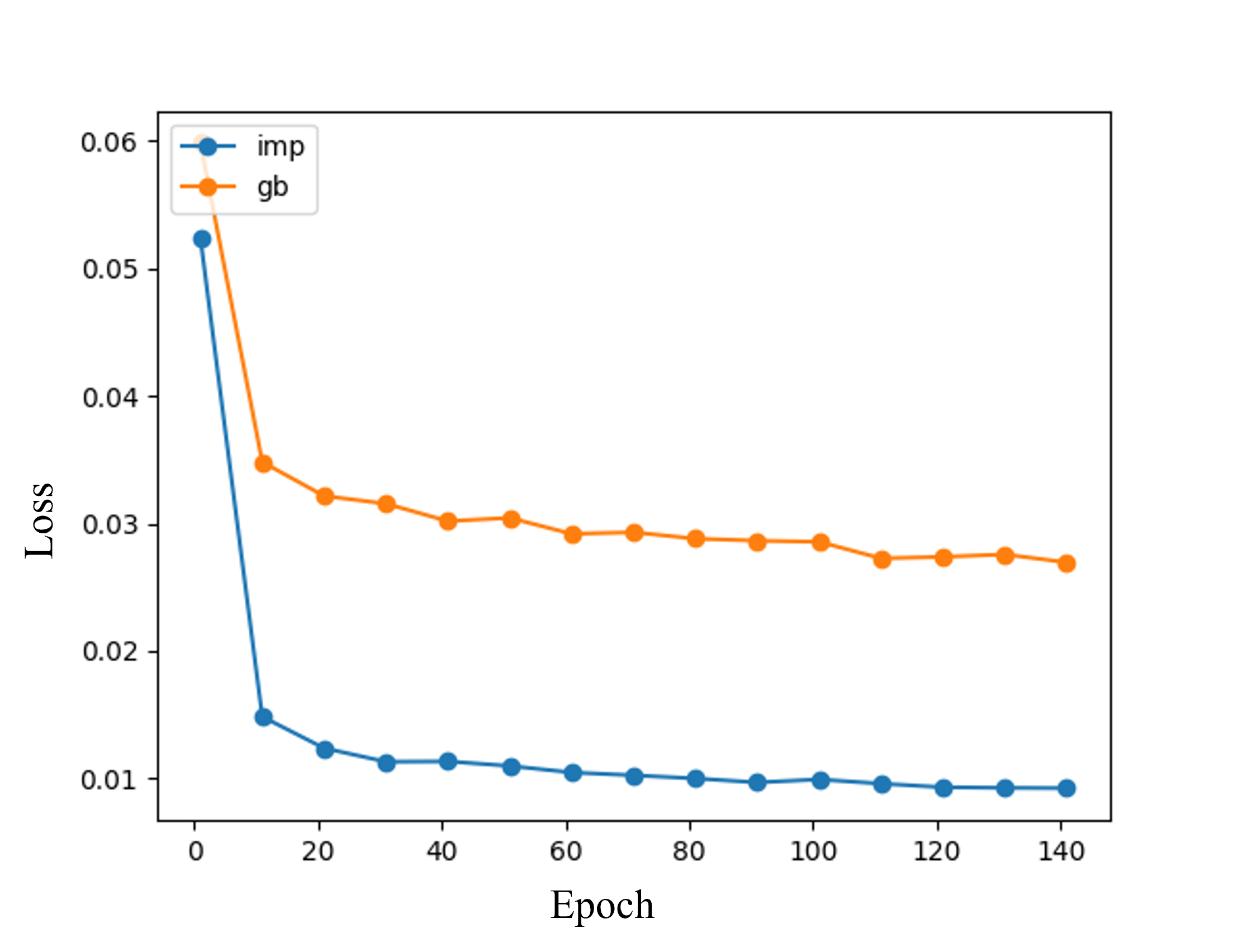}}
\end{minipage}
\begin{minipage}{0.49\textwidth}
\centering
  \subcaptionbox{Training accuracy. Impurities Segmentation in blue, GB segmentation in orange.\label{fig:seg_acc}}{\includegraphics[height=6cm]{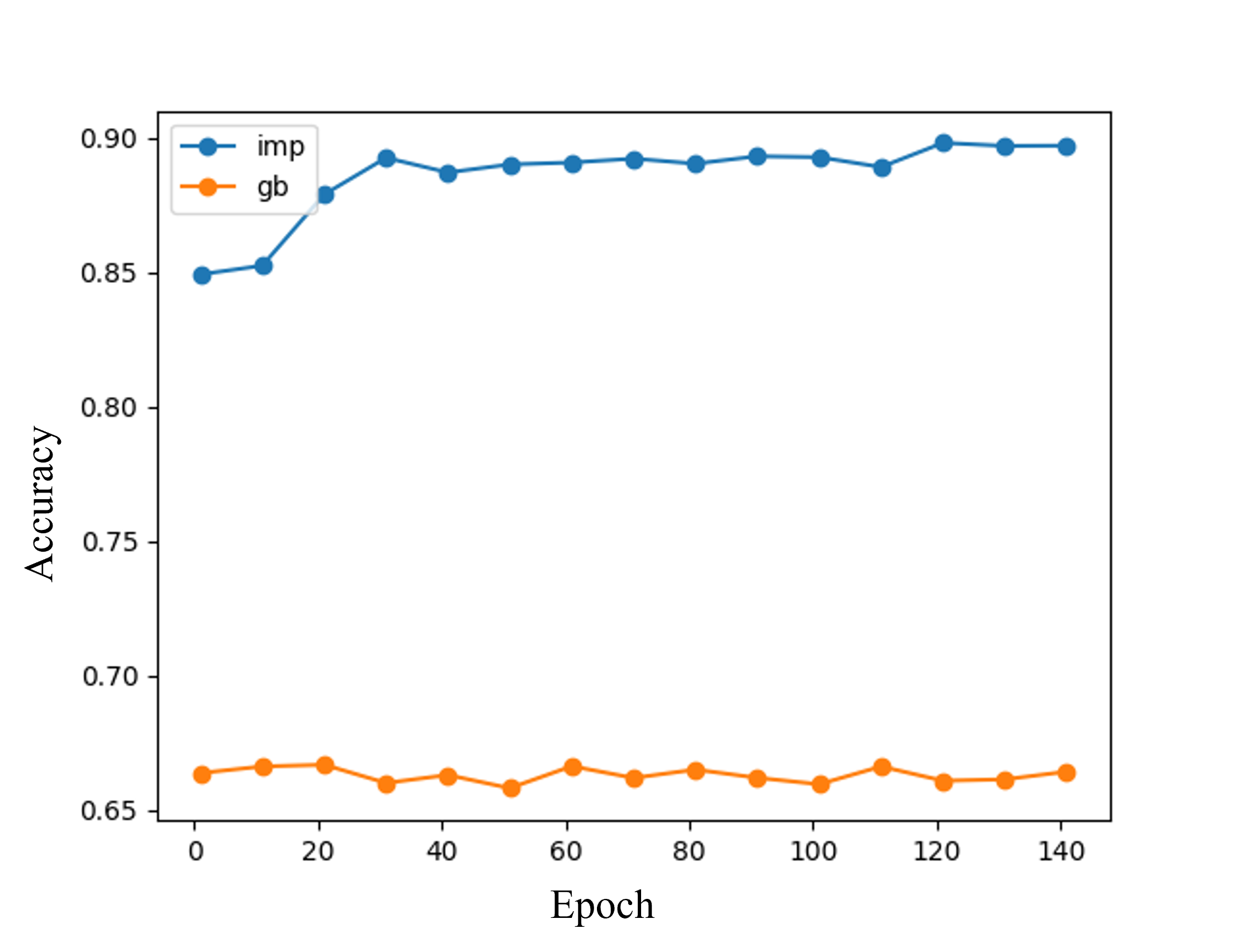}}
\end{minipage}
 \caption{Training loss and accuracy of the segmentation tasks.}
\end{figure}

U-Net highly inspires the architecture of our model, but it has some modifications. First, we use VGG16 \cite{simonyan2014very}, that was pre-trained on ImageNet as the encoder (similarly to \cite{iglovikov2018ternausnet}). It receives colored input images of size $128 \times 128$, and it passes the latent information to a symmetric decoder network, making the entire model look like a U-shaped network with a total number of 27 convolutional layers. It was optimized using Adam and with the focal loss function \cite{lin2017focal} with $\alpha = 0.2$ in order to penalize cases in which the model misclassifies a pixel as a background. We trained the model on small images ($128 \times 128$) that were labeled using \cite{maninis2018deep, papadopoulos2017extreme}, for 150 epochs -- 300 steps in each. Training loss and accuracy trends are presented in blue in Figures \ref{fig:seg_loss}, \ref{fig:seg_acc}.

\begin{figure}
    \centering
    \begin{minipage}{.33\linewidth}
    \centering
    \subcaptionbox{Window \#1.\label{fig:window1}} {%
      \includegraphics[height=4.2cm, frame]{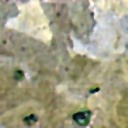}
    }%
    \end{minipage}
    \begin{minipage}{.33\linewidth}
    \centering
    \subcaptionbox{Window \#2.\label{fig:window2}} {%
      \includegraphics[height=4.2cm, frame]{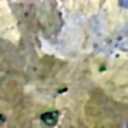}
    }%
    \end{minipage}
    \begin{minipage}{.33\linewidth}
    \centering
    \subcaptionbox{Window \#3.\label{fig:window3}} {%
      \includegraphics[height=4.2cm, frame]{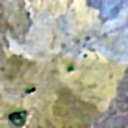}
    }%
    \end{minipage}
    \par\bigskip
    \begin{minipage}{.33\linewidth}
    \centering
    \subcaptionbox{Segmentation of Window \#1.\label{fig:segmented_window1}} {%
      \includegraphics[height=4.2cm, frame]{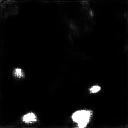}
    }%
    \end{minipage}
    \begin{minipage}{.33\linewidth}
    \centering
    \subcaptionbox{Segmentation of Window \#2.\label{fig:segmented_window2}} {%
      \includegraphics[height=4.2cm, frame]{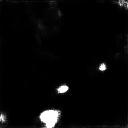}
    }%
    \end{minipage}
    \begin{minipage}{.33\linewidth}
    \centering
    \subcaptionbox{Segmentation of Window \#3.\label{fig:segmented_window3}} {%
      \includegraphics[height=4.2cm, frame]{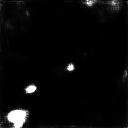}
    }%
    \end{minipage}
    \caption{Squares of impurities and their segmentation.}
\end{figure}

In order to be able to segment images bigger than the small images that the network was trained on, we propose the following methodology that extends the compatibility of the model to images of any dimensions and improves the performance of the model significantly. We suggest using a sliding window with the same size as the inputs of the network, with \textit{high-overlap} in order to minimize noise and to increase the certainty of the segmentation. This way, each window is segmented using the network, and each pixel in the output of the full segmented image is determined via an average of all pixels in the corresponding segmented windows that contain him. As we can see in Fig. \ref{fig:segmented_window2} the network can not infer the true nature of the objects in the borders of the images since it does not have the 'full picture.' Therefore, we see that the networks have small confidence that impurities may reside in the borders. Using the high-overlapping sliding-window methodology helps with this issue: the right edge of an impurity can be seen in the left border of Fig. \ref{fig:segmented_window2}. Then, since several windows will contain this impurity in its entirety (e.g., Fig. \ref{fig:segmented_window1}), they should agree on segmenting it, and thus after averaging the pixels of all segmented windows, this impurity should be classified as an impurity. Following a similar logic, noise reduction might be achieved since if there is no real impurity in the left border of the window, the same windows should agree via \textit{majority vote} on not segmenting any of the corresponding pixels. The results of the model on Several examples are presented in Figures \ref{fig:window1}-\ref{fig:segmented_window3}. For fast applications, we suggest tuning the stride of the sliding windows, reducing the overlap, and lowering the running time of the application. However, this running-time reduction might come with performance degradation, since \textit{majority vote} is crucial, especially if the network was trained upon such a small dataset as in our case. Therefore, we suggest the user address this trade-off and set the stride parameter according to the application.

\begin{figure}[H]
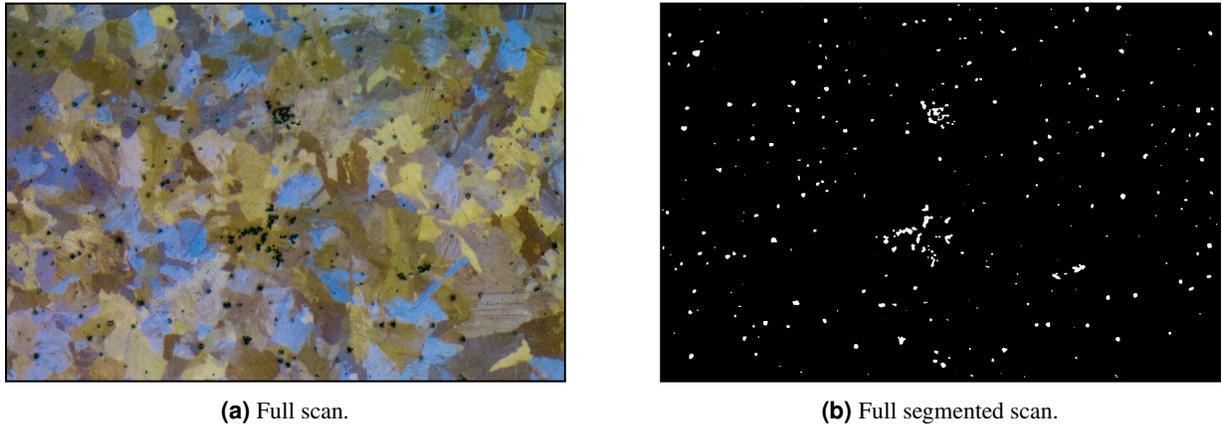

\centering
\begin{minipage}{0.49\textwidth}
\centering
  \subcaptionbox{Full scan.\label{fig:full}}{\includegraphics[height=5cm, frame]{Segmentation/big/25.jpg}}
\end{minipage}
\begin{minipage}{0.49\textwidth}
\centering
  \subcaptionbox{Full segmented scan.\label{fig:segmented_full}}{\includegraphics[height=5cm, frame]{Segmentation/big/s25.jpg}}
\end{minipage}
\caption{Impurities segmentation.}
\end{figure}

In order to increase the model segmentation resolution, we suggest focusing the model on finer details by zooming-in the picture prior to the methodology described above. The final segmentation result using windows of size $128\times 128$, 8 pixels offset between every two consecutive windows, and a zoom-in factor of 3 on some input scan (Fig. \ref{fig:full}) can be seen in Fig. \ref{fig:segmented_full}.

\textit{\textbf{Evaluation.}} In order to evaluate the proposed U-Net architecture, we trained another network with the same architecture on 95\% of the dataset (247 squares), leaving a test set of 14 squares. The Receiver Characteristic Operator -- Area Under the Curve (ROC-AUC) score is 0.96, implying that the network classifies pixels very well. The mean Intersection over Union (IoU) score of the impurities' bounding boxes is 0.73. IoU severely penalizes misclassifications of a few pixels since the final score is normalized over the union of segmented pixels, and the number of pixels that correspond to impurities is usually tiny compared to the size of the image. Moreover, in most cases of QM, it is much more crucial not to neglect big impurities than to suffer from misclassifications of a few pixels. We present an alternative measure that counts the number of intersections of the bounding boxes of the impurities over the number of impurities (maximum between the number of impurities in the prediction and the ground truth). This measure can be interpreted as Object Intersection over Union (OIoU), suitable for object localization. The mean OIoU on our test set is 0.85. The mean percentage of impurities from the squares in the predicted images is 8, while the ground truth is: 7.

\subsection{Impurities' Inpainting} \label{imps-inpaint}
Once the exact placements of the impurities from a given metallographic scan are achieved, one can try to fill them with synthetic 'normal' parts such that the new image will resemble a realistic full metallographic scan without any impurities. Several works purpose solution for this task \cite{liu2018image, yu2018generative}. We tested both models and found the pre-trained generative inpainting model \cite{yu2018generative} to be very effective on the impurities masks (after applying a dilation filter for border expansion). The result on the full metallographic image is presented in Fig. \ref{fig:without_impurities}, while zoomed-in images representing a small portion of the image with and without impurities are presented in Fig. \ref{fig:inpaint_zoom}. Although the result seems to be satisfactory, the network struggles to complete small grains (areas of the same color) that are shadowed almost entirely by impurities. For example, a small blue grain in the center of the images in Fig. \ref{fig:inpaint_zoom} has disappeared and 'collided' into the yellow grain below it. This issue might be fixed by optimizing the network with a tailor-made annotated dataset. However, as we will later explain, small modifications in the borders of the grains, or even collisions of small grains into other grains, usually are not meaningful since metallographic examination of GBs often involves statistical analysis that neglects small differences.

\textit{\textbf{Evaluation.}} Since the generative inpainting model was not trained on our metallographic data, we could test its performance on it entirely. We applied the model on 32 couples of metallographic images, $I_i$, and their corresponding impurities masks, $M_i$. For each couple of $I_i, M_i$, the model generated clean metallographic images: $C_i$.
Then, random crops of $512 \times 512$ pixels were taken from $C_i$ and $M_i$, denote these crops as $C^{'}_{i}$ and $M^{'}_{i}$ respectively.
The model returned a prediction $P_i$ for each couple $C^{'}_{i}$, $M^{'}_{i}$.
We report the mean Peak Signal-to-Noise Ratio (PSNR) value between all $P_i$ and $C^{'}_{i}$ to be 34.68, while in \cite{yu2018generative} the reported PSNR value was 18.91. A possible explanation for the higher performance in our case is that the impurities mask spans a relatively much smaller region than in \cite{yu2018generative}.

\begin{minipage}{.48\textwidth}
  \centering
  \begin{figure}[H]
  \centering
      \includegraphics[height=5cm, frame]{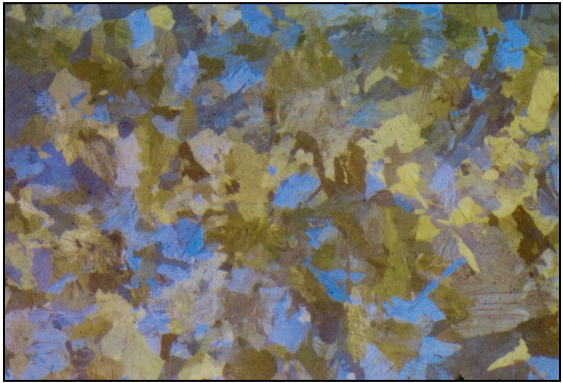}
      \caption{Inpainted impurities.}\label{fig:without_impurities}
  \end{figure}
\end{minipage}
\begin{minipage}{.5\textwidth}
\begin{figure}[H]
  \centering
    \subcaptionbox{With impurities.} {%
      \includegraphics[height=4cm, frame]{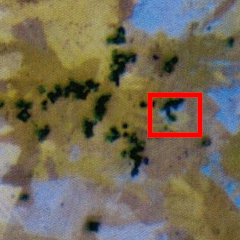}
    }%
    \subcaptionbox{Without impurities.} {%
      \includegraphics[height=4cm, frame]{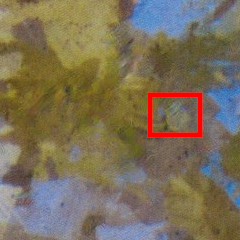}
    }%
                      
            
              
    \caption{Zoomed portion with and without impurities.The blue grain in the red square disappeared in the inpainted image.}\label{fig:inpaint_zoom} 
\end{figure}
\end{minipage}%

\subsection{Grains' boundary Segmentation} \label{gb_seg}
Another critical aspect of metallography is the study of the spatial distribution of grains.
Similar to the task of anomaly detection for impurities, one must first detect the grains or their GBs as a preliminary stage. However, differently from the former case, impurities' segmentation deals with localization and segmentation of pixels belonging to the class of impurities. Impurities might be characterized by several discriminative properties, e.g., continuous untextured zones of pixels with a dark homogeneous color in relation to grains' heterogeneous (color-wise and texture-wise) background. On the other hand, GB segmentation deals with a much more complicated problem. Unlike learning discriminative features of impurities, the objective in GB segmentation requires learning and identifying differences between adjacent grains, specifically regarding their edges. Since grains come with a wide range of colors and textures, achieving a generalization is more challenging.

Edge detection is a well-formulated problem analogous to GB segmentation since, in both cases, the aim is to identify points in a digital image at which there are discontinuities or sharp changes in image intensity. However, grains might accommodate sharp intrinsic intensity changes that should not be considered as edges since they are defined as parts of the grain. Moreover, these intrinsic discontinuities can sometimes be even more intense than the discontinuities found on the GB. Therefore, na\"ive image processing approaches that search for intensity changes without a deep understanding of the image, such as Canny \cite{canny1986computational} were found to be not satisfactory for GB segmentation. On the contrary, W-net \cite{xia2017w} is a self-supervised deep neural network that is trained to segment edges based on labels that represent sharp intensity changes, who found to be not satisfactory for GB segmentation for the same reasons as above. We also tested DexiNed \cite{poma2020dense} which is a state-of-the-art edge detection model that was pre-trained on a dataset of natural images with human labels named BIPED. Although DexiNed is optimized to understand deep features from images for edge detection, we found it not suitable for GB segmentation.

\begin{figure}[H]
\centering
\begin{minipage}{0.495\textwidth}
\centering
  \subcaptionbox{Full raw GB segmentation.\label{fig:full_gb}}{\includegraphics[height=5cm, frame]{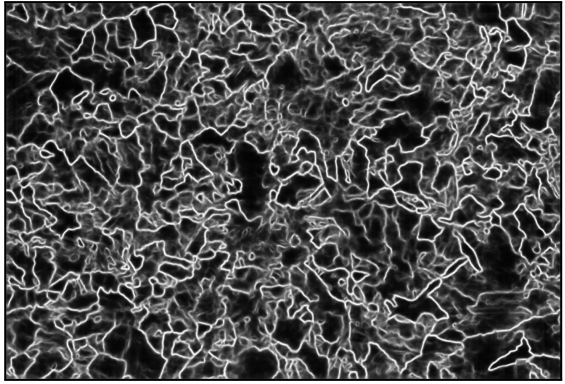}}
\end{minipage}
\begin{minipage}{0.495\textwidth}
\centering
  \subcaptionbox{Binarized raw GB segmentation.\label{fig:bin_full_gb}}{\includegraphics[height=5cm, frame]{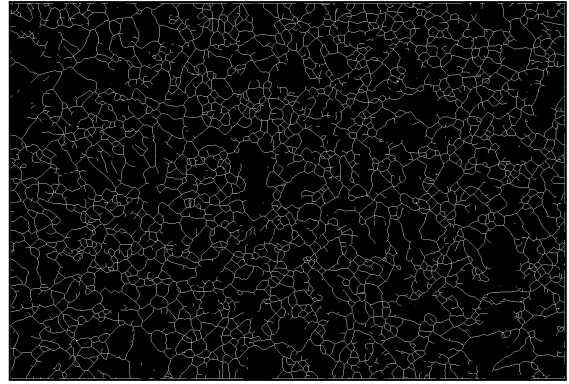}}
\end{minipage}
\par\bigskip
\begin{minipage}{0.495\textwidth}
\centering
  \subcaptionbox{Watershed contours after binarization.\label{fig:post_full_gb}}{\includegraphics[height=5cm, frame]{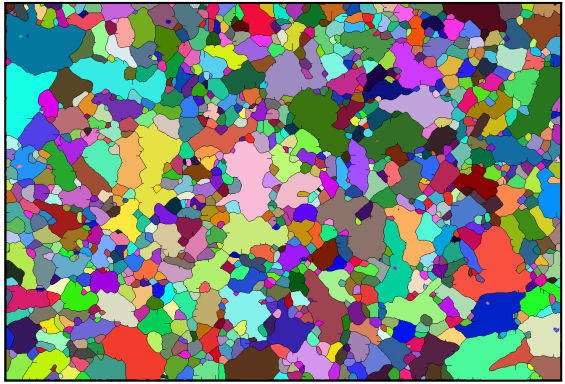}}
\end{minipage}
\begin{minipage}{0.495\textwidth}
\centering
  \subcaptionbox{Full GB segmentation.\label{fig:masked_full_gb}}{\includegraphics[height=5cm, frame]{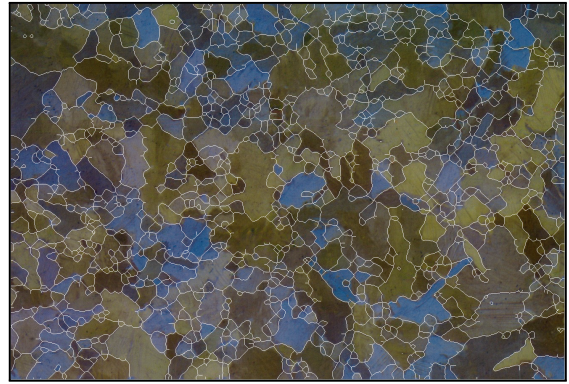}}
\end{minipage}
\caption{GB segmentation.}
\end{figure}

From the reasons above, we generated a dataset with manually annotated GBs and optimized a U-net model with the same architecture as in Section \ref{imps-seg} against it. Training loss and accuracy trends are presented in orange in Figures \ref{fig:seg_loss}, \ref{fig:seg_acc}. The raw output of the model on Fig. \ref{fig:without_impurities} as an input, is presented in Fig. \ref{fig:full_gb}. However, since the model's output is a prediction on each pixel, there is often a variance on the predicted values on different boundaries and even on the same consecutive boundary 'line.' As a result, some predicted 'boundaries' appear as incomplete lines. Since these incomplete lines represent the uncertainty of the model, we interpret them as noise. We stress that this decision is based on the working methodology of material scientists. That is to say, if material scientists were to mark any suspicious boundary (and not only the 'certain' boundaries) -- these incomplete lines would not have been neglected and treated as noise.

In order to suppress the noise, a few post-processing steps are required. A binarization, followed by Guo-Hall thinning \cite{guo1989parallel} steps are applied (Fig. \ref{fig:bin_full_gb}). Then, Watershed algorithm \cite{beucher1979use} is used to segment grains with complete boundaries. This step allows us to eliminate incomplete lines and persist only 'certain' boundaries. The contours that were generated from the Watershed algorithm are presented in Fig. \ref{fig:post_full_gb}, and the input image marked with the post-processed boundaries is presented in Fig. \ref{fig:masked_full_gb}.

We note that from the perspective of repeatable manufacturing and the need for reproducibility, uniformity in the metallographic examination process must be achieved. An agreed method of quantifying a metallographic sample is by inspecting the distribution of grains' size \cite{american2013e112}. Therefore, the benefits of the proposed procedure are quantitative and qualitative: It shortens the time-consuming, demanding, and most of the time \textit{subjective} task of manually annotating GBs, but it also regulates and standardizes the process based on 'few' commonly accepted training examples of annotated GB's. 

\textit{\textbf{Evaluation.}} Like the impurities segmentation task, we trained another network with the same architecture on 95\% of the dataset (304 squares), leaving a test set of 16 squares. The ROC-AUC score is 0.84, and the IoU is 0.68, showcasing the complexity of the GB segmentation task compared to the impurities segmentation.

\section{Segmentation Pipeline Evaluation} \label{Segmentation-Pipeline-Evaluation}

Different from standard evaluation procedures of classic machine learning tasks such as classification, segmentation, and inpainting -- that were previously described with their corresponding evaluations metrics individually -- there is also a need to evaluate the fusion of those models for the physical domain original purposes. Specifically, it would be of interest to evaluate the assembly of some or all of the presented algorithmic techniques as a QM process. That is, to test the decomposition of a given metallographic input, on which one or several algorithmic steps are applied (i.e., impurities and GB segmentation), followed by recomposition into some metallographic insight. For the segmentation task, we can examine the composition of the three models of Sections \ref{imps-seg}--\ref{gb_seg} and their post-process procedure to match the guidelines of current standards in the field \cite{astm20113, standarde112, astm199745, american2013e112, iso2012643, astm2015e1382}.

\begin{figure}
\begin{minipage}{0.5\textwidth}
  \subcaptionbox{Test input image.\label{fig:input_seg_test}}{\includegraphics[height=5cm, frame]{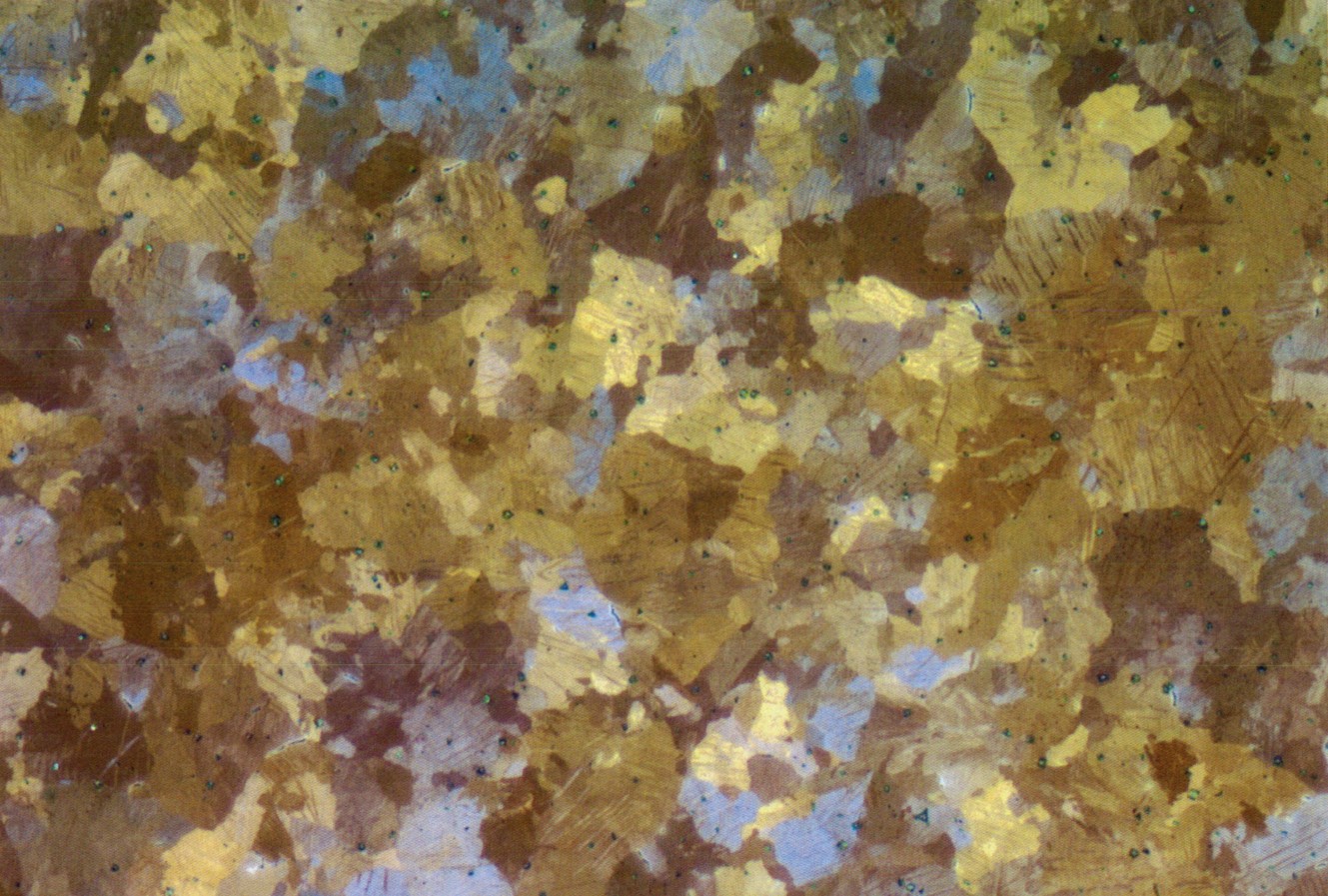}}
\end{minipage}
\begin{minipage}{0.5\textwidth}
  \subcaptionbox{Final output.\label{fig:masked_42}}{\includegraphics[height=5cm, frame]{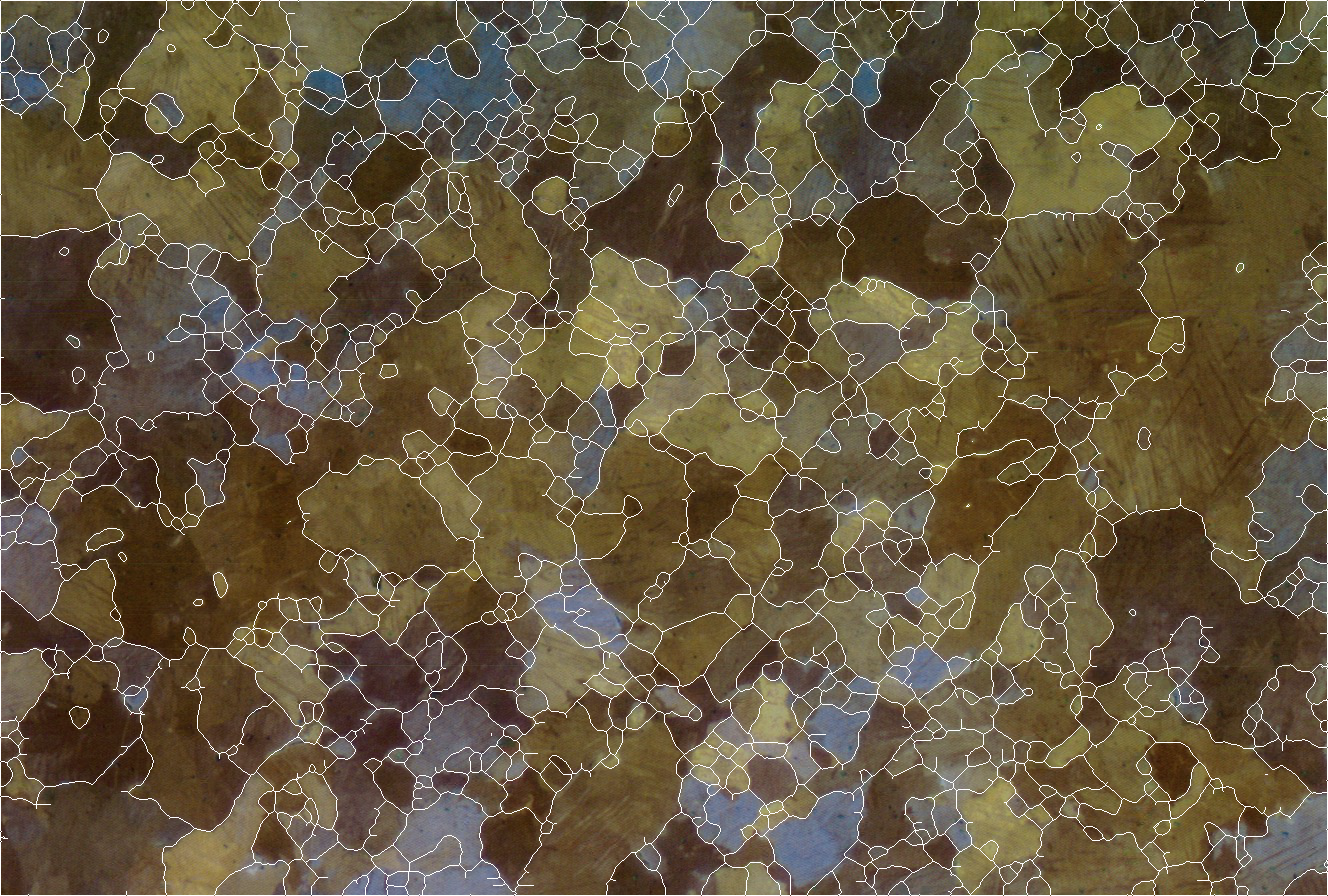}}
\end{minipage}
\par\bigskip
\begin{minipage}{0.5\textwidth}
  \subcaptionbox{Impurities ground truth.\label{fig:imp_gt}}{\includegraphics[height=5cm, frame]{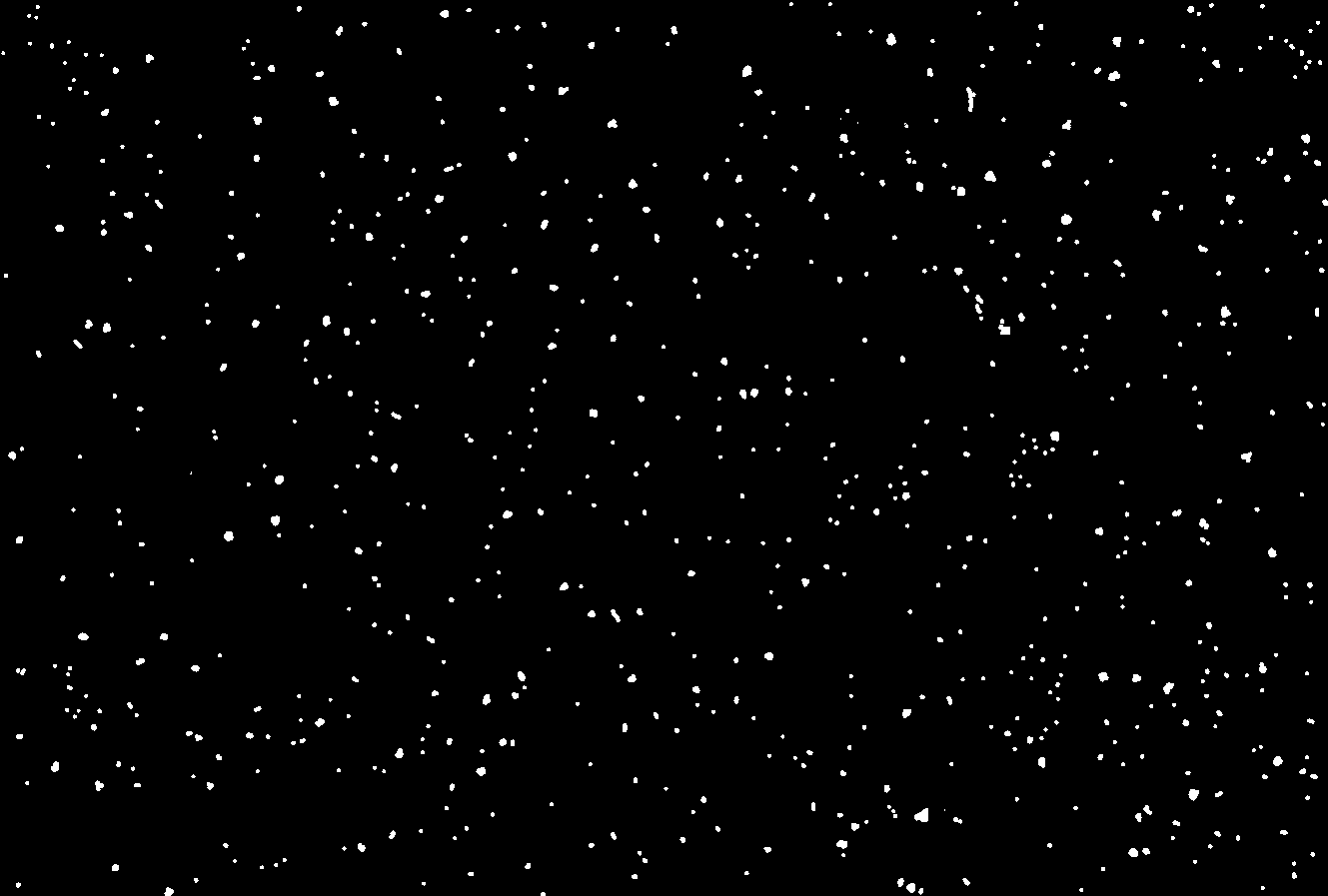}}
\end{minipage}
\begin{minipage}{0.5\textwidth}
  \subcaptionbox{Raw impurities segmentation.\label{fig:imp_pred}}{\includegraphics[height=5cm, frame]{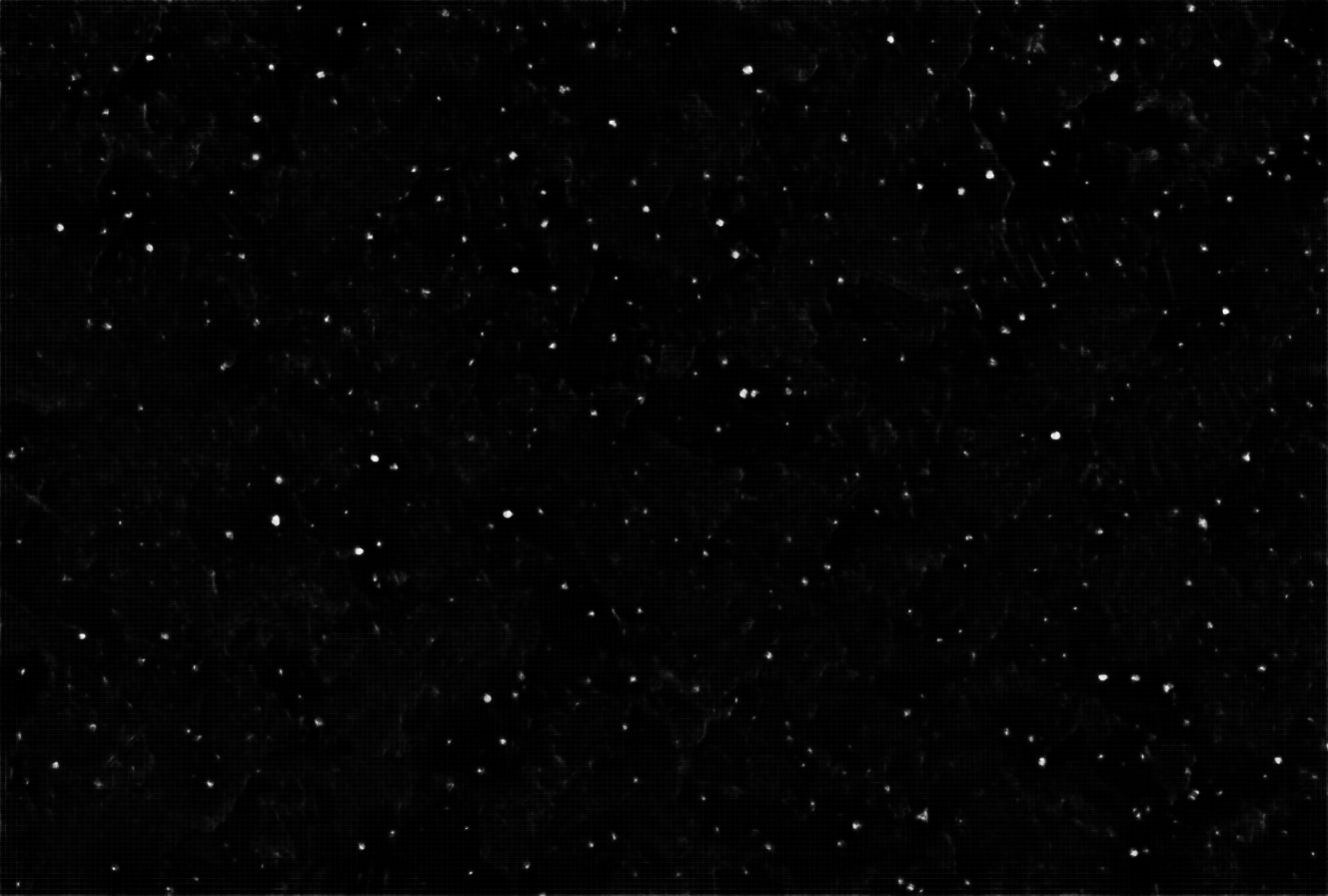}}
\end{minipage}
\par\bigskip
\begin{minipage}{0.5\textwidth}
  \subcaptionbox{GB ground truth.\label{fig:gb_gt}}{\includegraphics[height=5cm, frame]{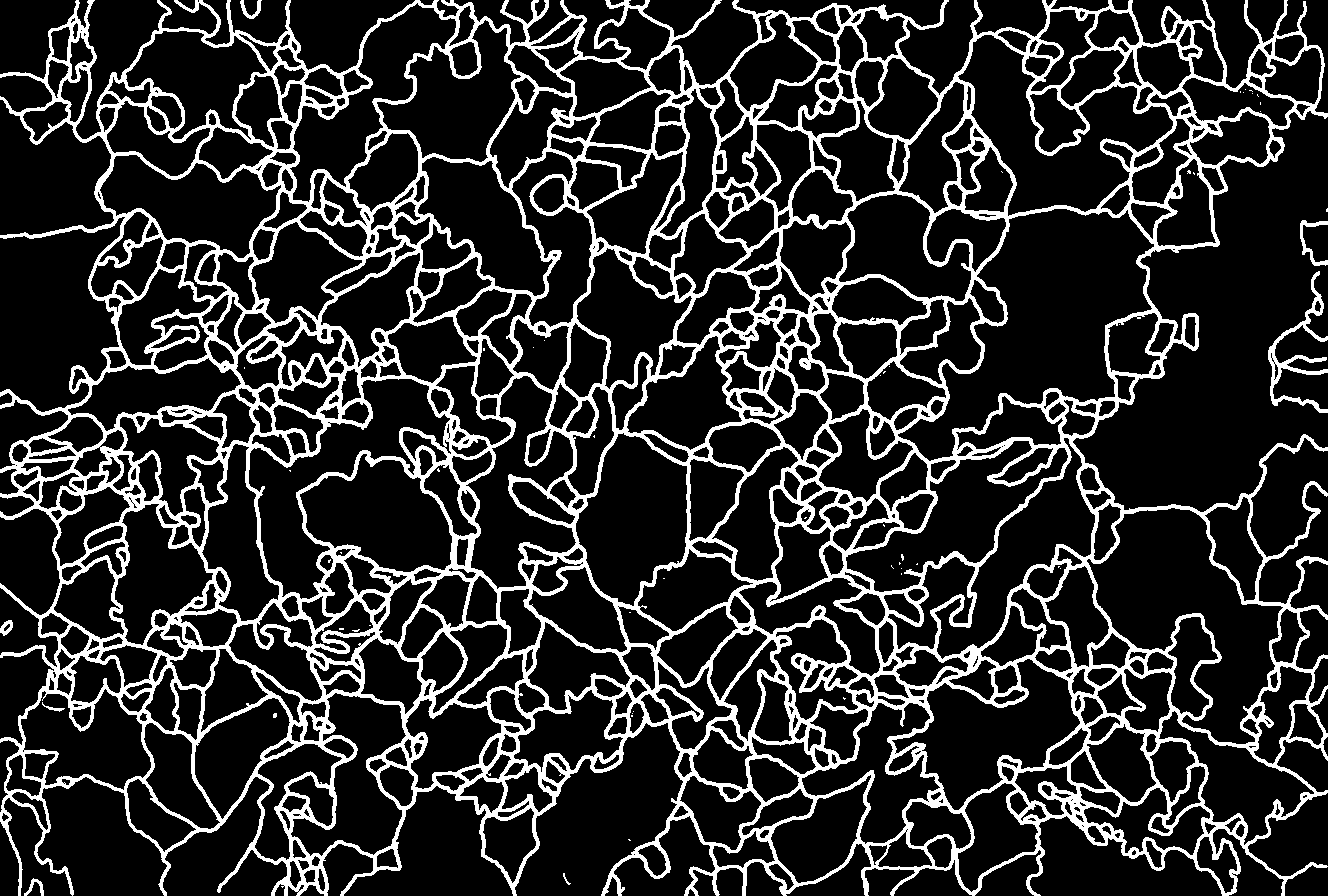}}
\end{minipage}
\begin{minipage}{0.5\textwidth}
  \subcaptionbox{Raw GB segmentation.\label{fig:gb_pred}}{\includegraphics[height=5cm, frame]{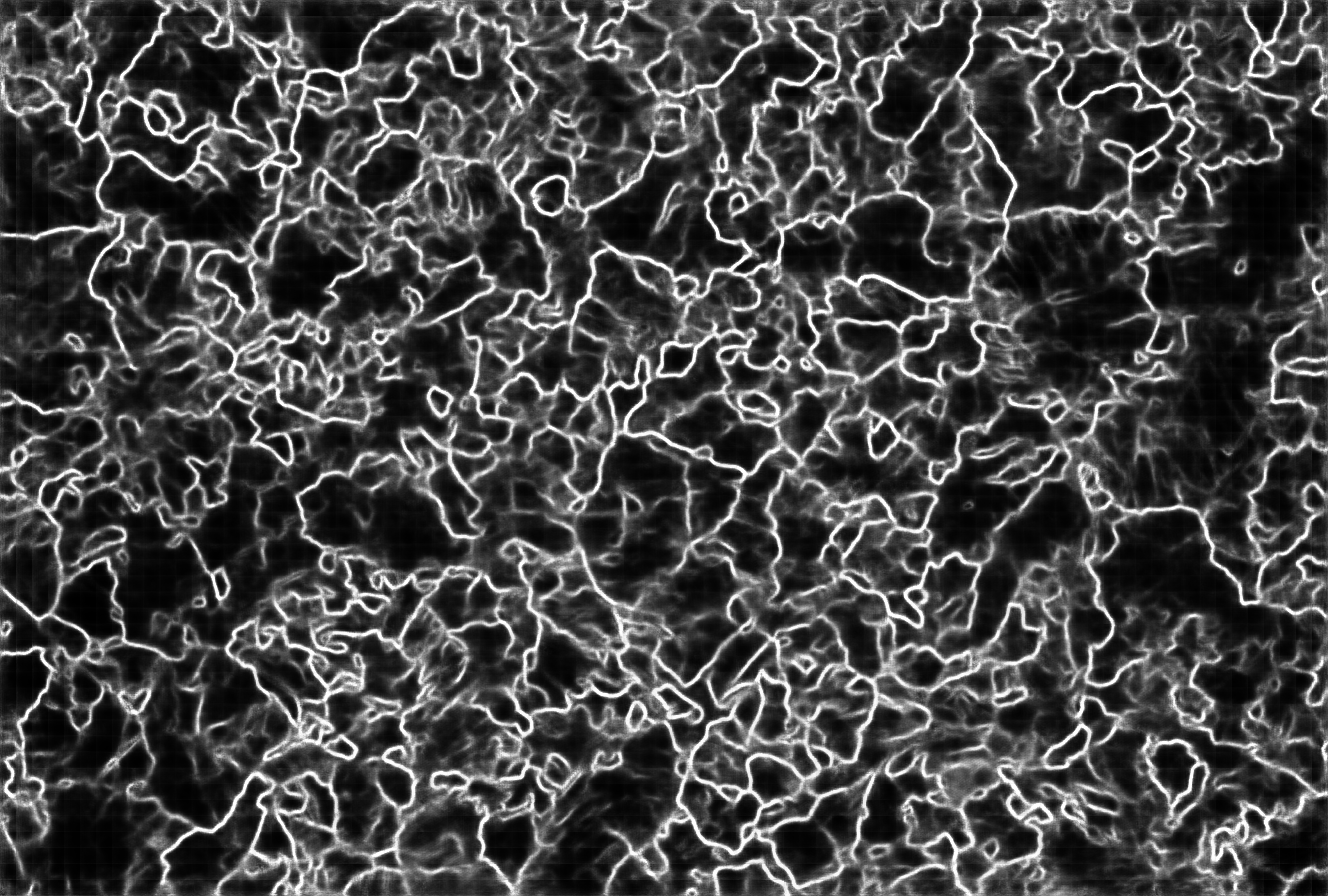}}
\end{minipage}
\caption{Segmentation pipeline applied on a test sample.}
\end{figure}

We evaluated both impurities and GB segmentation tasks with ROC-AUC statistic on a test scan of size 1328$\times$896 (Fig. \ref{fig:input_seg_test}). The outputs of the impurities and GB segmentation tasks are presented in Fig. \ref{fig:imp_pred}, Fig. \ref{fig:gb_pred} and their corresponding ground truths are presented in Fig. \ref{fig:imp_gt} and Fig. \ref{fig:gb_gt}. As can be seen in Fig. \ref{fig:imp_pred} our model tend to ignore small impurities. However, we note that these impurities are less important than the bigger ones and are neglected in our anomaly detection measures regardless. If desired, we suggest to introduce smaller impurities in the training set of the model. The ROC-AUC value for impurities segmentation is 0.944, while the value for GB segmentation is 0.868. The results emphasize the complexity gap between the two tasks, as was described in Section \ref{gb_seg}. 
To further test the impurities segmentation model, we compared the percentage of white pixels in the ground truth image as well as in the segmented impurities mask. In addition, for the GB segmentation model, we compared the average grain size via average grain diameter calculation based on a standard method \cite{astm2015e1382}, using the Image Pro Plus 6.0 \cite{plusv} software's dedicated module reporting the average length of 6 diameters at 5 degrees intervals around the centroid of each object.
We note that the ground truth image of impurities contained only 0.32\% more white pixels than the segmented mask of impurities and that the difference between the average sizes is under 8\%. These gaps are acceptable since different human experts tend to mark impurities and grains' boundaries in similar discrepancies. An essential advantage of an automated procedure as the one we propose is that it regulates and standardizes the segmentation process based on a few commonly accepted examples used for training. Additionally, since the same algorithm can be used for all segmentation tasks, fewer inter-inconsistencies between different samples are expected. This step is in contrast to manual impurities and grains' boundaries marking, which is usually very challenging to achieve by a single human expert.
The final output of the model, including impurities (segmentation and) inpainting and boundaries segmentation, is presented in Fig. \ref{fig:masked_42}.

\section{Anomaly Detection Measures} \label{anomaly_detection_measures}
\subsection{Spatial Anomaly Measure}\label{chap:spatial_anomaly_measure}

\begin{figure}
\begin{minipage}{.5\textwidth}
  \centering
  \subcaptionbox{Distance based on fixed point.\label{fig:fixed_point_dist}}{
      \resizebox{!}{3.5cm}{
          \begin{tikzpicture}
                \draw[] (0,0) rectangle (1.25,0.75) node[label={[label distance=0.25cm]60:$i_1$}, pos=.5] {};
                \draw[] (2,0.5) rectangle (3.1,1.25) node[label={[label distance=0.25cm]90:$i_2$}, pos=.5] {};
                \draw[] (-1.2,0.9) rectangle (0.75,2.5) node[label={[label distance=0.85cm]15:$i_3$}, pos=.5] {};
                \draw[] (-0.8, -0.6) rectangle (-0.5, -0.25) node[label={[label distance=0.02cm]180:$i_4$}, pos=.5] {};
                \draw[] (1.1,-0.5) rectangle (2.35,0.15) node[label={[label distance=0.45cm]0:$i_5$}, pos=.5] {};
                
                \node[circle, scale=0.2, draw, fill] (i1) at (0.625,0.375) {};
                \node[circle, scale=0.2, draw, fill] (i2) at (2.55,0.875) {};
                \draw[-, anchor=north] (i1) -- node[above, scale=0.8, sloped] {$d_2$} (i2);
                
                \node[circle, scale=0.2, draw, fill] (i3) at (-0.225,1.7) {};
                \draw[-, anchor=east] (i1) -- node[above, scale=0.8, sloped] {$d_3$} (i3);
                
                \node[circle, scale=0.2, draw, fill] (i4) at (-0.65, -0.425) {};
                \draw[-, anchor=north] (i1) -- node[above left, scale=0.8, sloped] {$d_4$} (i4);
                
                \node[circle, scale=0.2, draw, fill] (i5) at (1.725, -0.175) {};
                \draw[-, anchor=north] (i1) -- node[below right, scale=0.8, sloped] {$d_5$} (i5);
            \end{tikzpicture}
        }
        }
  \end{minipage}
  \begin{minipage}{.35\textwidth}
  \subcaptionbox{Suggested distance function. \label{fig:imp_dist}}{
      \resizebox{!}{3.5cm}{
      \begin{tikzpicture}
            
            \draw[] (0,0) rectangle (1.25,0.75) node[label={[label distance=0.25cm]60:$i_1$}, pos=.5] {};
            \draw[] (2,0.5) rectangle (3.1,1.25) node[label={[label distance=0.25cm]90:$i_2$}, pos=.5] {};
            \draw[] (-1.2,0.9) rectangle (0.75,2.5) node[label={[label distance=0.85cm]15:$i_3$}, pos=.5] {};
            \draw[] (-0.8, -0.6) rectangle (-0.5, -0.25) node[label={[label distance=0.02cm]180:$i_4$}, pos=.5] {};
            \draw[] (1.1,-0.5) rectangle (2.35,0.15) node[label={[label distance=0.45cm]0:$i_5$}, pos=.5] {};
            
            \node[circle, scale=0.2, draw, fill] (i1i2) at (1.25,0.6) {};
            \node[circle, scale=0.2, draw, fill] (i2i1) at (2,0.6) {};
            \draw[-, anchor=north] (i1i2.east) -- node[above, scale=0.8] {$d_2$} (i2i1.west);
            
            \node[circle, scale=0.2, draw, fill] (i1i3) at (0.05,0.75) {};
            \node[circle, scale=0.2, draw, fill] (i3i1) at (0.05,0.9) {};
            \draw[-, anchor=east] (i1i3.north) -- node[left, scale=0.8] {$d_3$} (i3i1.south);
            
            \node[circle, scale=0.2, draw, fill] (i1i4) at (0,0) {};
            \node[circle, scale=0.2, draw, fill] (i4i1) at (-0.5, -0.25) {};
            \draw[-, anchor=north] (i1i4) -- node[above, scale=0.8, sloped] {$d_4$} (i4i1);
    \end{tikzpicture}
    } 
    }
    \end{minipage}
\caption{Possible distance functions.}
\end{figure}
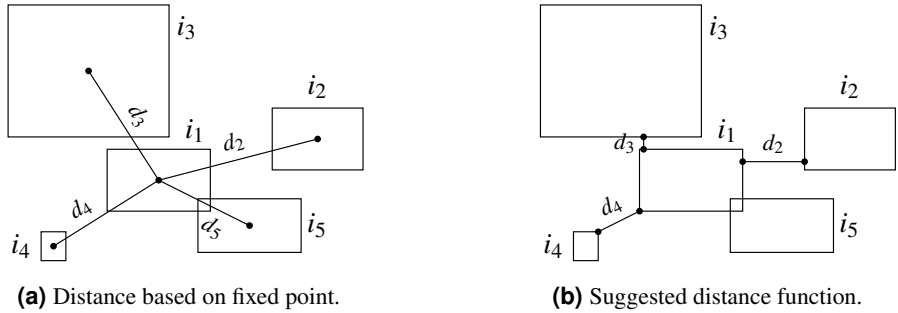

Unsupervised Distance-based is one of the most common setups for anomaly detection \cite{goldstein2016comparative}. In this approach, an object is considered as an \textit{outlier} based on its spatial properties. Most common among those properties is how distant the object is from its neighborhood. A unified distance-based notion of anomaly presented in \cite{knox1998algorithms}:
\textit{An object $O$ in dataset $T$ is a $DB(p,D)-outlier$ if at least fraction $p$ of the objects in $T$ are $\geq$ distance $D$ from $O$}, where $DB$ stands for Distance-Based.
Although this notion is applicable for generalizing statistical anomaly detection in distributions such as Normal, Exponential, and Poisson distributions, it lacks few crucial properties: It is not able to produce scores of an anomaly; It requires the user to provide the distance $D$; And most importantly it does not treat objects with shapes of a positive area, as the impurities in our study. 
Another common distance-based anomaly detection approach, K$^{th}$-Nearest-Neighbour \cite{ramaswamy2000efficient, zhang2009new} henceforth, K$^{th}$-NN, defines outliers by their distance from their $k^{th}$ nearest neighbor and sorts them by that measure. Indeed, this approach allows one to order an object by a measure that indicates how that object is \textit{distant} from its neighborhood. K$^{th}$-NN was compared to other 18 different unsupervised anomaly detection algorithms on ten datasets, and it was found to outperform all other algorithms with regard to the accuracy, determinism, and the ability to detect global anomaly \cite{goldstein2016comparative}. However, in this study, we focus on anomaly detection for geometric objects with a positive area with a high emphasis on their size, i.e., the desired spatial measure should consider the areas of the impurities in order to score each impurity by how it is \textit{distant} and \textit{big} compared to its neighborhood. To that end, we present a novel approach for \textit{spatial anomaly detection for positive area geometric objects}. Our spatial anomaly detection approach first defines a pseudo-semi-metric distance function between two geometric objects by the distance between their \textit{Straight Bounding Rectangles}. We use rectangles since they are simplistic and computationally easy to calculate for each impurity, yet accurate enough -- both in terms of Contour Approximation as well as edges distance calculation. We note that a trivial approach for implementing this distance function might be using the euclidean distance between any two points on the objects, e.g., the centers of the objects as presented in Fig. \ref{fig:fixed_point_dist}. Nevertheless, in the case of almost-intersecting two big objects, this approach will yield a much higher distance than what is expected since their borders are much closer than their centers (e.g., the distance between $i_1$ and $i_3$ in Fig. \ref{fig:fixed_point_dist}). A similar argument can be made on any fixed points residing on the objects.
Our function is summarized with the following 4 representative cases in Fig. \ref{fig:imp_dist}.
The distance between an object $i_1$ to another object, in the case of non-intersecting rectangles ($i_2,i_3,i_4$), is defined by the shortest Euclidean distance between the boundaries of the two enclosing rectangles (i.e., the distance between the closest two edges in the first two cases, and the distance between the closest corner vertices in the last case respectively). The distance between two intersecting enclosing rectangles ($i_1, i_5$) is simply defined as 0. It can be shown that this distance measure satisfies the symmetry axiom, and that for each two objects {$o_1, o_2$} the distance is $d(o_1,o_2) \geq 0$ but the triangle inequality axiom is not met and not necessarily $d(o_1,o_2) = 0$ means that $o_1=o_2$.

Next, we present a modified version of the classical K$^{th}$-NN algorithm \cite{cover1967nearest}: \textit{Weighted}-K$^{th}$-Nearest-Neighbor henceforth, WK$^{th}$-NN, in which each object $i$ refers to the distance between $i$ to its neighborhood (defined above), along with the proportion between $i$'s area to its neighbors. This modification allows having the spatial anomaly score to be calculated as a function of how the object $i$ is \textit{distant} from its neighbors, and also as how it is \textit{big} compared to its neighbors. We now describe the algorithm, which is summarized in Algorithm \ref{WKthNN}. As in K$^{th}$-NN, the algorithm is parameterized by \textit{k} -- a constant that states how far is the neighbor from which we calculate the distance from. 
The procedure \textsc{WeightedDist} calculates the weighted distance measure between the impurity $i$ and the other impurity $o$. This measure combines the proportion between the areas of $i$ and $o$ and the distance between them (Fig. \ref{fig:imp_dist}). The main procedure, \textsc{W$K^{th}$NN}, iterates over all objects (impurities in our case), $\mathcal{I}$, in line \ref{s_alg:all_imps}. For each object $i$, it calculates for all other objects $o$, the weighted distance in line \ref{s_alg:calc_others}. Then, it sorts the returned distances in line \ref{s_alg:sort}, and adds a factor of $\mbox{\sc Area}(i)$ in line \ref{s_alg:self_area}, in order to emphasize the significance of that object. Finally, when the iteration over all objects completes, we normalize and save the spatial scores of each object $i$ in $SS_i$. The constants $c_1, c_2$ were set to $4,2$ respectively, but we encourage users to determine the values of the constants $c_1, c_2$, to suit best to their datasets. The output of the spatial anomaly detection algorithm on the input image with $k=50$ is presented in Fig. \ref{fig:spatial}. The anomaly scores are normalized to $[0,1]$, while the most anomalous impurities are with scores close to 1 and are colored in red, and the most non-anomalous impurities are with scores close to 0 and are colored in blue.

\begin{figure}
    \centering
    \begin{algorithm}[H]
		\footnotesize
		\begin{flushleft}
        	$SS$: list of size $|\mathcal{I}|$, $SS$ as $SpatialScores$
        \end{flushleft}
        \begin{procedure}[H]
        \caption{() \small \mbox{\sc WeightedDist}\ ($i, o$)}
          \textbf{return} $\left(\tfrac{\mbox{\sc Area}(i)}{\mbox{\sc Area}(o)}\right) ^ {c_1} * \mbox{\sc ImpurityDist}(i,o)$ \label{s_alg:calc_w_dist} \;
         \end{procedure}
        \begin{procedure}[H]
        \caption{() \small \mbox{\sc WK$^{th}$NN}\ ($k$)}
            \For{$i\in \mathcal{I}$ \label{s_alg:all_imps}} {
                $l = [\mbox{\sc WeightedDist}(i,o); \forall o \in \mathcal{I} \setminus \{i\}]$ \label{s_alg:calc_others} \;
                $\mbox{\sc Sort}(l)$ \label{s_alg:sort} \;
                $SS_i = \mbox{\sc Area}(i) * l[k]^{c_2}$ \label{s_alg:self_area} \;
          }
          $\forall ss \in SS, \frac{ss - min(SS)}{max(SS) - min(SS)}$ \label{s_alg_norm} \tcp*{Min-Max norm} 
          \textbf{return} $SS$ \;
         \end{procedure}
    \caption{Weighted-K$^{th}$NN}\label{WKthNN}
    \end{algorithm}
\end{figure}

\subsection{Shape Anomaly Measure}
Another crucial geometric object property is its shape, or how \textit{close to some objective shape} is it, which in our case, how symmetric and how close the impurity is to a circle.
Examining the output of the spatial anomaly detection algorithm may give the idea that spatial anomaly detection is sufficient for describing the degree of the anomaly in each object since it successfully marks objects that are clear to be anomalous with a high anomaly score.
However, the spatial anomaly measure is not able to distinguish between an object that is not that big and distant compared to its neighborhood and does not have \textit{anomalous} shape (e.g., an 'O' shape impurity), with an object of the same distance and size compared to its neighborhood, but with a much more anomalous shape (e.g., an 'X' shape impurity). Therefore, a consideration of the actual shape of each impurity is necessary to determine whether it is an outlier or not.
A trivial measure for non-symmetric shape anomaly detection might be finding for each object $i$ its smallest enclosing circle object, $c$ (or some other basic geometric shape as in \cite{igathinathane2008shape}) and setting $i$'s shape anomaly score as:
$\tfrac{\mbox{\sc Area}(c) - \mbox{\sc Area}(i)}{\mbox{\sc Area}(c)} \inlineeqnum\label{eq_circle_diff}$.
This measure indeed catches the most anomalous and non-anomalous objects based on their shape (i.e., impurities of a shape with area far smaller than their smallest enclosing circle's area, and impurities of a shape very close to a circle, respectively), but it fails to classify objects properly in the middle of the scale, as can be seen in Fig. \ref{fig:circle_diff}.

\begin{figure} [H]
\centering
  \includegraphics[height=5cm]{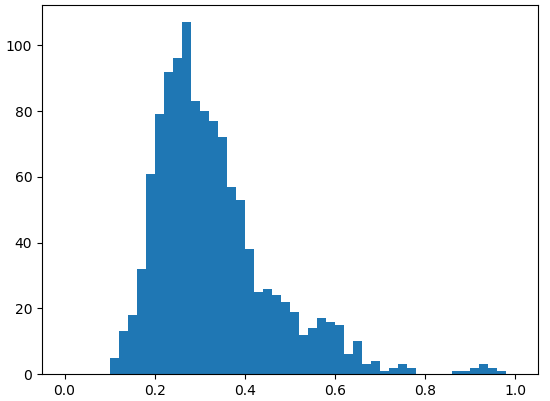}
  \caption{Histogram of circle difference scores in Fig. \ref{fig:circle_diff}.} \label{fig:circle_hist}
\end{figure}
\begin{figure}[H]
\centering
\resizebox{!}{7.5cm}{
\begin{tikzpicture}
\tikzstyle{connection}=[ultra thick,every node/.style={sloped,allow upside down},draw=\edgecolor,opacity=0.7]
\tikzstyle{copyconnection}=[ultra thick,every node/.style={sloped,allow upside down},draw={rgb:blue,4;red,1;green,1;black,3},opacity=0.7]

\node[canvas is zy plane at x=0] (temp) at (-3,0,0) {\includegraphics[width=8cm,height=8cm, frame]{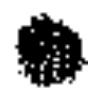}};

\pic[shift={(0,0,0)}] at (0,0,0) 
    {Box={
        name=conv1,
        caption= ,
        xlabel={{256, }},
        zlabel=100,
        fill=\ConvColor,
        height=40pt,
        width=3pt,
        depth=40,
        }
    };

\pic[shift={ (0,0,0) }] at (conv1-east) 
    {Box={
        name=pool1,
        caption= ,
        fill=\PoolColor,
        opacity=0.5,
        height=25pt,
        width=1pt,
        depth=25
        }
    };

\pic[shift={(1,0,0)}] at (pool1-east) 
    {Box={
        name=conv2,
        caption= ,
        xlabel={{256, }},
        zlabel=50,
        fill=\ConvColor,
        height=25pt,
        width=3pt,
        depth=25
        }
    };

\draw [connection]  (pool1-east)    -- node {\midarrow} (conv2-west);

\pic[shift={ (0,0,0) }] at (conv2-east) 
    {Box={
        name=pool2,
        caption= ,
        fill=\PoolColor,
        opacity=0.5,
        height=13pt,
        width=1pt,
        depth=13
        }
    };

\pic[shift={(1,0,0)}] at (pool2-east) 
    {Box={
        name=conv3,
        caption= ,
        xlabel={{256, }},
        zlabel=25,
        fill=\ConvColor,
        height=13pt,
        width=3pt,
        depth=13
        }
    };

\draw [connection]  (pool2-east)    -- node {\midarrow} (conv3-west);

\pic[shift={ (0,0,0) }] at (conv3-east) 
    {Box={
        name=pool3,
        caption= ,
        fill=\PoolColor,
        opacity=0.5,
        height=7pt,
        width=1pt,
        depth=7
        }
    };

\pic[shift={(1,0,0)}] at (pool3-east) 
    {Box={
        name=conv4,
        caption= ,
        xlabel={{256, }},
        zlabel=13,
        fill=\ConvColor,
        height=7pt,
        width=3pt,
        depth=7
        }
    };

\draw [connection]  (pool3-east)    -- node {\midarrow} (conv4-west);

\pic[shift={ (0,0,0) }] at (conv4-east) 
    {Box={
        name=pool4,
        caption= ,
        fill=\PoolColor,
        opacity=0.5,
        height=3pt,
        width=1pt,
        depth=3
        }
    };

\pic[shift={(1,0,0)}] at (pool4-east) 
    {Box={
        name=conv5,
        caption= ,
        xlabel={{256, }},
        zlabel=7,
        fill=\ConvColor,
        height=3pt,
        width=3pt,
        depth=3
        }
    };

\draw [connection]  (pool4-east)    -- node {\midarrow} (conv5-west);

\pic[shift={ (0,0,0) }] at (conv5-east) 
    {Box={
        name=unpool1,
        caption= ,
        fill=\UnpoolColor,
        opacity=0.5,
        height=7pt,
        width=1pt,
        depth=7
        }
    };

\pic[shift={(1,0,0)}] at (unpool1-east) 
    {Box={
        name=conv6,
        caption= ,
        xlabel={{256, }},
        zlabel=10,
        fill=\ConvColor,
        height=5pt,
        width=3pt,
        depth=5
        }
    };

\draw [connection]  (unpool1-east)    -- node {\midarrow} (conv6-west);

\pic[shift={ (0,0,0) }] at (conv6-east) 
    {Box={
        name=unpool2,
        caption= ,
        fill=\UnpoolColor,
        opacity=0.5,
        height=10pt,
        width=1pt,
        depth=10
        }
    };

\pic[shift={(1,0,0)}] at (unpool2-east) 
    {Box={
        name=conv7,
        caption= ,
        xlabel={{256, }},
        zlabel=16,
        fill=\ConvColor,
        height=8pt,
        width=3pt,
        depth=8
        }
    };

\draw [connection]  (unpool2-east)    -- node {\midarrow} (conv7-west);

\pic[shift={ (0,0,0) }] at (conv7-east) 
    {Box={
        name=unpool3,
        caption= ,
        fill=\UnpoolColor,
        opacity=0.5,
        height=16pt,
        width=1pt,
        depth=16
        }
    };

\pic[shift={(1,0,0)}] at (unpool3-east) 
    {Box={
        name=conv8,
        caption= ,
        xlabel={{256, }},
        zlabel=28,
        fill=\ConvColor,
        height=14pt,
        width=3pt,
        depth=14
        }
    };

\draw [connection]  (unpool3-east)    -- node {\midarrow} (conv8-west);

\pic[shift={ (0,0,0) }] at (conv8-east) 
    {Box={
        name=unpool4,
        caption= ,
        fill=\UnpoolColor,
        opacity=0.5,
        height=28pt,
        width=1pt,
        depth=28
        }
    };

\pic[shift={(1,0,0)}] at (unpool4-east) 
    {Box={
        name=soft1,
        caption= ,
        xlabel={{" ","dummy"}},
        zlabel=500,
        fill=\SoftmaxColor,
        opacity=0.8,
        height=3pt,
        width=1.5pt,
        depth=10
        }
    };

\draw [connection]  (unpool4-east)    -- node {\midarrow} (soft1-west);

\pic[shift={(1,0,0)}] at (soft1-east) 
    {Box={
        name=soft2,
        caption= ,
        xlabel={{" ","dummy"}},
        zlabel=10000,
        fill=\SoftmaxColor,
        opacity=0.8,
        height=3pt,
        width=1.5pt,
        depth=30
        }
    };

\draw [connection]  (soft1-east)    -- node {\midarrow} (soft2-west);

\node[canvas is zy plane at x=0] (temp) at (18,0,0) {\includegraphics[width=8cm,height=8cm,frame]{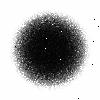}};

\draw[dashed, ultra thick,every node/.style={sloped,allow upside down},draw=blue,opacity=0.7] (16,0,0) --  (18,0,0);

\draw[dashed, ultra thick,every node/.style={sloped,allow upside down},draw=blue,opacity=0.7] (-3,0,0) --  (0,0,0);

\end{tikzpicture}
}
\captionof{figure}{The AE architecture.}
\label{AE_layers}
\end{figure}
\begin{figure}[H]
\begin{minipage}{0.5\textwidth}
  \subcaptionbox{Training and validation loss.\label{fig:ae_loss}}{\includegraphics[height=5cm]{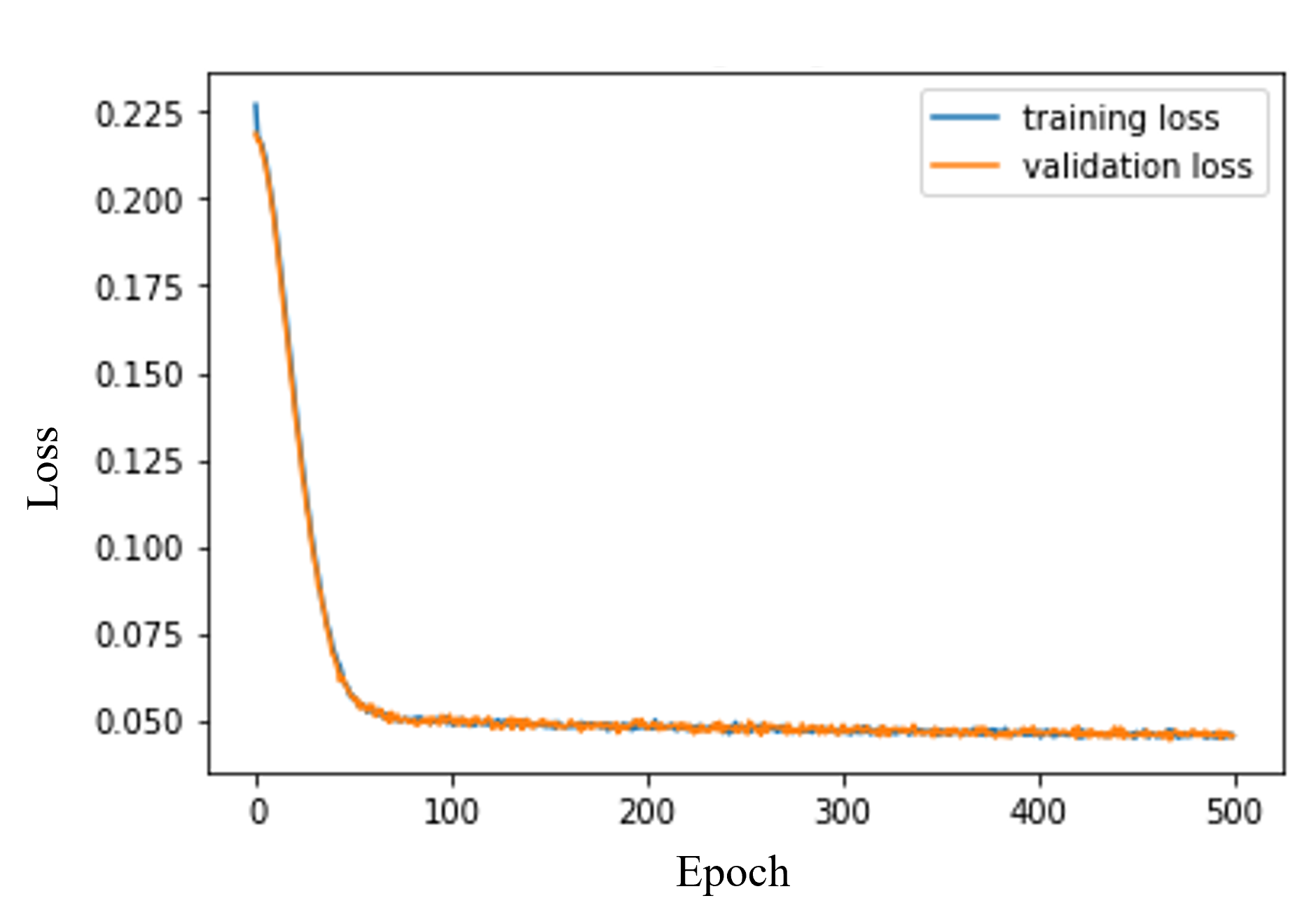}}
\end{minipage}
\begin{minipage}{0.5\textwidth}
  \subcaptionbox{Training and validation accuracy.\label{fig:ae_acc}}{\includegraphics[height=5cm]{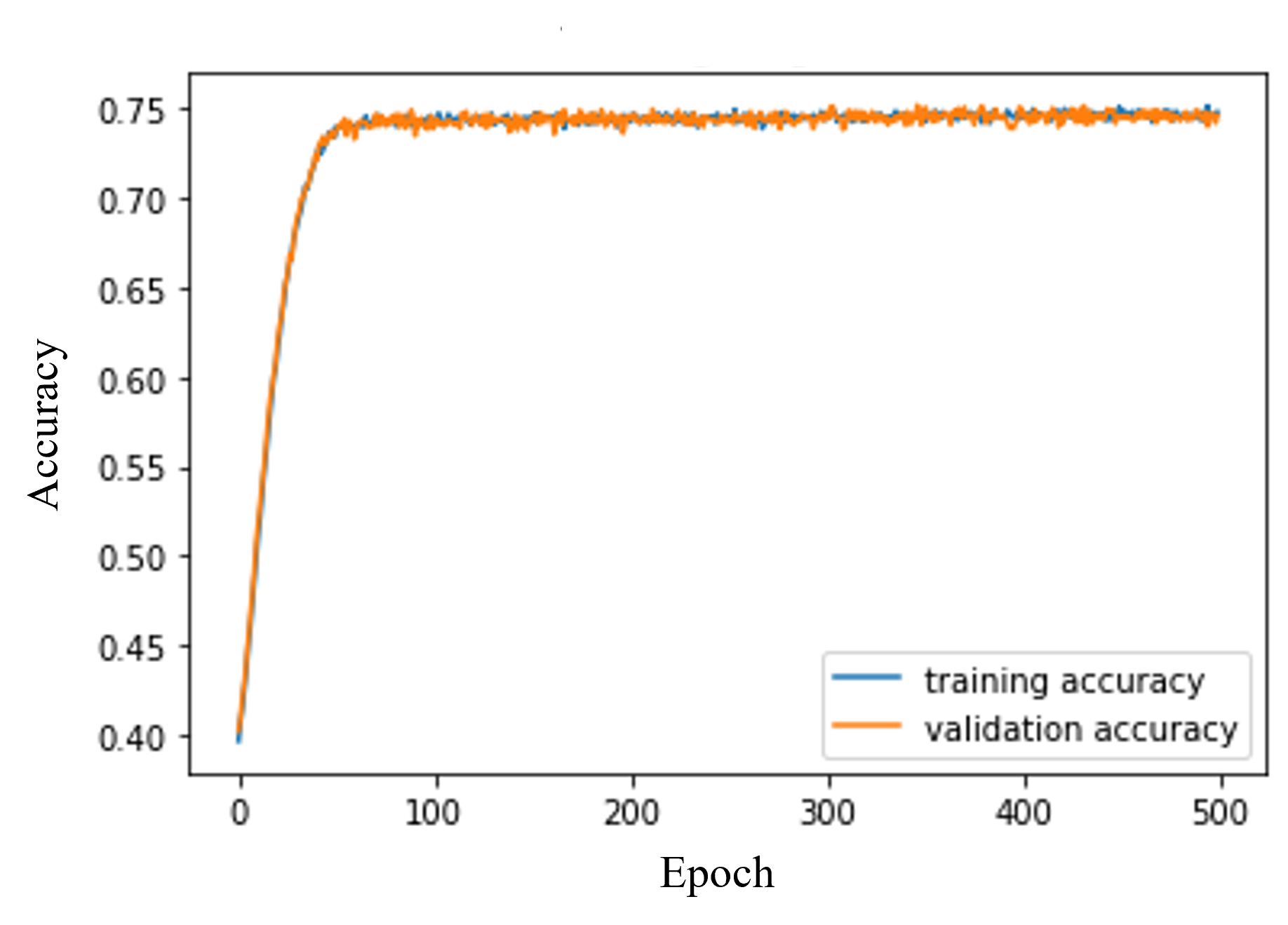}}
\end{minipage}
 \caption{Loss and accuracy of the training and validation of the AE.}
\end{figure}

\begin{figure}
    \centering
        \includegraphics[height=6.5cm]{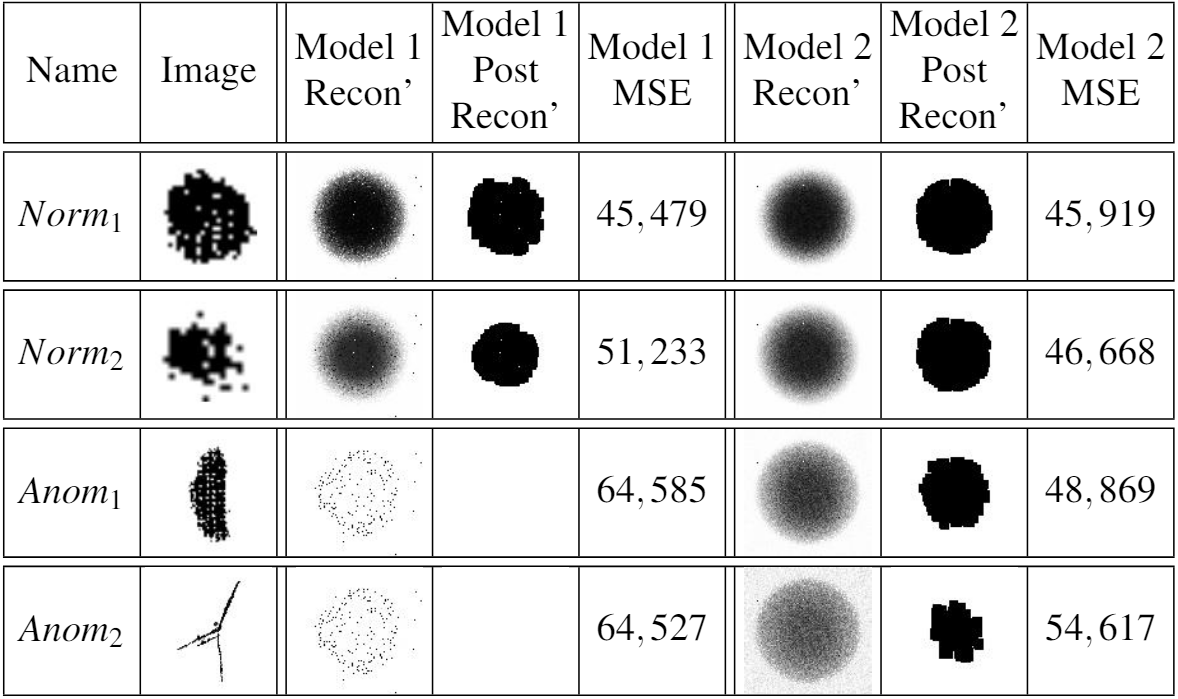}
        \captionof{table}{Reconstruction of impurities in both AEs.}
        \label{fig:recon}
\end{figure}

 For example, impurities of the anomalous shape 'X' are marked only in the middle of the scale (e.g., the impurity within the black rectangle in Fig. \ref{fig:circle_diff}, which should get a higher shape anomaly score), together with not-that-anomalous ellipse-shaped impurities, as their shape's area is not that far from their smallest enclosing circle's area, although they should have appeared higher in the shape anomaly score scale. Indeed, Fig. \ref{fig:circle_hist} shows that there is a decent separation between the most anomalous impurities (scores $\geq 0.6$ ) and the rest of the impurities (scores around $0.3$), but the right tail of the distribution is quite long, which imposes noise to the model.
 Thus, from the non-linear nature of the problem at hand, we turned to train a Deep Convolutional Auto-Encoder Neural Network (henceforth, AE) for shape anomaly detection (also called Replicator Neural Networks) \cite{hawkins2002outlier, dau2014anomaly}. This method enforces the network to reconstruct images similar to the images from the training set and, hopefully, to fail to reconstruct images that are not similar to the images in the training set.
 As we already stated, the circle difference measure from Equation \ref{eq_circle_diff} is a reasonable estimate for shape anomaly in the two ends of the scale -- the most anomalous impurities and most non-anomalous impurities, which we will denote normal impurities from now on. Thus, one can train an AE network in an \textit{unsupervised} manner by providing the network, in its training phase, couples of all normal impurities as the training samples and copies of themselves as their labels. That is, fix a threshold for normal impurities and take all impurities with an anomaly score lower than that threshold.
 \pgfplotsset{every tick label/.append style={font=\big}}

We present in this work a novel approach to empower the separation capability of AE networks, in which, together with the normal couples of input and label images, the network is provided with couples of all the most anomalous images as input and blank images as labels. This method will urge the network to reconstruct the normal images successfully and to return a noisy-blank image upon an anomalous input, or in our case, anomalous impurity. This mechanism, in turn, will yield higher reconstruction loss for anomalous impurities, thus normalizing the reconstruction loss and using it as a shape anomaly measure will offer a higher separation between normal and abnormal impurities. We stress that one significant advantage of using neural networks is that it requires no assumption about the data. Therefore, one can employ the presented technique on any predefined 'normal' and 'abnormal' objects (in our case, difference from a circle).
We set the threshold for normal impurities to $0.3$ and anomalous impurities to $0.55$, normalized and scaled all input images into the same size of $100 \times 100$ pixels. We note that albeit the size feature is not preserved in this measure, we still consider it in the spatial anomaly measure in chapter \ref{chap:spatial_anomaly_measure}. Then we trained an AE network of the architecture presented in Fig. \ref{AE_layers}. We used different paddings for the convolutional layers in the decoder in order to reduce memory footprint and fit the model into 32GB Tesla V100 GPU \cite{negevhpc}, using TensorFlow \cite{abadi2016tensorflow}. Loss and accuracy trends are presented in Figures \ref{fig:ae_loss}, \ref{fig:ae_acc}. Yellow layers are Convolutional layers (5x5), Orange ones are Max-Pooling layers, Blue ones are Up-Sampling layers, Purple ones are Fully Connected layers, and the left and right-most images are example input and output images, respectively. In Table \ref{fig:recon} we present the achieved reconstruction results on several use-cases, consisting of two normal-shaped impurities, $Norm_1, Norm_2$, and two anomalous-shaped impurities, $Anom_1, Anom_2$.
Column \textit{Image} holds the representative image of each impurity, and \textit{Model 1 Recon'} holds the reconstructed image from the AE model trained on both normal and anomalous impurities. As we can see, there is a strong separation between the normal and anomalous impurities' reconstruction in the first model, as in the first two impurities, the reconstruction is a well-formed circle (with a varying intensity with respect to the degree of anomaly), and for the last ones, the reconstruction is a noisy-circle. For even sharper separation, we applied post-processing (threshold, erode-dilate) on the output of the AE, which is shown in \textit{Model 1 Post-Recon'} column, and by that obtaining circles of different sizes for each of the normal impurities, and a blank image for the anomalous impurities. The column \textit{Model 1 MSE} shows the Mean Squared Error (MSE) as the reconstruction loss between the input image and the reconstructed image after post-processing. As we can see, the first impurity is the most 'normal' impurity, and the last two impurities are much more anomalous.
Conversely, the reconstruction results of the same input impurities on an AE trained only on normal set of impurities are presented in the columns \textit{Model 2 Recon', Model 2 Post-Recon'} and \textit{Model 2 MSE}.
As we can see, the intensity of the reconstructed circle scales negatively with the degree of anomaly, thus again yielding circles of different sizes in the post-processed reconstructions. Additionally, the MSE of the most anomalous impurity, $Anom_2$, is significantly higher than that of the most symmetric impurity, $Norm_1$, but the difference between the errors of $Norm_2$ and $Anom_1$ is mild. Thus the separation between the normal and anomalous impurities is flawed. 
Figures \ref{fig:shape_blank}, \ref{fig:shape_same} present the output of both models. In each, the normalized reconstruction losses serve as the shape anomaly measure.
The model that utilizes blank images as labels for anomalous impurities in the training phase significantly outperforms the second one since it has a more acute separation between normal-shaped and anomalous-shaped impurities, and it marks the anomalous 'X'-shaped impurities with a high anomaly score. We, therefore, use this model.
The previously purposed spatial anomaly measure -- combined by simple multiplication and normalization with the shape anomaly measure -- is presented in Fig. \ref{fig:shape_spatial}. This measure significantly reduces the noise we had in the spatial anomaly measure while emphasizing the degree of the anomaly of anomalous impurities based on their shape and compared to their neighborhood.

\subsection{Area Anomaly Measure}
As previously explained, an essential application for anomaly detection in materials sciences is detecting defects. These defects usually span an anomalous \textit{area} of several objects, rather than just a single anomalous object \cite{deepak2016anomaly}. For this reason, we present a novel clustering algorithm, which we call \textit{Market-Clustering}, that divides impurities into anomalous areas, based on the anomaly scores of the impurities from the previous anomaly measures. The algorithm's name is inspired by the 'purchasing power' of each area/cluster and the economic decisions it should take to grow and merge with other significant clusters. In fact, each cluster's size, reach, and anomaly score is determined based on the anomaly score of the objects from the previous measures.
The returned clusters are then ranked based on a measure that we later describe, and the anomalous areas beyond some pre-determined threshold are suggested for further physical tests.
We next present the algorithm in Algorithm \ref{mc} and then describe its actions.

\begin{figure*}[h!]
	\begin{algorithm}[H]
		\footnotesize
		\begin{flushleft}
            	$\mathcal{A}$: for $i\in \mathcal{I}$ that participated in some auction, stores the highest bid for $i$ from some cluster \\
            	$s$: anomaly scores based on spatial and shape anomaly measures
            \end{flushleft}
		\begin{multicols*}{2}
            \begin{procedure}[H]
            
            \caption{() \small \mbox{\sc MarketClustering}\ ($k, scores$)}
                $clusters = \mbox{\sc InitClusters}(k, scores)$ \label{a_alg:init_clusters} \;
                $status =$ not converged  \;
                \While{$status \neq$ converged \label{a_alg:mc_while}} {
                    $status =$ converged  \;
                    sort $clusters$ by their $wallet$  \;
                    \For{$c \in clusters$ \label{a_alg:mc_for}} {
                        $(i,o) \leftarrow$ cheapest couple, $i\in c.\mathcal{I}$, $o \notin c.\mathcal{I}$ based on \mbox{\sc Price}($i,o$) \label{a_alg:find_cheapest_couple} \;
                        \uIf{$o \in \mathcal{A}$ and $c.\mathcal{W} \leq \mathcal{A}[o]$ \label{a_alg:higher_bid}} { 
                            \textbf{goto} \ref{a_alg:find_cheapest_couple} \; 
                        }
                        $status = \mbox{\sc AttemptToExpand}(c, i, o, \mathcal{A})$  \label{a_alg:attempt_to_expand}  \;
                        \uIf{$status = $ merged \label{a_alg:if_merged}} {
                            \textbf{goto} \ref{a_alg:mc_while} \;
                        }
                  }
              }
              \textbf{return} $clusters$ \;
             \end{procedure}
             \begin{procedure}[H]
             \caption{() \small \mbox{\sc InitClusters}\ (k, scores)}
              \For{$i\in [1,k]$} { \label{a_alg:init_clusters_start}
                $core$ = imp' with the $i$-highest anomaly score \;
                $c$:  New Cluster \;
                $c.\mathcal{C}= [coreImpurity]$ \tcp*{core imps'} \label{a_alg:core}
                $c.\mathcal{I} = [coreImpurity]$  \label{a_alg:inside} \tcp*{imps' inside}
                $c.\mathcal{W} = \mbox{\sc F} \left(scores[coreImpurity]\right)$ \label{a_alg:wallet} \;
                $clusters.$\mbox{\sc Append}$\left(c\right)$ \;
              }
              \textbf{return} $clusters$ \label{a_alg:init_clusters_end} \;
            \end{procedure}
			
			\columnbreak
			\begin{procedure}[H]
            \caption{() \small \mbox{\sc AttemptToExpand}\ ($c, i, o, \mathcal{A}$)}
            \uIf{$\exists c' \in clusters$ s.t. $o \in c'.\mathcal{C}$ \label{a_alg:other_is_core}} {
                $\mathcal{A}[o] = c.\mathcal{W}$ \tcp*{place bid}
                $c.\mathcal{W} = c.\mathcal{W} + c'.\mathcal{W}$ 
                $c.\mathcal{C} = \mbox{\sc Concatenate}(c.\mathcal{C}, c'.\mathcal{C})$ \;
                $c.\mathcal{I} = \mbox{\sc Concatenate}(c.\mathcal{I}, c'.\mathcal{I})$ \;
                $clusters.\mbox{\sc Remove}(c')$ \;
                \textbf{return} merged \;
            }
            \ElseIf{$c.\mathcal{W} \geq \mbox{\sc Price}(i,o)$ \label{a_alg:there_is_money}} {
                $\mathcal{A}[o] = c.\mathcal{W}$ \tcp*{place bid}
                $c.\mathcal{W} = c.\mathcal{W} - \mbox{\sc Price}(i,o)$  \tcp*{pay}
                $c.\mathcal{I}.\mbox{\sc Append}(o)$ \;
                \uIf{$\exists c' \in clusters$ s.t. $o \in c'.\mathcal{I}$} {
                    $c'.\mathcal{I}.\mbox{\sc Remove}(o)$ \;
                }
            }
            \textbf{return} converged \;
            \end{procedure}
            \begin{procedure}[H]
            \caption{() \small \mbox{\sc Price}\ ({$i, o$)}}
            $d = \left(\exp{\left(\sqrt{\mbox{\sc ImpurityDist}(i,o)}\right)}\right)^{c_1}$ \label{price:d} \;
            $s = \left(1 - \left(s[i] * c_2^1 \right)^{c_3^1} * \left(s[o] * c_2^2 \right)^{c_3^2} \right)^{c_4}$ \label{price:s}  \;
            $price = d * s$ \label{price:d*s} \;
            \uIf{$\exists c' \in clusters$ s.t. $o \in c'.\mathcal{C}$ \label{price:if_core}} {
                $dis = \left(1 - \left(s[i] * c_2^3 \right)^{c_5^1} * \left(s[o] * c_2^4 \right)^{c_5^2} \right)^{c_6}$ \label{price:discount} \;
                $pen = \left(2 - \left|s[i] - s[o] \right| \right)^{c_7}$ \label{price:penalty} \;
                $price = price * dis * pen$ \label{price:discount*penalty} \;
            }
            \textbf{return} $price$ \;
            \end{procedure}
            
		\end{multicols*}
		\caption{Market-Clustering}
		\label{mc}
	\end{algorithm}
\end{figure*}

\begin{figure}
\begin{minipage}{0.47\textwidth}
\centering
  \subcaptionbox{Spatial Anomaly measure. \label{fig:spatial}}{\includegraphics[height=6cm, frame]{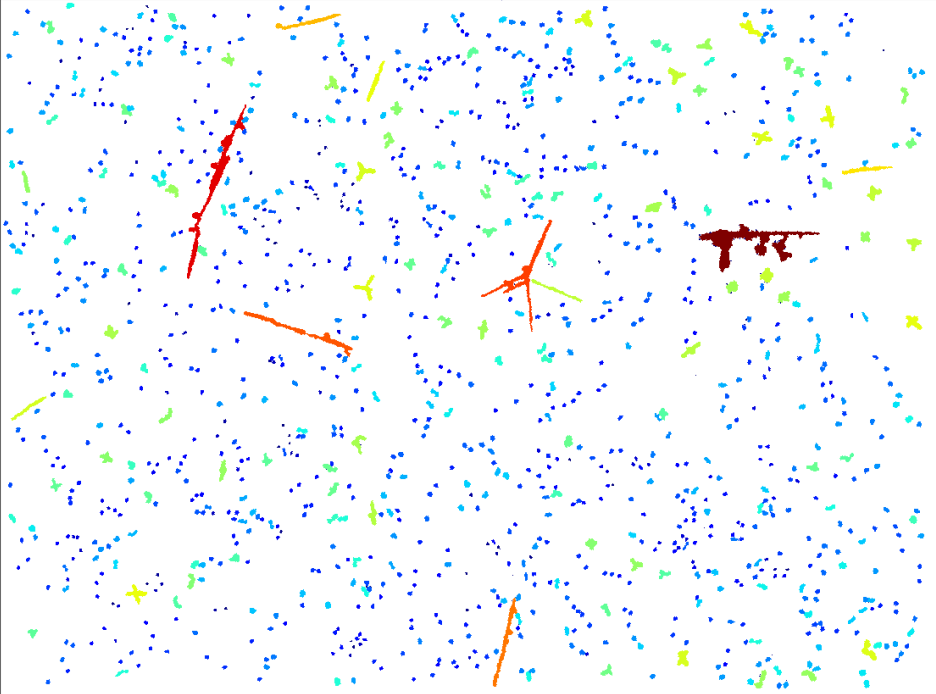}}
\end{minipage}
\begin{minipage}{0.47\textwidth}
  \centering
  \subcaptionbox{Shape Anomaly with Eq. \ref{eq_circle_diff}.
    \label{fig:circle_diff}}{\includegraphics[height=6cm, frame]{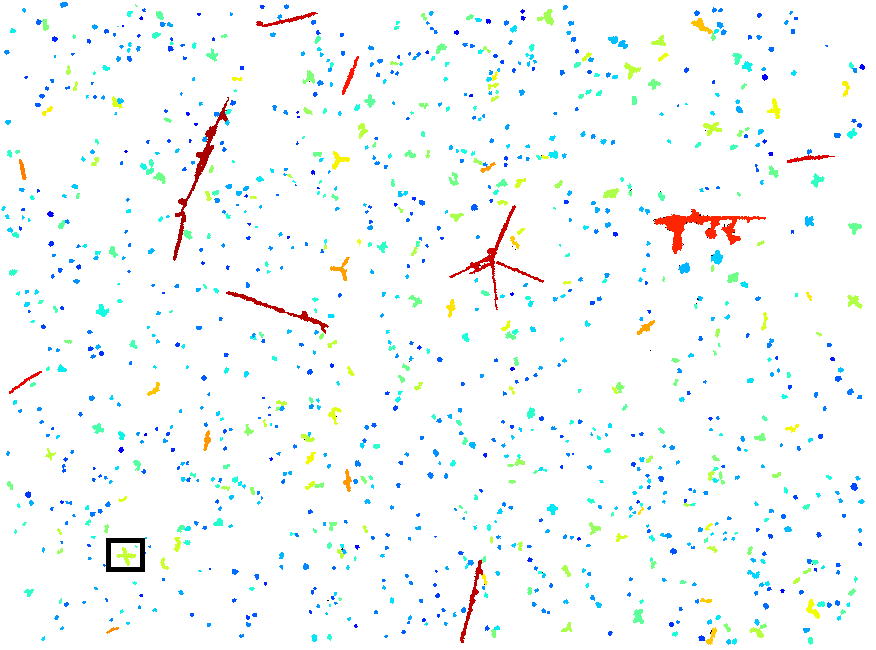}}
\end{minipage}
\begin{minipage}{.05\textwidth}
\centering
\subcaptionbox*{\newline}{\vspace{-0.3cm}\includegraphics[height=6cm]{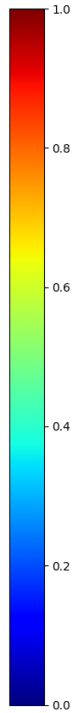}}
\end{minipage}
\par\bigskip 
\begin{minipage}{0.47\textwidth}
\centering
  \subcaptionbox{AE trained with blank images as labels for anomalous impurities.\label{fig:shape_blank}}{\includegraphics[height=6cm, frame]{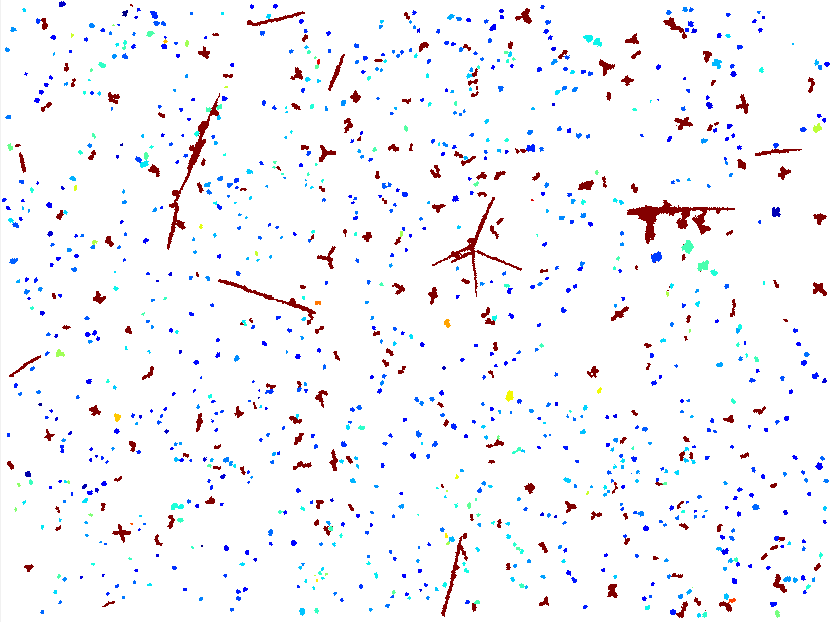}}
\end{minipage}
\begin{minipage}{0.47\textwidth}
  \centering
  \subcaptionbox{AE trained with only normal images as labels. \newline \label{fig:shape_same}}{\includegraphics[height=6cm, frame]{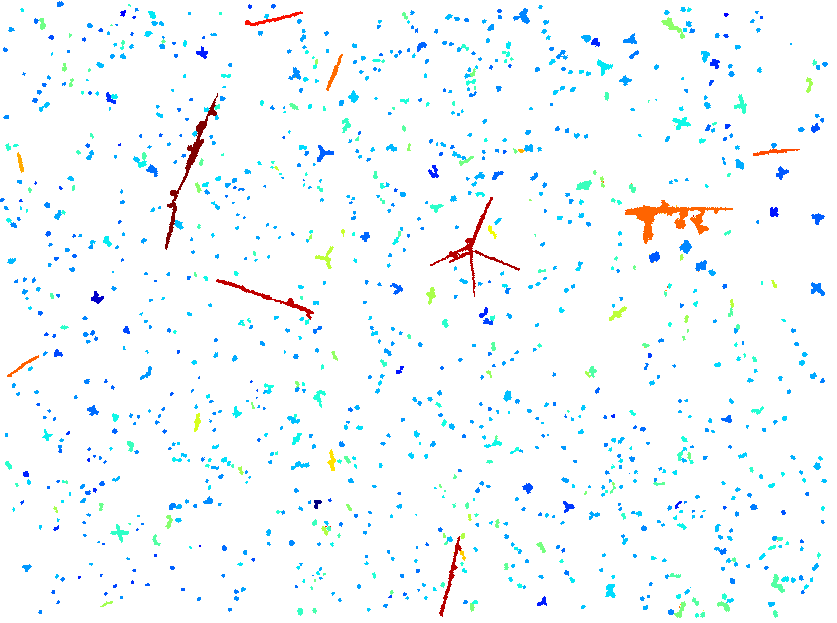}}
\end{minipage}
\begin{minipage}{.05\textwidth}
\centering
\subcaptionbox*{\newline}{\includegraphics[height=6cm]{colormap.PNG}}
\end{minipage}
\par\bigskip
\begin{minipage}{0.47\textwidth}
\centering
  \subcaptionbox{Shape\&Spatial Anomaly. \label{fig:shape_spatial}}{\includegraphics[height=6cm, frame]{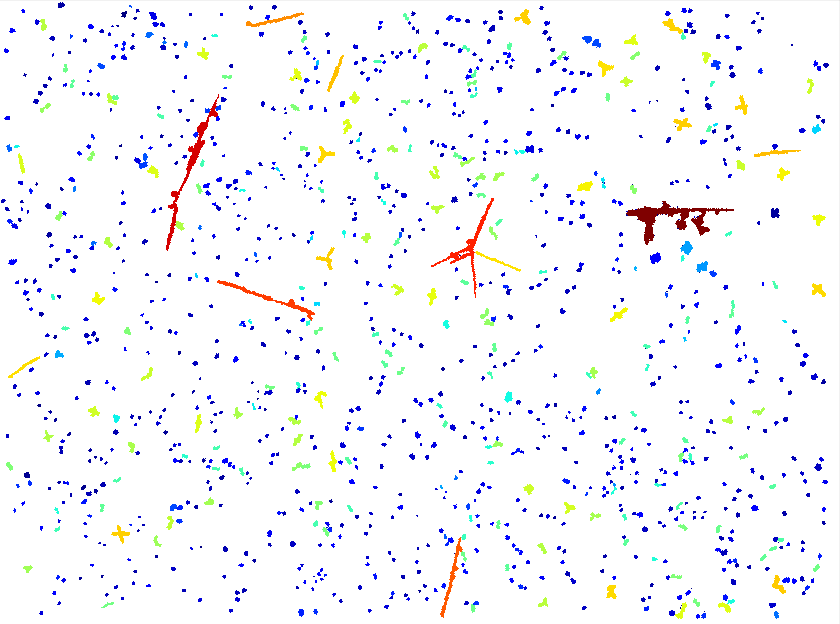}}
\end{minipage}
\begin{minipage}{0.47\textwidth}
\centering
  \subcaptionbox{Area Anomaly measure. \label{fig:area}}{\includegraphics[height=6cm, frame]{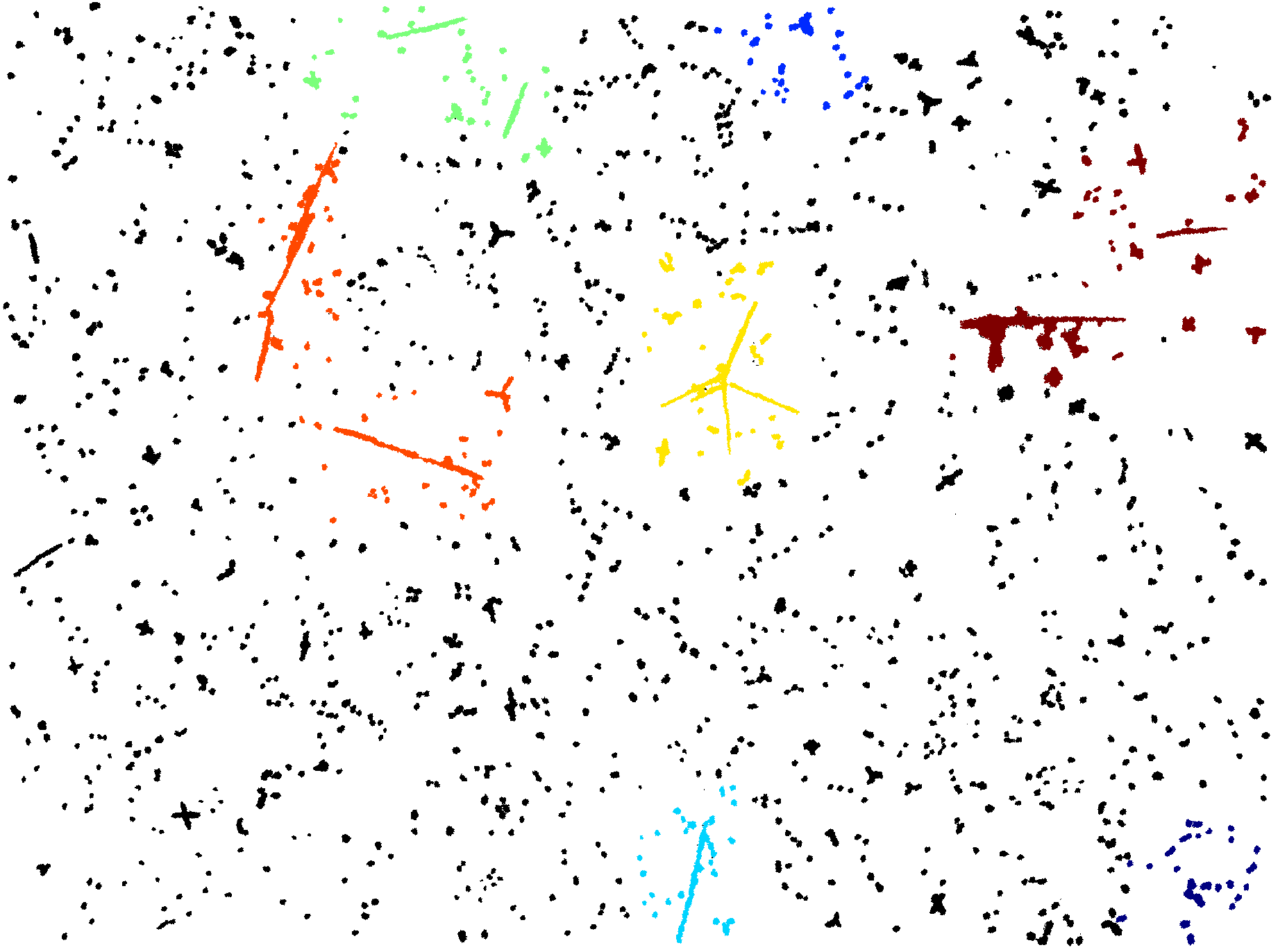}}
\end{minipage}
\begin{minipage}{.05\textwidth}
\centering
\subcaptionbox*{\newline}{\vspace{-0.35cm}\includegraphics[height=6cm]{colormap.PNG}}
\end{minipage}
\par\bigskip
 \caption{Anomaly detection measures.}
\end{figure}

\begin{figure}[h!t]
\begin{minipage}{0.5\textwidth}
\centering 
  \subcaptionbox{Line \ref{price:s}.\label{fig:price_s}}{\begin{tikzpicture}[scale=0.6]
        \begin{axis}[
            xtick={0, 1},
            ztick={0, 1},
            x tick label style={inner sep=0, font=\large,},
            z tick label style={inner sep=0, font=\large,},
            xtick pos=left,
            ztick pos=left,
            ymajorticks=false
        ]
        \addplot3[
            mesh,
            samples=20,
            domain=0:1,
            colormap/jet,
        ]
        {(1-(x*0.95)^0.5*(y*0.95)^0.5)^1.6};
        
        \end{axis}
        \end{tikzpicture}}
\end{minipage}
\begin{minipage}{0.5\textwidth}
\centering 
  \subcaptionbox{Line \ref{price:discount}.\label{fig:price_discount}}{\begin{tikzpicture}[scale=0.6]
        \begin{axis}[
            xtick={0, 1},
            ztick={0, 1},
            x tick label style={inner sep=0, font=\large,},
            z tick label style={inner sep=0, font=\large,},
            xtick pos=left,
            ztick pos=left,
            ymajorticks=false
        ]
        \addplot3[
            mesh,
            samples=20,
            domain=0:1,
            colormap/jet,
        ]
        {(1-(x*0.95)^0.05*(y*0.95)^0.05)^2.5};
        \end{axis}
        \end{tikzpicture}}
\end{minipage}
 \caption{Proposed functions from Algorithm \ref{mc}.}
\end{figure}
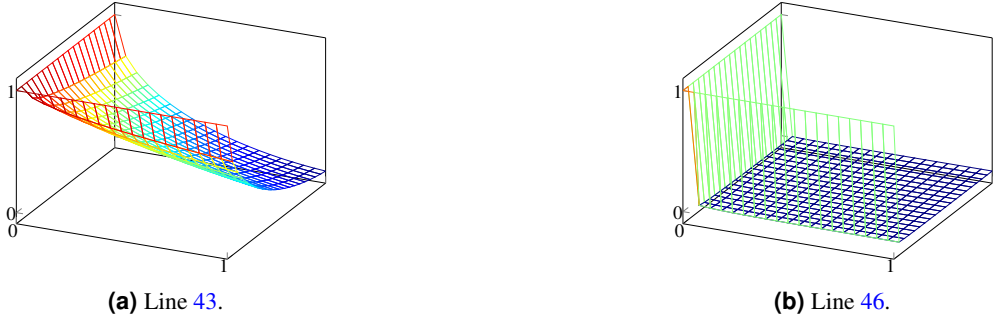

        

The \textsc{MarketClustering} procedure is the main procedure of the algorithm. It receives as input a parameter $k$ -- number of initial clusters, and a list $scores$ -- impurities' anomaly scores, which in our case are the combination of the spatial and shape measures. First we initialize the list $clusters$ in line \ref{a_alg:init_clusters}, by defining in \textsc{InitClusters} for each cluster $c$ its \textit{Core impurities} list -- $c.\mathcal{C}$, \textit{Impurities inside} list -- $c.\mathcal{I}$, and \textit{Wallet balance} variable -- $c.\mathcal{W}$. 
The procedure in lines \ref{a_alg:init_clusters_start} -- \ref{a_alg:init_clusters_end} does that by setting the core impurities of $k$ clusters with the $k$ most anomalous impurities (lines \ref{a_alg:core}, \ref{a_alg:inside}).
Core impurities represent each cluster; thus, the degree of the anomaly of each cluster is first defined by the anomaly score of the initial core impurity, and it is stored in the wallet of the cluster in line \ref{a_alg:wallet}.
The loop in line \ref{a_alg:mc_while} iterates until convergence, and in each iteration, it sorts the clusters by their wallet and initiates the main loop in line \ref{a_alg:mc_for}. This loop iterates over all clusters, and for each cluster $c$ it finds in line \ref{a_alg:find_cheapest_couple} the cheapest couple $(i,o)$, s.t. $i \in c.\mathcal{I}$ is an impurity inside of it, and $o\in \mathcal{I}\setminus c.\mathcal{I}$ is an impurity outside of it, under some parametric price function \textsc{Price}($i,o$). This couple, in fact, implies that the cheapest impurity for cluster $c$ to append is $o$, and the price for it is \textsc{Price}($i,o$). We suggest that the price will be a function of distance and anomaly score, i.e., the price should decrease as the distance decreases to encourage clusters to be continuous, and as the anomaly scores increase to instruct clusters to expand towards anomalous impurities and cover a larger anomalous area. We later present our parametric price function in \textsc{Price} procedure.
In line \ref{a_alg:higher_bid} we make use of $\mathcal{A}$ which is a list that stores for each impurity $i\in \mathcal{I}$ what is the cluster with the highest wallet balance that tried to append $i$ to himself. In order to prevent clusters from fighting and emptying their wallets over impurities, we allow $c$ to proceed to line \ref{a_alg:attempt_to_expand} only if it is the cluster with the highest balance that has attempted to append $o$ until now. In line \ref{a_alg:attempt_to_expand}, $c$ attempts to expand its reach by calling the procedure \textsc{AttemptToExpand}. In line \ref{a_alg:other_is_core} we check if the other impurity $o$ is a core impurity of some other cluster $c'$. If it is, first $\mathcal{A}$ is updated with the new bid on $o$, and then the clusters $\{c,c'\}$ are merged into $c$. Otherwise, as a utilization of \textit{credit with no overdraft} policy, if there is enough credit in the wallet of $c$, again $\mathcal{A}$ is updated with the new bid on $o$, and $c$ pays for and appends $o$. Then we check in line \ref{a_alg:if_merged} if a merge has occurred, and if it did, we sort the clusters again by their wallet balance and proceed to iterate over all new clusters in line \ref{a_alg:mc_for}.
The parametric price function that we used in line \ref{a_alg:find_cheapest_couple} is presented in \textsc{Price} procedure.
The parameters that we used are: $c_1 = 1.7,$ for $i \in [1,4]$ $c_2^i=0.95,$ $c_3^1=c_3^2=0.5,$ $c_4=1.6,$ $c_5^1=c_5^2=0.05,$ $c_6=2.5,$ $c_7=8$. The price between two impurities $i,o$ is determined by a function of the distance between them in line \ref{price:d} and by a function of how anomalous are they in line \ref{price:s}. $d$ grows with the distance, i.e. the price gets higher as the impurities are more distant from each other. However, $s$ scales negatively with the anomaly scores of $i$ and $o$. The behavior of the function can be seen at Fig. \ref{fig:price_s}. 
Since we want clusters to merge (line \ref{a_alg:other_is_core}) and span a larger anomalous area, we check in line \ref{price:if_core} if $o$ is a core impurity of some cluster, and if it is, there is a price reduction in line \ref{price:discount}. The value $dis$ falls much more drastically than $s$. This is because we encourage clusters merging, therefore, giving a relatively low price to core impurities despite the distance to them. The behavior of the function in line \ref{price:discount} is presented in Fig. \ref{fig:price_discount}. Line \ref{price:penalty} penalizes cluster merging of similar sizes in order to encourage big and anomalous clusters to absorb smaller and less anomalous clusters at a lower price.

After marking the anomalous areas, we now quantify the degree of the anomaly of each area. First, we suggest the following area anomaly measure for cluster $c$, and we next prove that it indeed indicates how $c$ is anomalous.
\begin{theorem}
$am(c)$ is monotonically increasing with the degree of anomaly of $c$, based on \textit{Market-Clustering} algorithm and on $c$'s Spatial and Shape anomaly score, where
\end{theorem}
\begin{equation}
am(c) \coloneqq \sum_{i\in c.\mathcal{I}} \left(\mbox{\sc Score}(i)*\mbox{\sc Area}^{2}(i)\right) * \mbox{\sc Diameter}(c) * |c.\mathcal{I}| \label{eq_area_measure} 
\end{equation}
\begin{proof}
Appending lots of non-anomalous (and not core) impurities is an expensive procedure, compared to appending lots of anomalous impurities because of the discount in line \ref{price:s} for anomalous impurities. Thus, clusters with a large number of impurities apparently have included cheaper, more anomalous impurities. Moreover, clusters that have appended core impurities, which are the most anomalous, clearly should be considered as more anomalous. Indeed, cluster merging imposes a higher wallet balance for future impurities addition, in addition to a concatenation of impurities in each cluster. 
Thus, as the number of impurities in the cluster, $|c.\mathcal{I}|$, grows, the degree of the anomaly of $c$ grows correspondingly.
Similarly, appending some far impurity $o$ (line \ref{price:d}) is naturally an expensive operation, as long as $o$ is not anomalous (line \ref{price:s}). Thus, if a cluster overcame the expenses of appending distant impurities, it is because it has presumably appended anomalous impurities. Therefore, as $\mbox{\sc Diameter}(c)$ grows, the degree of the anomaly of $c$ grows as well.
Finally, big and anomalous impurities highly imply that the cluster has a high spatial anomaly score. Therefore the component $\sum_{i\in c.\mathcal{I}} \left(\mbox{\sc Score}(i)*\mbox{\sc Area}^{2}(i)\right)$ grows with the degree of anomaly of $c$.
Since all components are monotonically increasing with the degree of the anomaly of $c$, and because multiplication preserves monotonicity, $am(c)$ is monotonically increasing with the degree of the anomaly of $c$.
\end{proof}
We also note that similarly to all other anomaly measures presented in this work, the usage of multiplication enhances the anomaly scores of clusters with high scores in each of the components, compared to clusters with lower scores on some of the components.
We present the output of the algorithm ($k=10$), on top of spatial and shape anomaly measures, after ordering the clusters based on Eq. \ref{eq_area_measure}, in Fig. \ref{fig:area}. The most anomalous cluster is the red one.



\section{Anomaly Detection Evaluation -- Tuning via Feedback} \label{ex-co}

The promise of an end-to-end model for a diverse field such as QM is only viable in the presence of tuning via feedback. For the task of data segmentation, there is a clear link between the number of labeled data pieces and the model's accuracy. As previously stated, unlike evaluation benchmarks for typical machine learning tasks, in this work, neither the data nor the labeling is trivial, as the data is not natural, extremely unique (even in the electron microscopy arena \cite{ede2021deep}), and inherently problematic to classify binary. As a result, in case of exceptional data to the model or indecisive outcomes, transferring the current model understating should be relatively easy using more labeled data. However, for the anomaly detection task, there are no specific mathematical guidelines for determining anomaly in the aspects of QM a priori (even in cases that benchmarks are present, an anomaly is still subjected to perspective \cite{chalapathy2019deep}). Moreover, from the physical perspective, and even in the presence of big data, there is no guarantee that a mathematical \textit{relative} anomaly is also a \textit{actual physical} anomaly, even when the said model takes into consideration all of the relevant properties of the material. Furthermore, different materials can result in different mathematical and physical anomaly thresholds, and as such, for each case, there is a need to ratify the model results by experiments, or rather re-tune it via experimental feedback. As a result, we introduce a physical method for feedback and re-tune the anomaly detection model to improve our big-data analysis.

\begin{figure}[H]
\begin{minipage}{0.47\textwidth}
  \subcaptionbox{The output of the model on $img_1$. \label{fig:out_img_1}}{\includegraphics[height=6cm, frame]{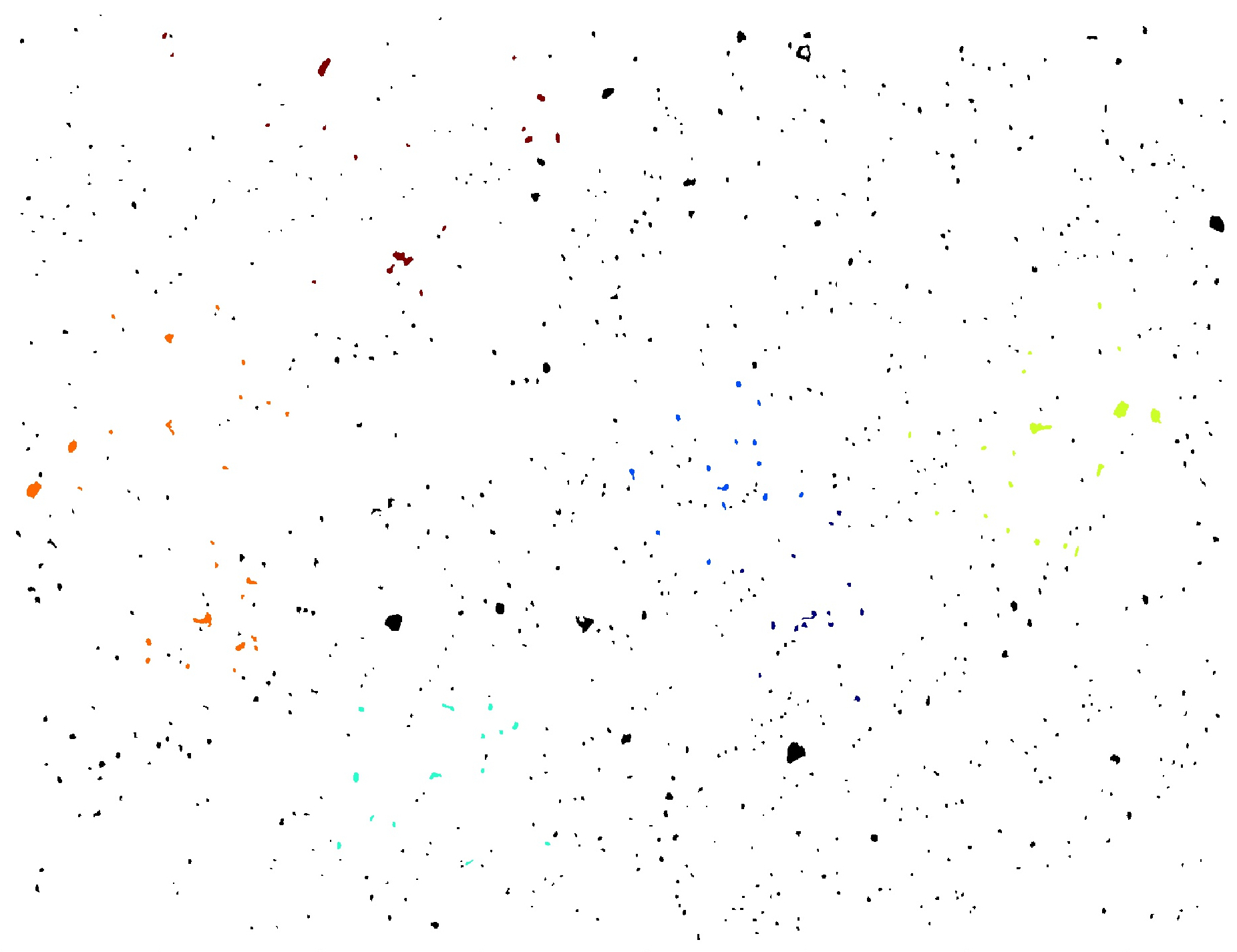}}
\end{minipage}
\begin{minipage}{0.47\textwidth}
  \subcaptionbox{The output of the model on $img_2$. \label{fig:out_anom}}{\includegraphics[height=6cm, frame]{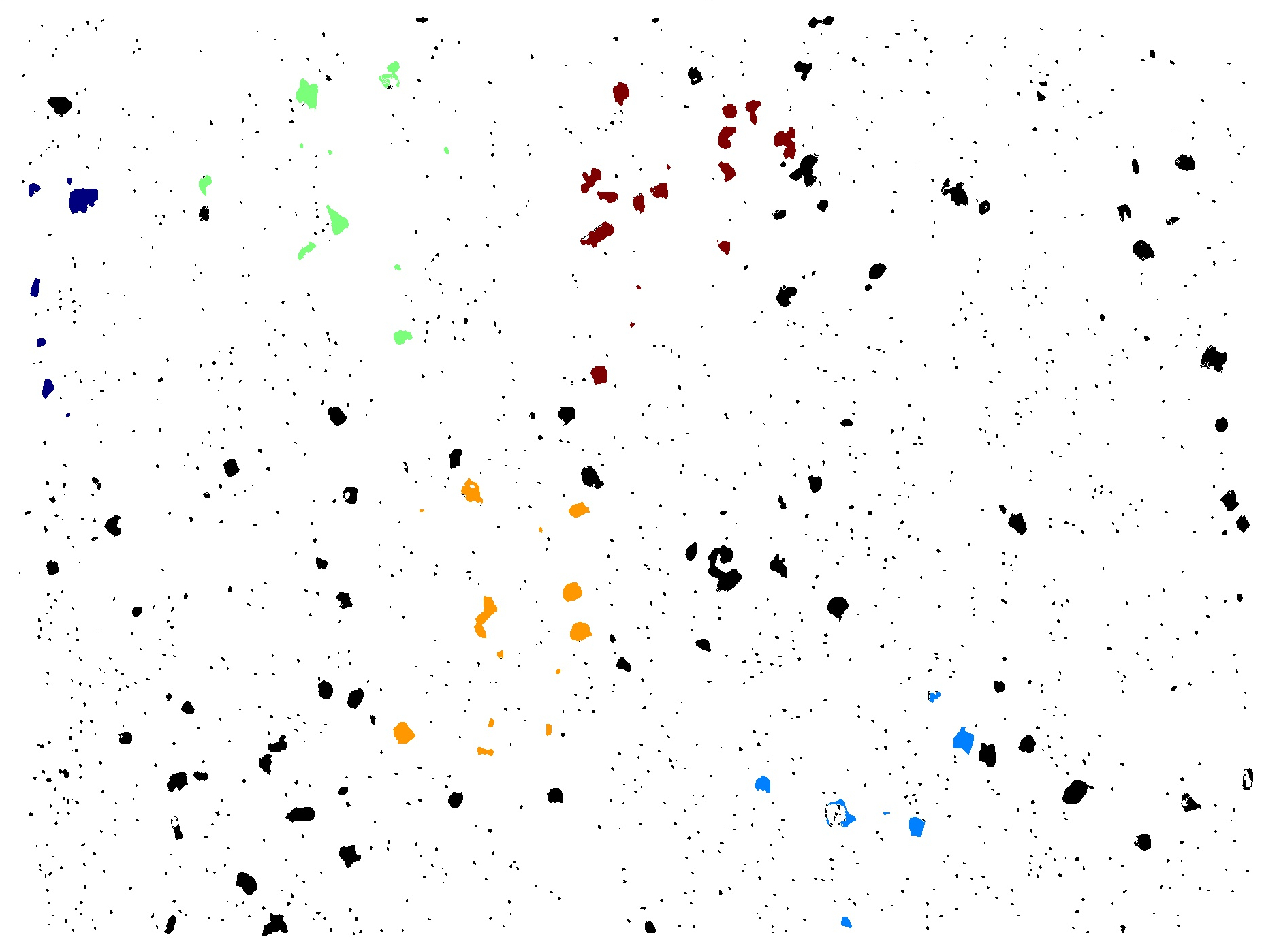}}
\end{minipage}
\begin{minipage}{.05\linewidth}
\subcaptionbox*{\newline}{\vspace{-0.3cm}\includegraphics[height=6cm]{colormap.PNG}}
\end{minipage}
 \caption{Anomaly detection pipeline applied on a test sample.}
\end{figure}

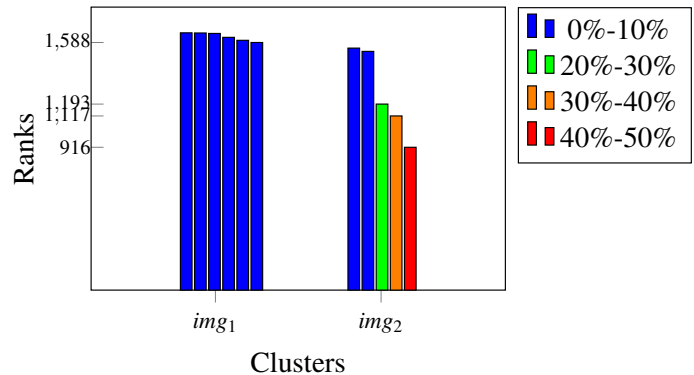
\begin{figure}[H]
    \centering
    \pgfplotsset{width=7cm,compat=1.8}
    \begin{figure}[H]
    \centering
    \resizebox{!}{5cm}{
    \begin{tikzpicture}
        \begin{axis}[
            ybar=0.85pt,
            height=50mm, 
            width=500mm,
            xlabel= Clusters,
            ylabel= Ranks,
            bar width=4pt,
            x = 0.5cm,
            legend pos=outer north east,
            legend entries={0\%-10\%,
                  20\%-30\% ,
                  30\%-40\% ,
                  40\%-50\%
                  },
            enlarge x limits={abs=2cm},
            ymin=0,
            ytick={916,1117, 1193, 1588},
            xtick={-1,3},
            xticklabels={{$img_1$},{$img_2$}},
            y tick label style={inner sep=0, font=\scriptsize,},
            x tick label style={align=center,font=\footnotesize,},
            xtick pos=left,
            ytick pos=left,
        ]
        \addlegendimage{fill=blue}
        \addlegendimage{fill=green}
        \addlegendimage{fill=orange}
        \addlegendimage{fill=red}
        \addplot [fill=blue] coordinates {(0,1650)};
        \addplot [fill=blue] coordinates {(0,1649)};
        \addplot [fill=blue] coordinates {(0,1646)};
        \addplot [fill=blue] coordinates {(0,1621)};
        \addplot [fill=blue] coordinates {(0,1602)};
        \addplot [fill=blue] coordinates {(0,1588)};
        %
        %
        \addplot [fill=blue] coordinates {(2,1552)};
        \addplot [fill=blue] coordinates {(2,1531)};
        \addplot [fill=green] coordinates {(2,1193)};
        \addplot [fill=orange] coordinates {(2,1117)};
        \addplot [fill=red] coordinates {(2,916)};
        \end{axis}
    \end{tikzpicture}
    }
    \caption{Ranks of all clusters. Each group is a single test scan.} \label{fig:all_ranks}
    \end{figure}
\end{figure}

We suggest testing the anomaly detection model with the following procedure: Prepare fresh metallographic samples and use the model to locate and quantify the anomaly scores of the most anomalous areas of impurities inside them. The model outputs are then used in order to determine whether and where there were physical defects in the materials, specifically in the areas of interest (see Section \ref{anomaly_detection_measures}). For example, the results of two samples are shown in Figures \ref{fig:out_img_1} and \ref{fig:out_anom}. The most anomalous test scans areas were ordered together with all other scans in the dataset, and they were placed in places 1588 and 916, respectively, out of a total of 1653 clusters. Cluster 1588 resulted under the first decile, while 916 was placed between the fourth and fifth deciles. All the results are shown in Fig. \ref{fig:all_ranks}.

After getting the outputs from the model, two examinations should be made: 1. Microhardness Vickers (MHV) \cite{buckle1959progress} test in the vicinity of the most anomalous inclusions (the inclusions size and the microhardness trace are both on the same scale of few to tens micrometers) and in normal areas; 2. EDS (energy dispersive spectroscopy) analysis in a SEM \cite{goldstein2017scanning} to evaluate the inclusions composition. For example, for the two said samples, we performed an examination by MHV and EDS to determine whether the mathematical \textit{relative} anomaly is also a \textit{actual physical} anomaly and found in this case that there was no difference between normal and anomalous inclusions. Although it is clear that the most anomalous area in Fig. \ref{fig:out_anom} looks much more anomalous than the others. Since we observed that there is no difference between normal and anomalous areas, we conclude that while our mathematical \textit{relative} anomaly detection model is capable of quantifying successfully how each area of inclusions is anomalous compared to all other areas, it is the task of the experienced user of the model to determine the threshold from which the area is considered anomalous enough to be defective and to re-tune the mathematical model accordingly.

\section*{Acknowledgments}

This work was supported by the Pazy foundation. Computational support was provided by the NegevHPC project~\cite{negevhpc}. The authors would like to thank Dr. Eyal Yahel for fruitful collaboration, and Prof. Shai Avidan for constructive suggestions.
\bibliography{egbib}
\newpage
\section*{Appendix A -- Data Examples} \label{app}
\begin{figure}[H]
\centering
\begin{minipage}{.48\textwidth}
  \centering
  \begin{figure}[H]
  \centering
      \includegraphics[height=5cm, frame]{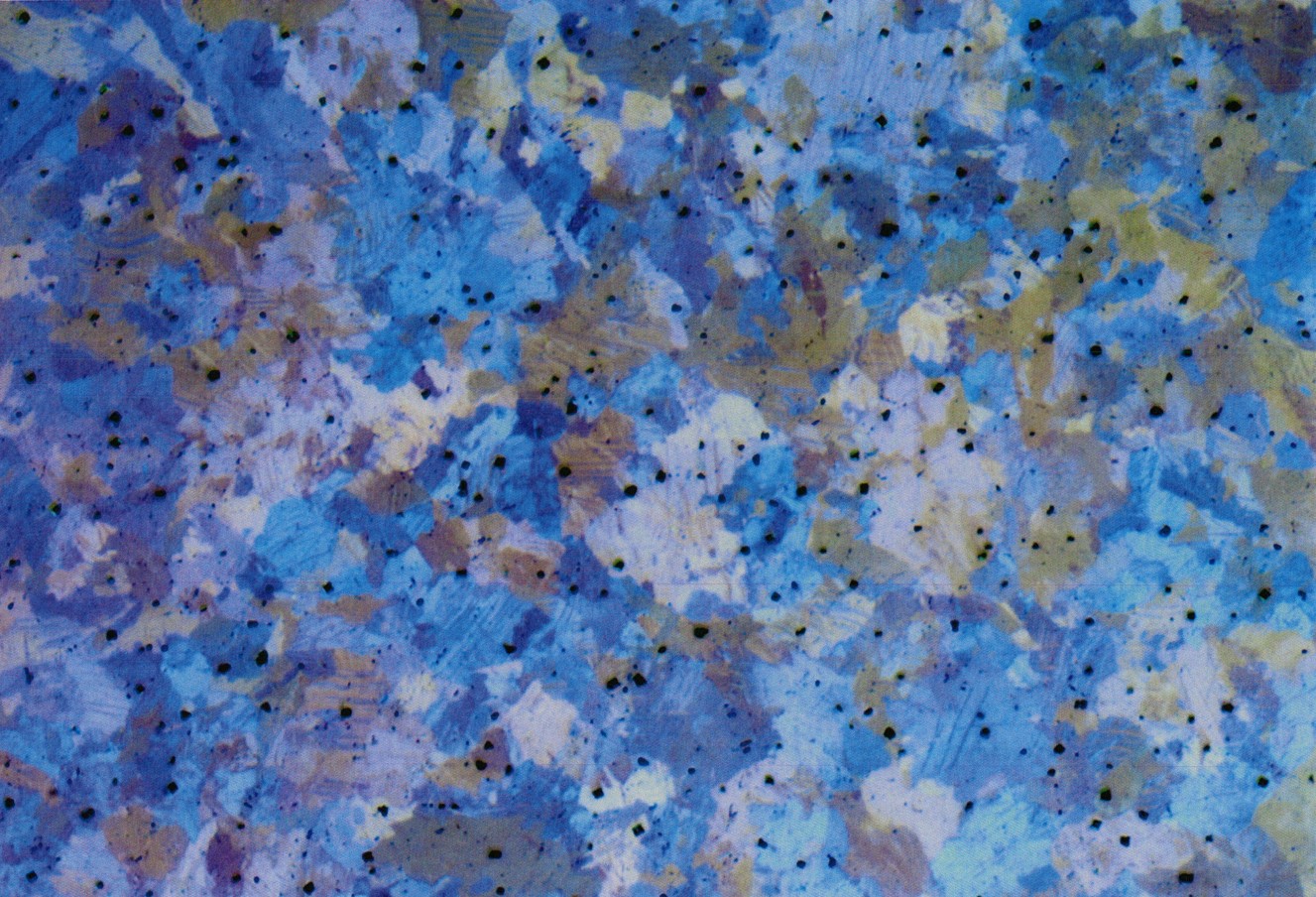}
      \caption*{Input image}
  \end{figure}
\end{minipage}%
\begin{minipage}{.48\textwidth}
  \centering
  \begin{figure}[H]
  \centering
        \includegraphics[height=5cm, frame]{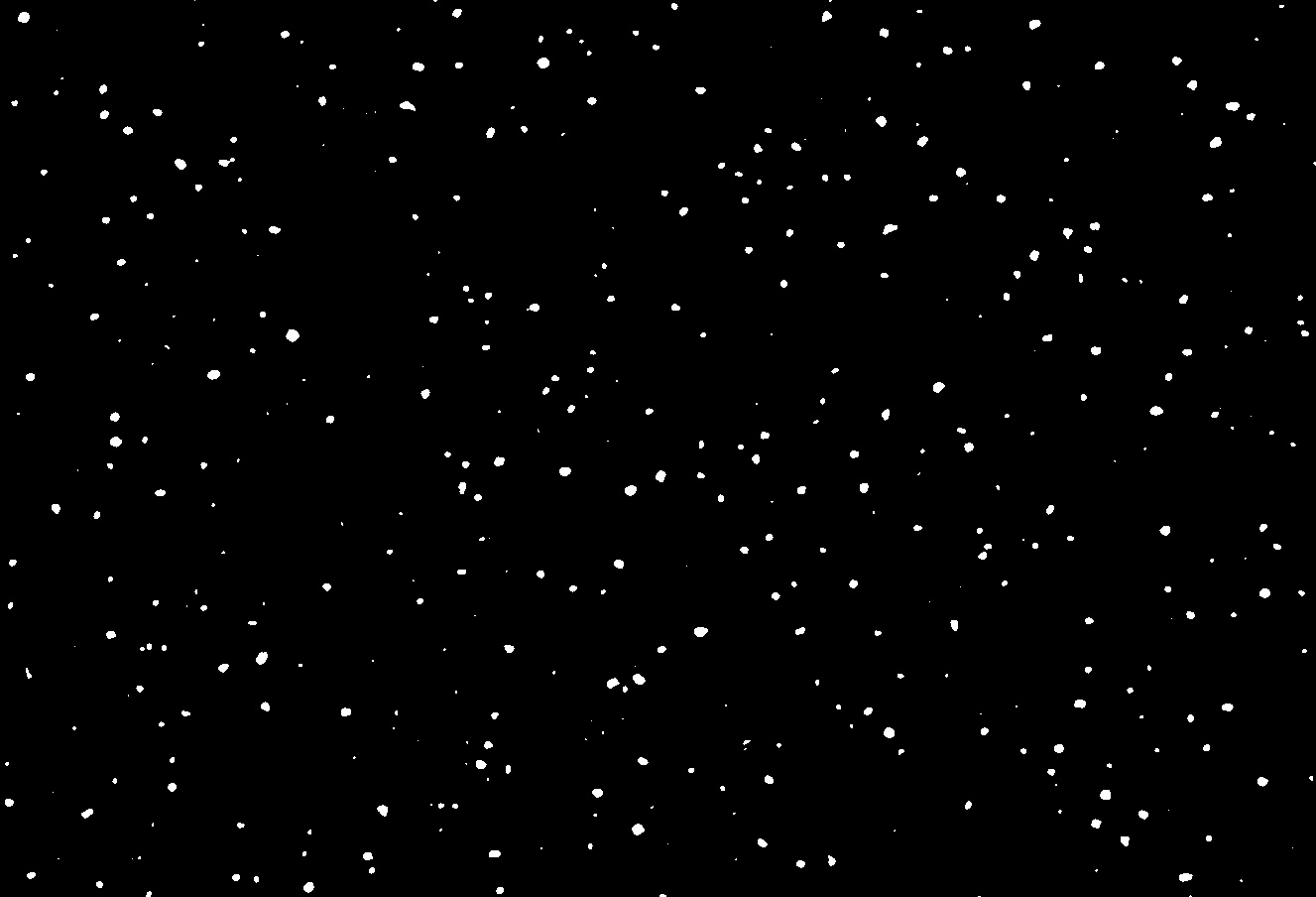}
        \caption*{Impurities segmentation}
  \end{figure}
\end{minipage}%
\end{figure}
\begin{figure}[H]
\centering
\begin{minipage}{.48\textwidth}
  \centering
  \begin{figure}[H]
  \centering
        \includegraphics[height=5cm, frame]{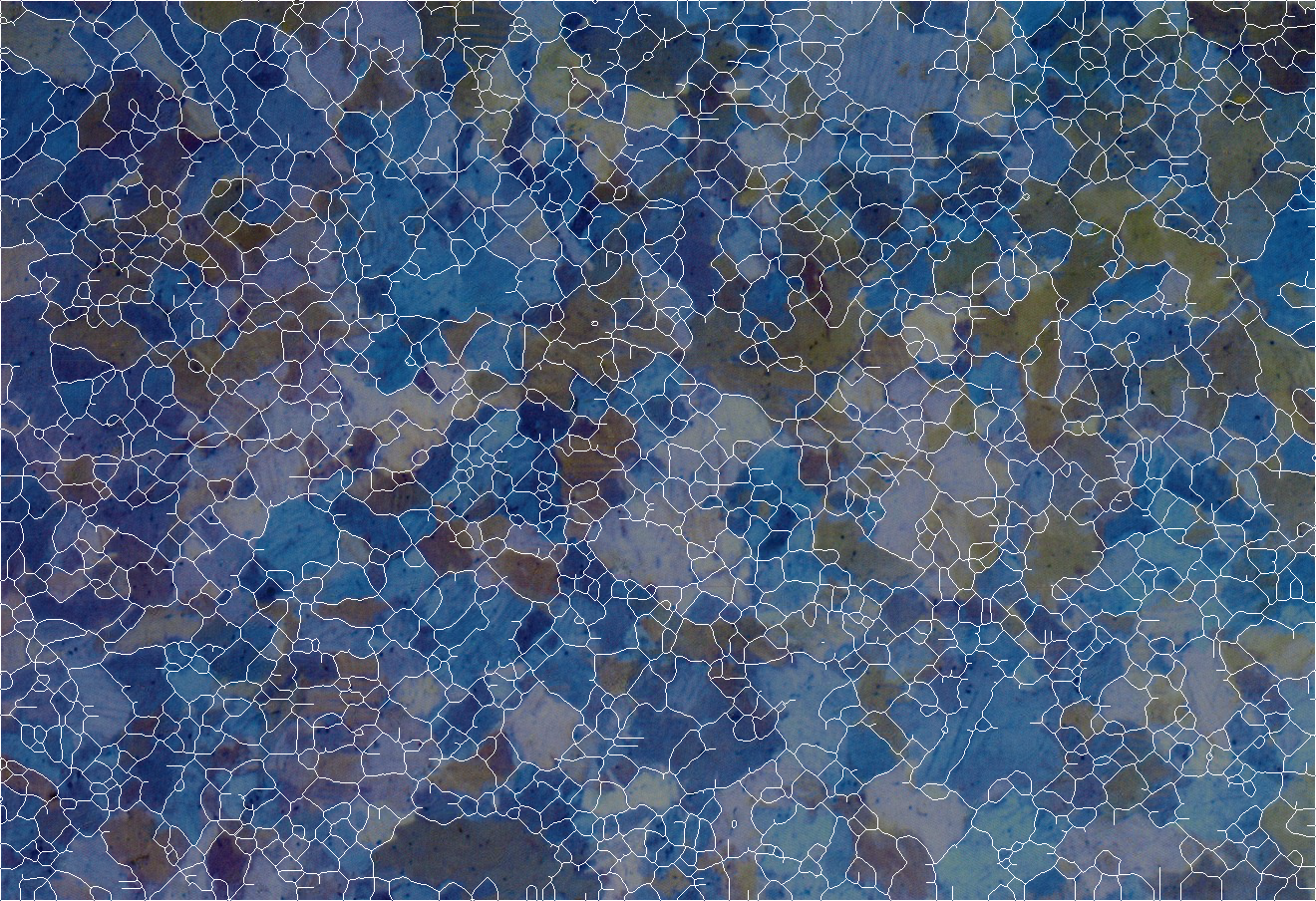}
        \caption*{Inpainting and GB segmentation}
  \end{figure}
\end{minipage}%
\begin{minipage}{.48\textwidth}
  \centering
  \begin{figure}[H]
  \centering
      \includegraphics[height=5cm, frame]{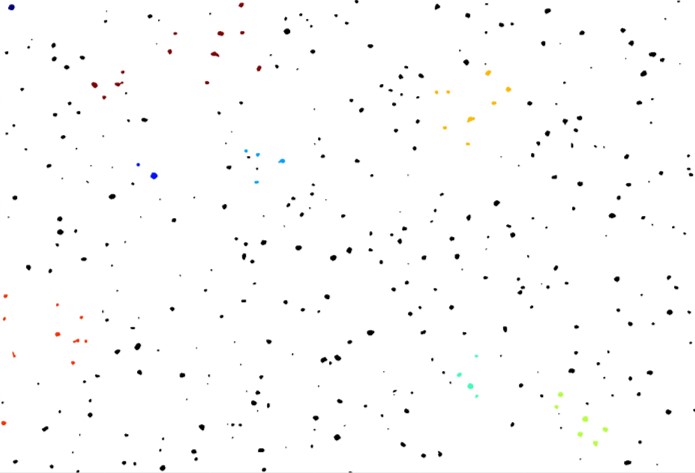}
      \caption*{Area anomaly detection}
  \end{figure}
\end{minipage}%
\caption{Example \#1.}\label{appendix:1}
\end{figure}

\newpage
\begin{figure}[H]
\centering
\begin{minipage}{.48\textwidth}
  \centering
  \begin{figure}[H]
  \centering
      \includegraphics[height=5cm, frame]{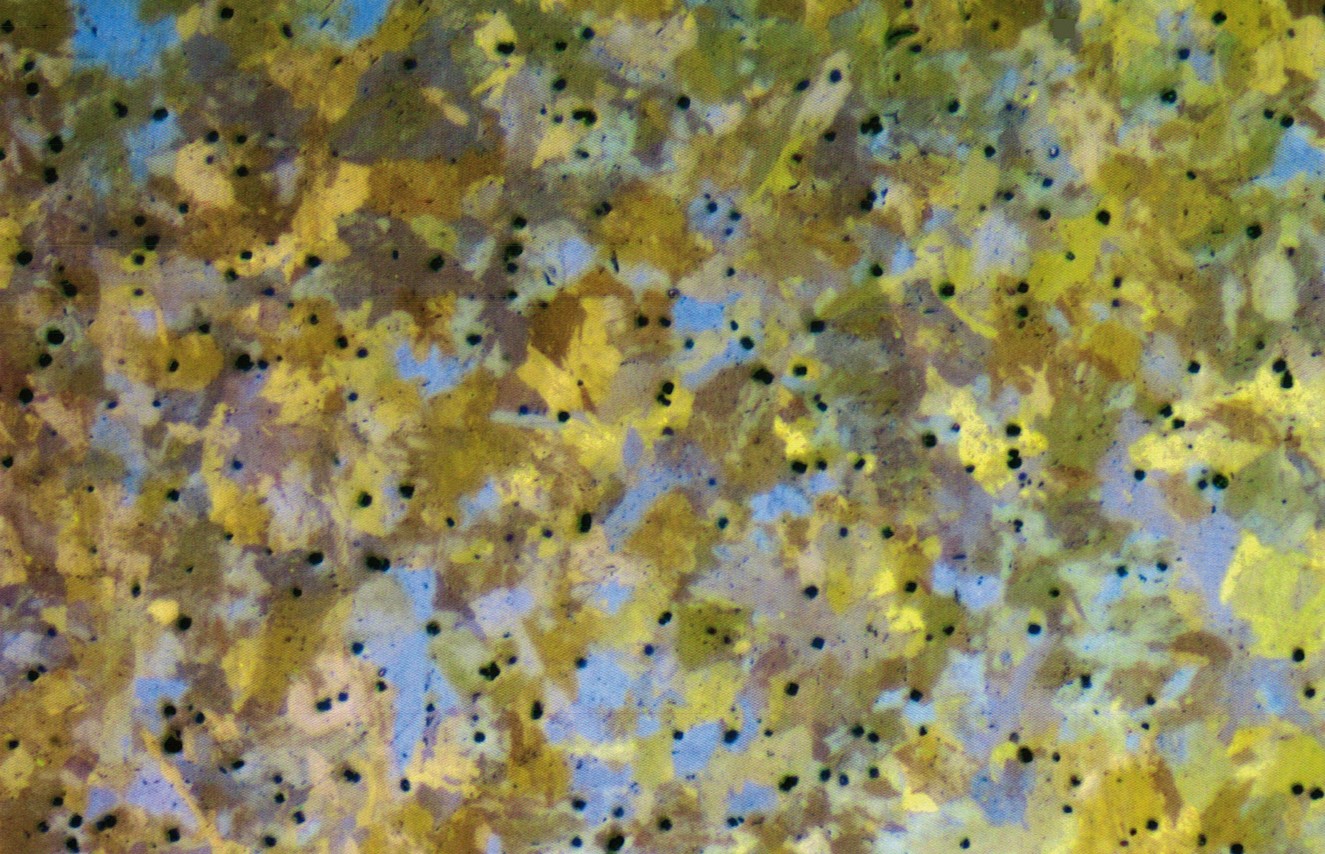}
      \caption*{Input image}
  \end{figure}
\end{minipage}%
\begin{minipage}{.48\textwidth}
  \centering
  \begin{figure}[H]
  \centering
      \includegraphics[height=5cm, frame]{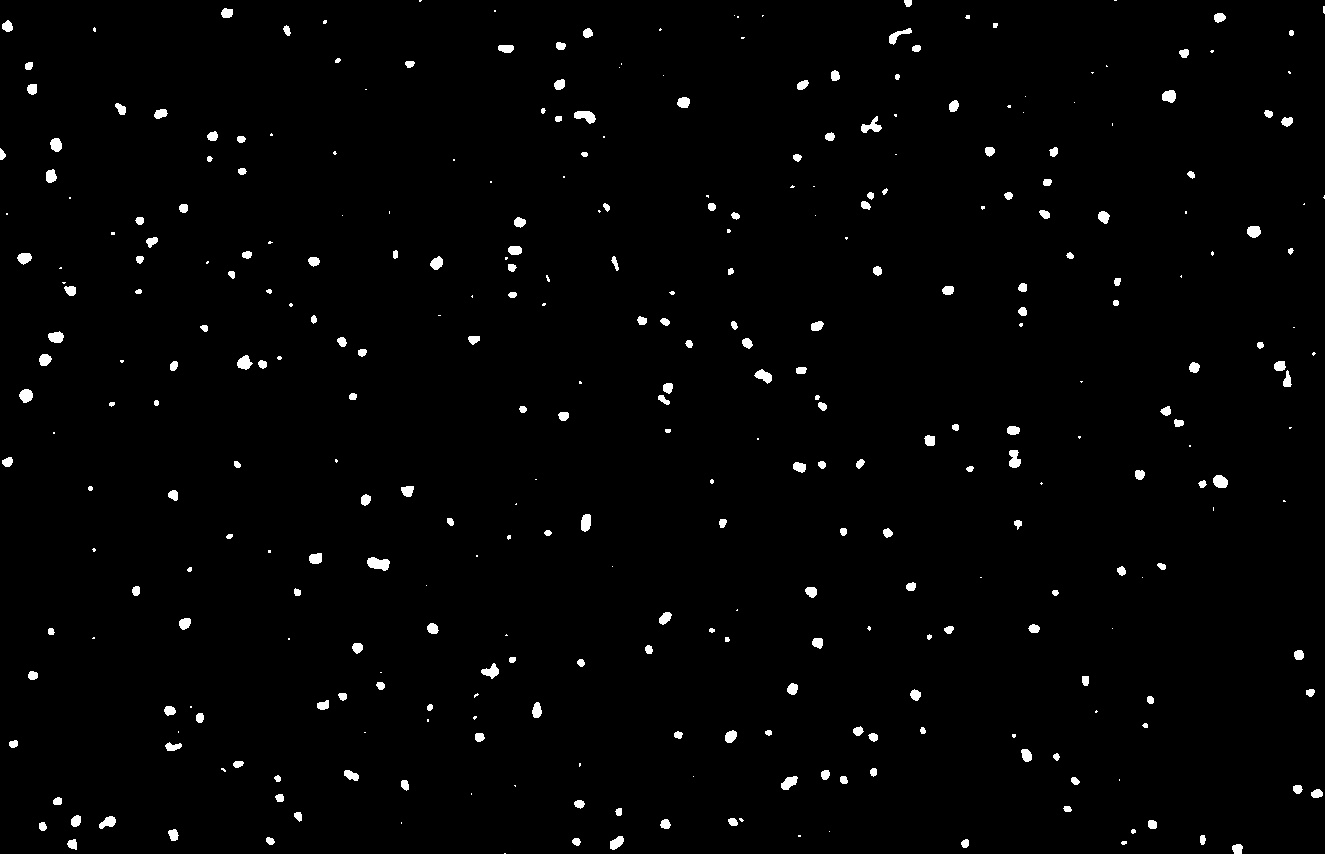}
      \caption*{Impurities segmentation}
  \end{figure}
\end{minipage}%
\end{figure}
\vspace{-0.5cm}
\begin{figure}[H]
\centering
\begin{minipage}{.48\textwidth}
  \centering
  \begin{figure}[H]
  \centering
      \includegraphics[height=5cm, frame]{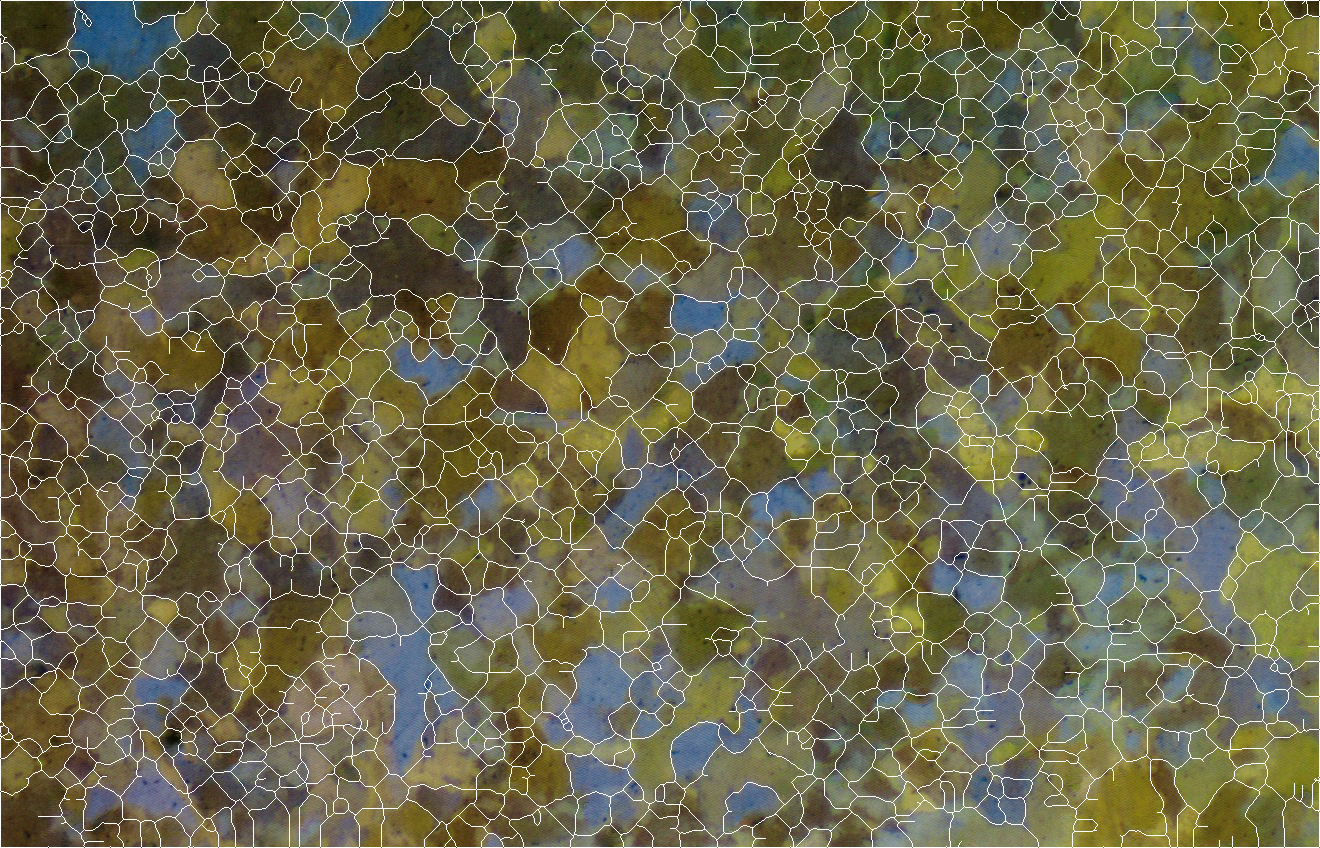}
      \caption*{Inpainting and GB segmentation}
  \end{figure}
\end{minipage}%
\begin{minipage}{.48\textwidth}
  \centering
  \begin{figure}[H]
  \centering
      \includegraphics[height=5cm, frame]{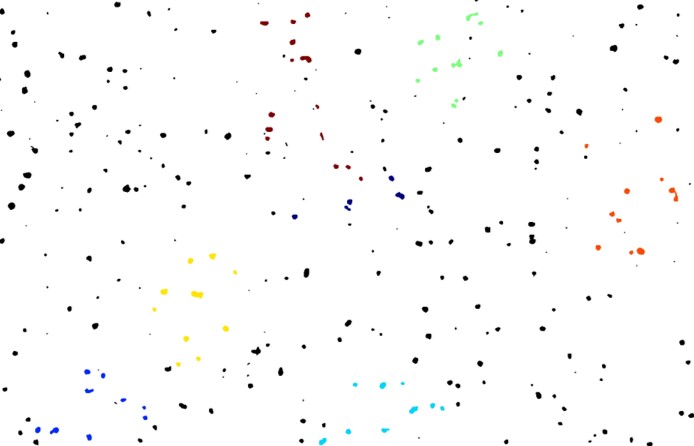}
      \caption*{Area anomaly detection}
  \end{figure}
\end{minipage}%
\caption{Example \#2.}\label{appendix:2}
\end{figure}

\newpage
\begin{figure}[H]
\centering
\begin{minipage}{.48\textwidth}
  \centering
  \begin{figure}[H]
  \centering
      \includegraphics[height=5cm, frame]{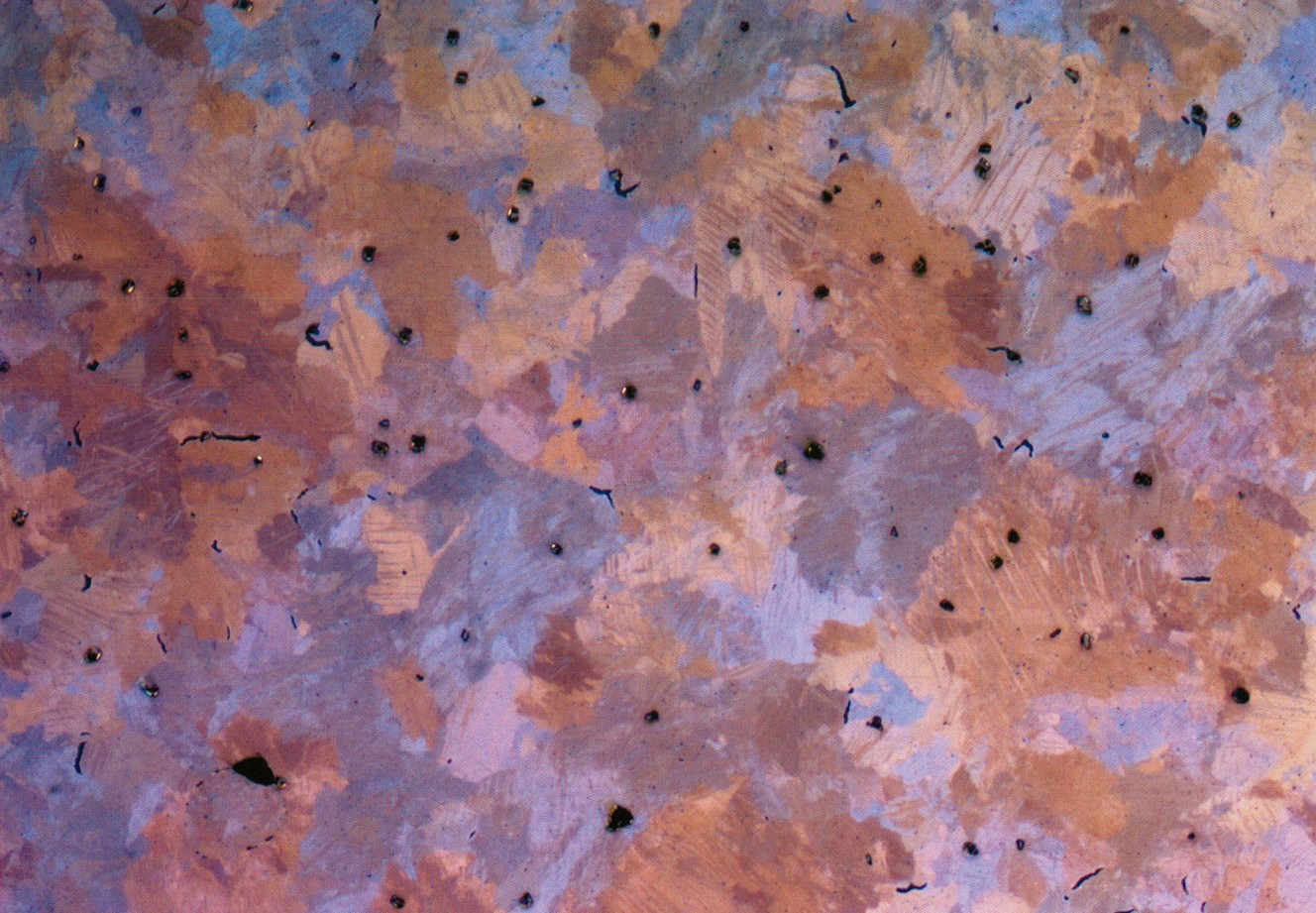}
      \caption*{Input image}
  \end{figure}
\end{minipage}%
\begin{minipage}{.48\textwidth}
  \centering
  \begin{figure}[H]
  \centering
      \includegraphics[height=5cm, frame]{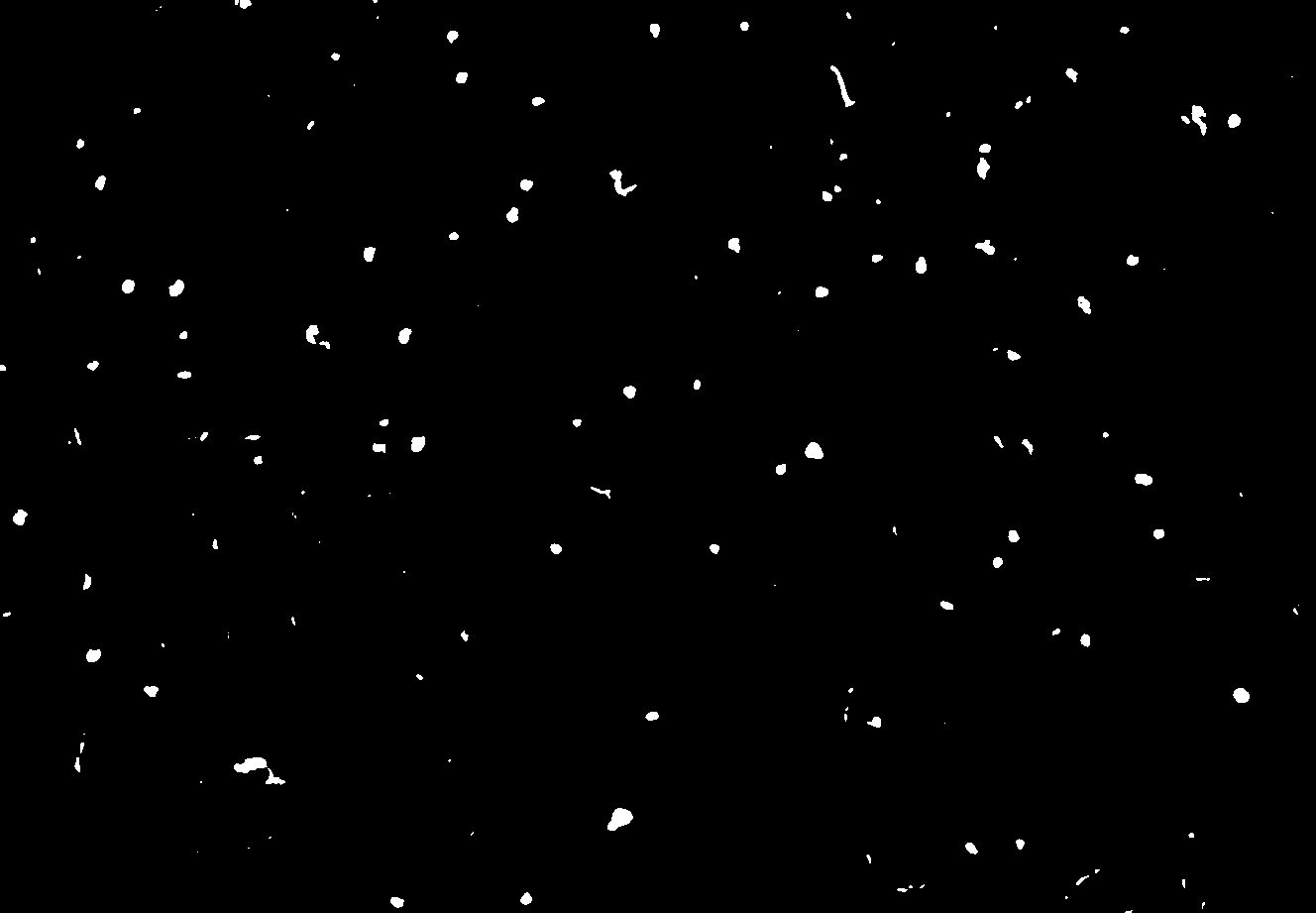}
      \caption*{Impurities segmentation}
  \end{figure}
\end{minipage}%
\end{figure}
\vspace{-0.5cm}
\begin{figure}[H]
\centering
\begin{minipage}{.48\textwidth}
  \centering
  \begin{figure}[H]
  \centering
      \includegraphics[height=5cm, frame]{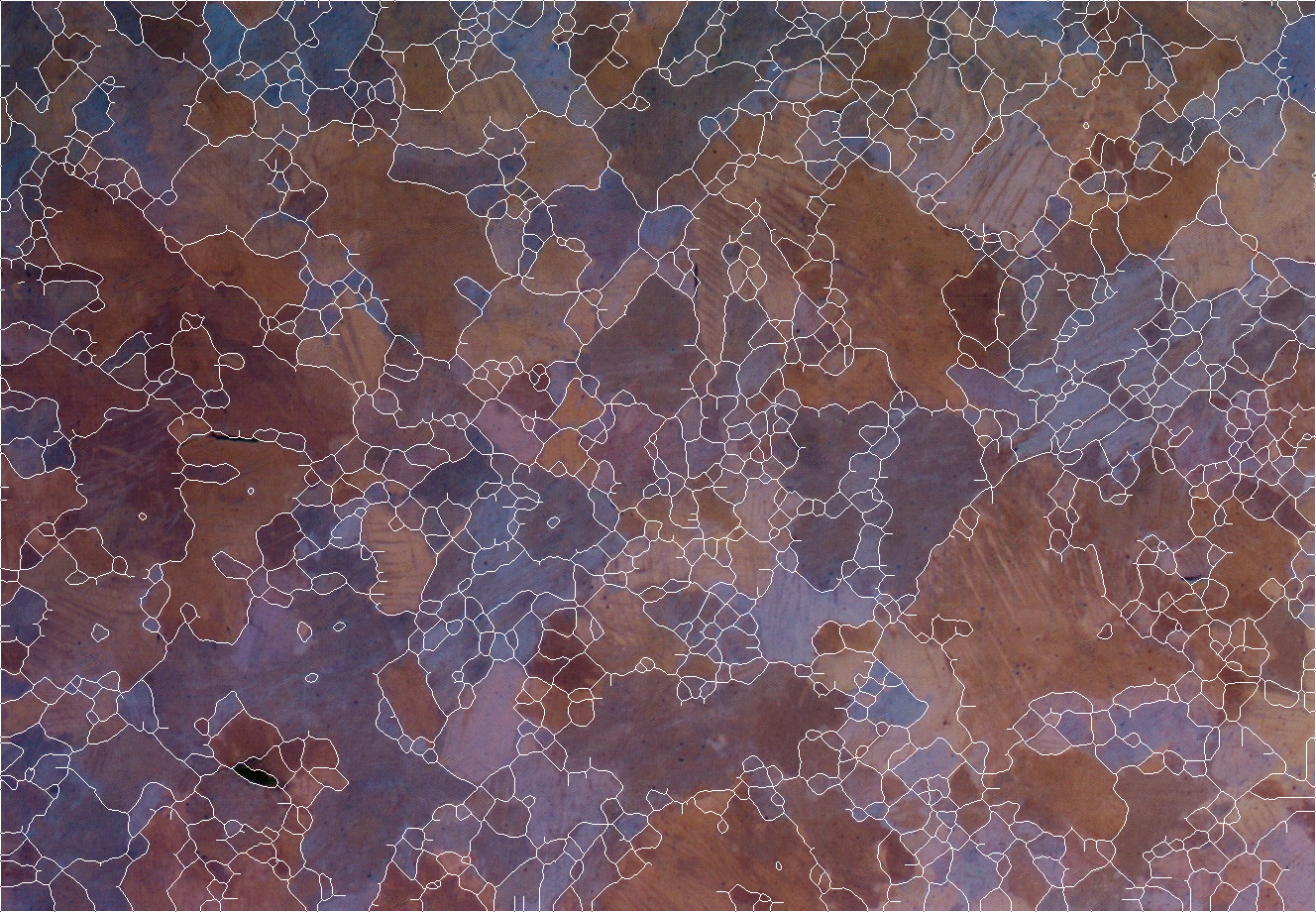}
      \caption*{Inpainting and GB segmentation}
  \end{figure}
\end{minipage}%
\begin{minipage}{.48\textwidth}
  \centering
  \begin{figure}[H]
  \centering
      \includegraphics[height=5cm, frame]{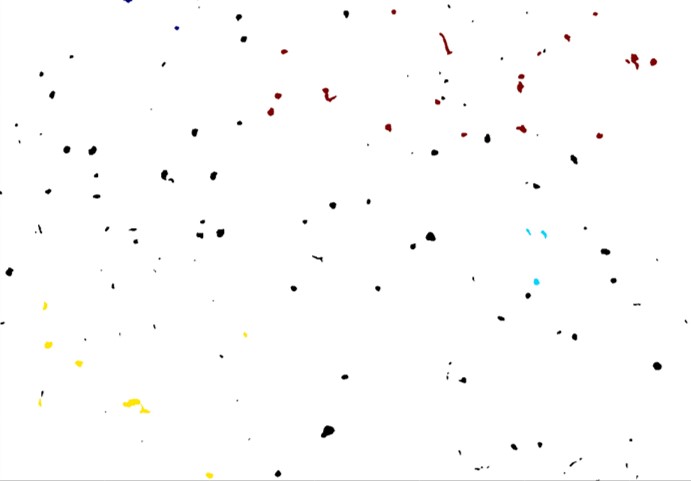}
      \caption*{Area anomaly detection}
  \end{figure}
\end{minipage}%
\caption{Example \#3.}\label{appendix:3}
\end{figure}

\newpage
\begin{figure}[H]
\centering
\begin{minipage}{.48\textwidth}
  \centering
  \begin{figure}[H]
  \centering
      \includegraphics[height=5cm, frame]{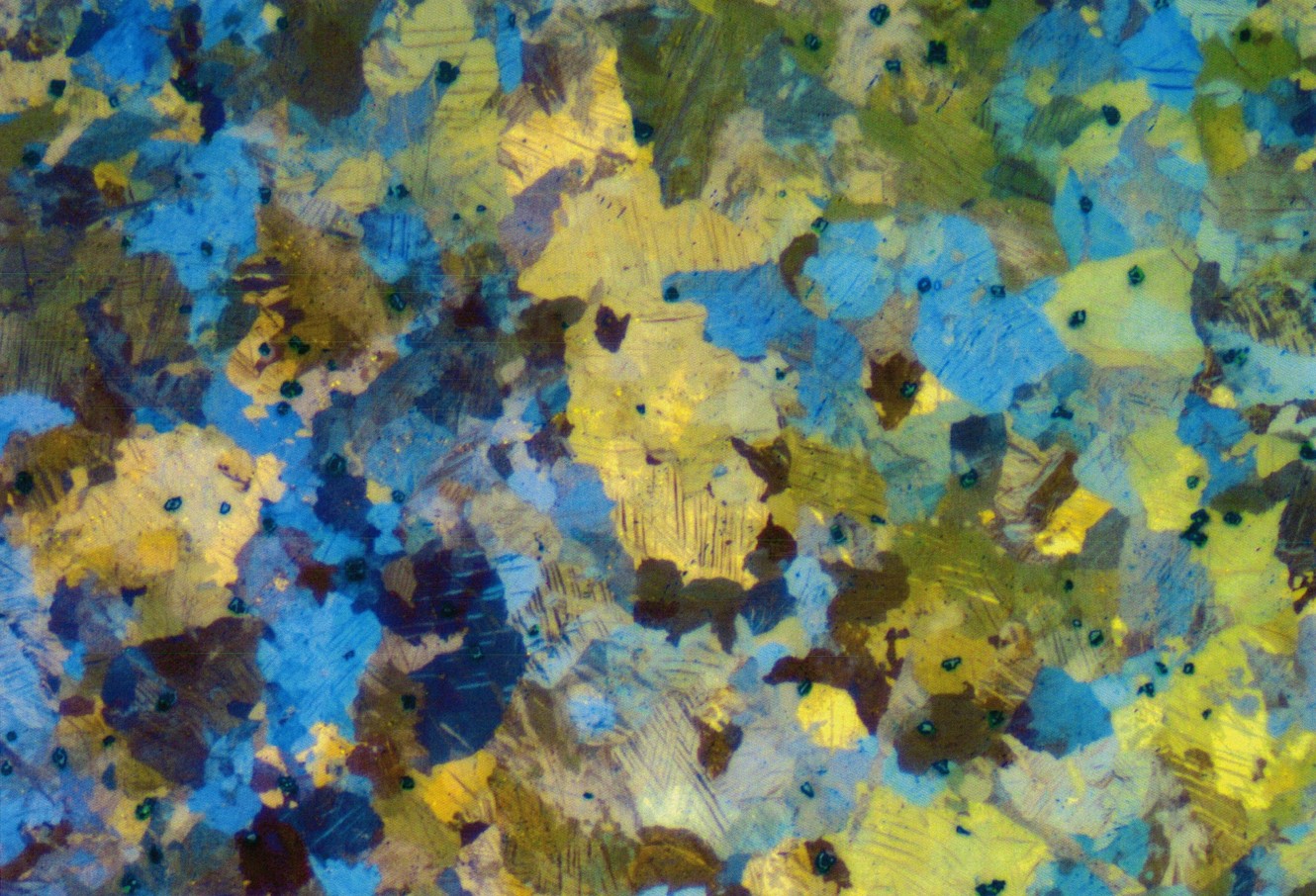}
      \caption*{Input image}
  \end{figure}
\end{minipage}%
\begin{minipage}{.48\textwidth}
  \centering
  \begin{figure}[H]
  \centering
      \includegraphics[height=5cm, frame]{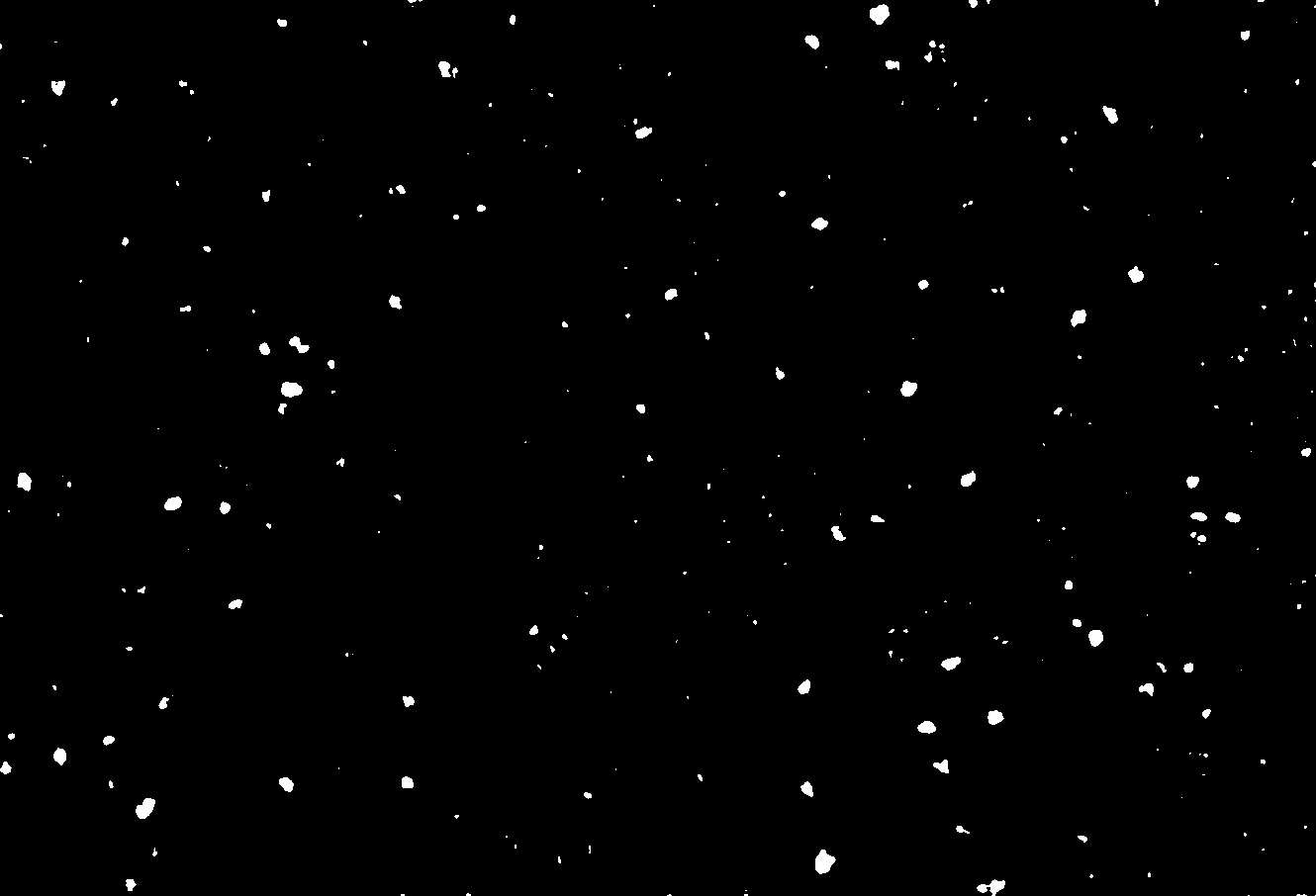}
      \caption*{Impurities segmentation}
  \end{figure}
\end{minipage}%
\end{figure}
\vspace{-0.5cm}
\begin{figure}[H]
\centering
\begin{minipage}{.48\textwidth}
  \centering
  \begin{figure}[H]
  \centering
      \includegraphics[height=5cm, frame]{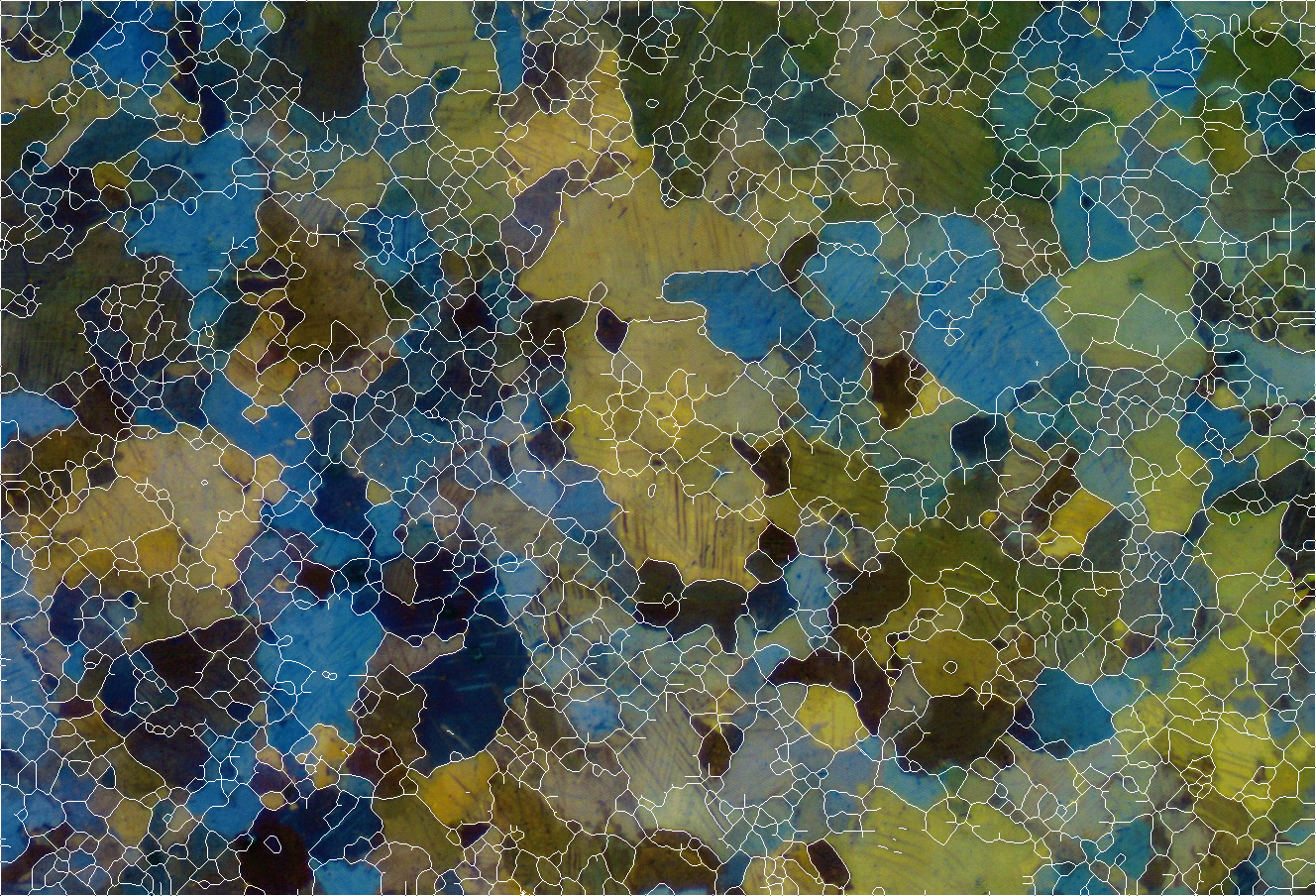}
      \caption*{Inpainting and GB segmentation}
  \end{figure}
\end{minipage}%
\begin{minipage}{.48\textwidth}
  \centering
  \begin{figure}[H]
  \centering
      \includegraphics[height=5cm, frame]{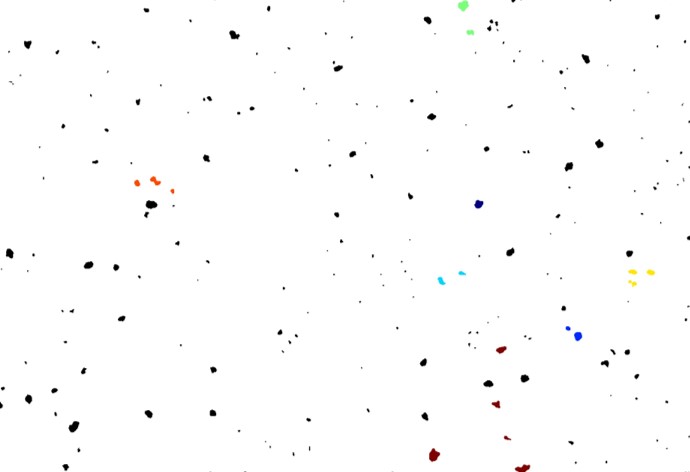}
      \caption*{Area anomaly detection}
  \end{figure}
\end{minipage}%
\caption{Example \#4.}\label{appendix:4}
\end{figure}

\newpage
\begin{figure}[H]
\centering
\begin{minipage}{.48\textwidth}
  \centering
  \begin{figure}[H]
  \centering
      \includegraphics[height=5cm, frame]{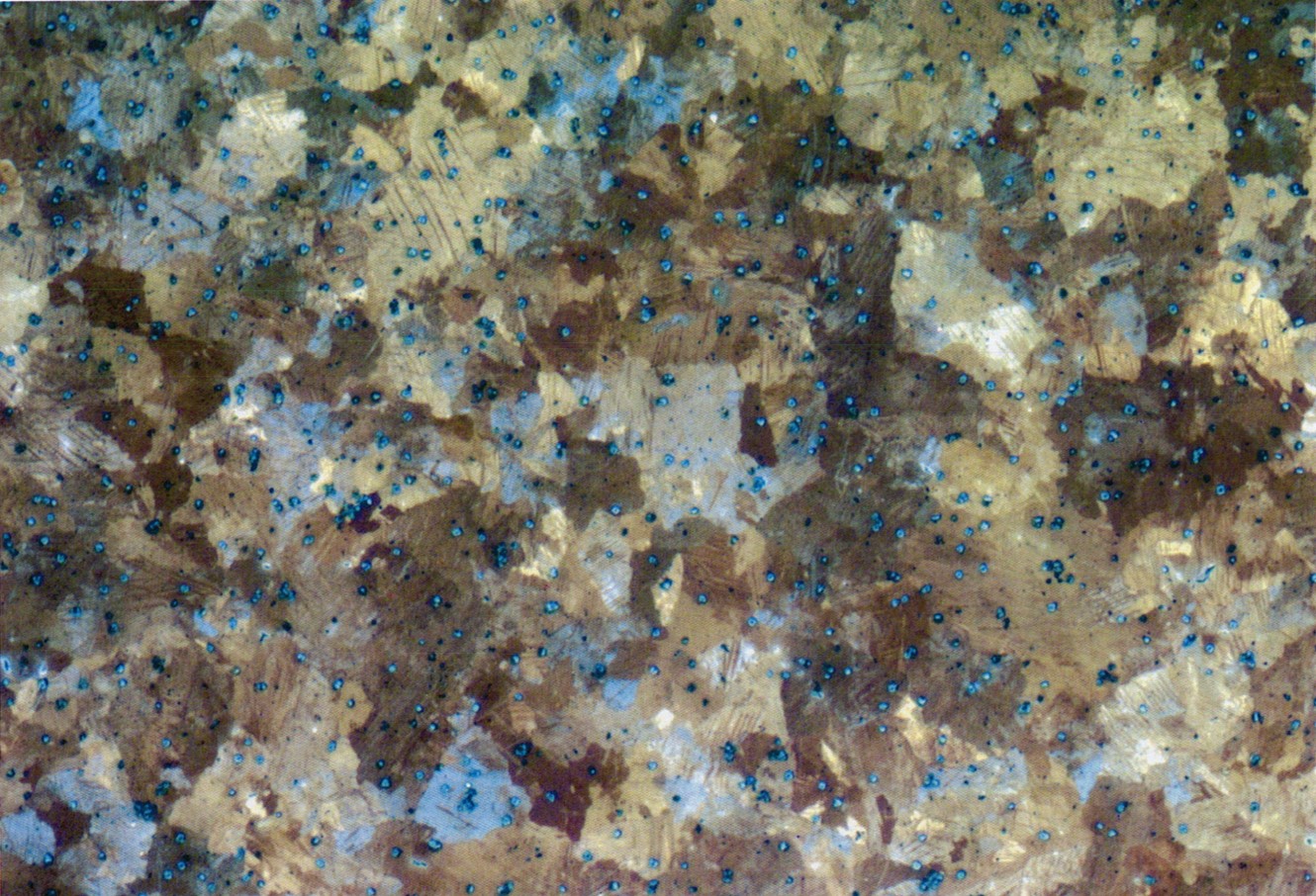}
      \caption*{Input image}
  \end{figure}
\end{minipage}%
\begin{minipage}{.48\textwidth}
  \centering
  \begin{figure}[H]
  \centering
      \includegraphics[height=5cm, frame]{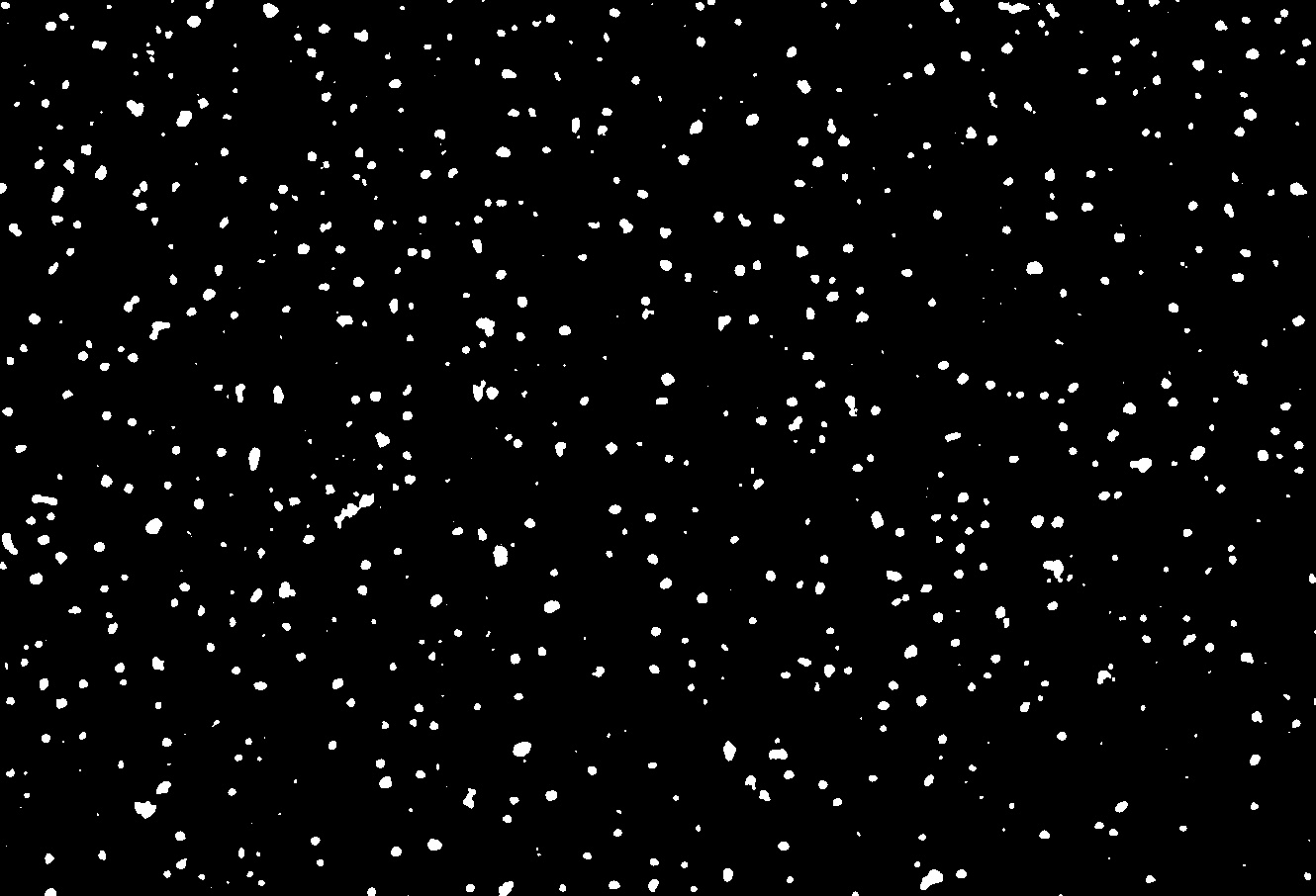}
      \caption*{Impurities segmentation}
  \end{figure}
\end{minipage}%
\end{figure}
\vspace{-0.5cm}
\begin{figure}[H]
\centering
\begin{minipage}{.48\textwidth}
  \centering
  \begin{figure}[H]
  \centering
      \includegraphics[height=5cm, frame]{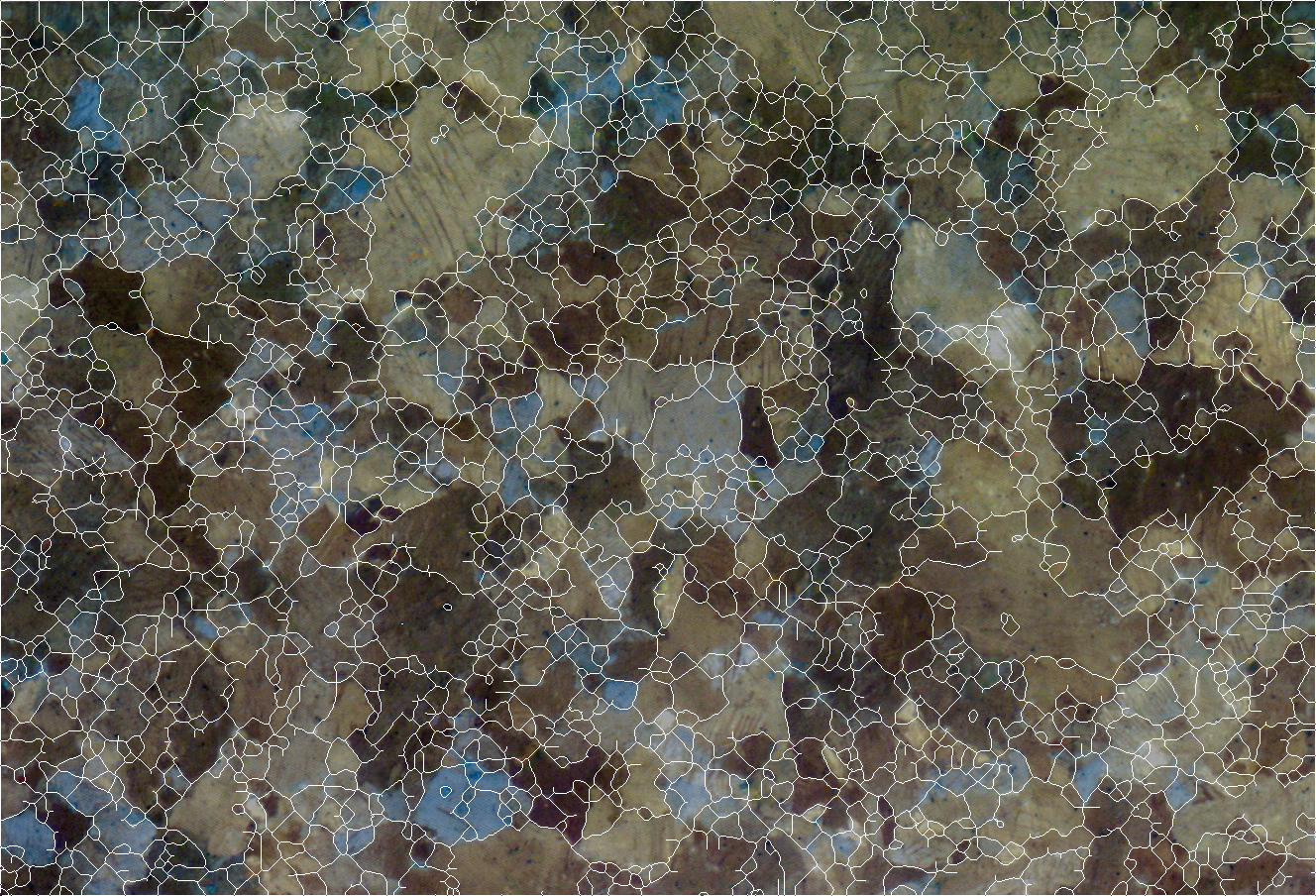}
      \caption*{Inpainting and GB segmentation}
  \end{figure}
\end{minipage}%
\begin{minipage}{.48\textwidth}
  \centering
  \begin{figure}[H]
  \centering
      \includegraphics[height=5cm, frame]{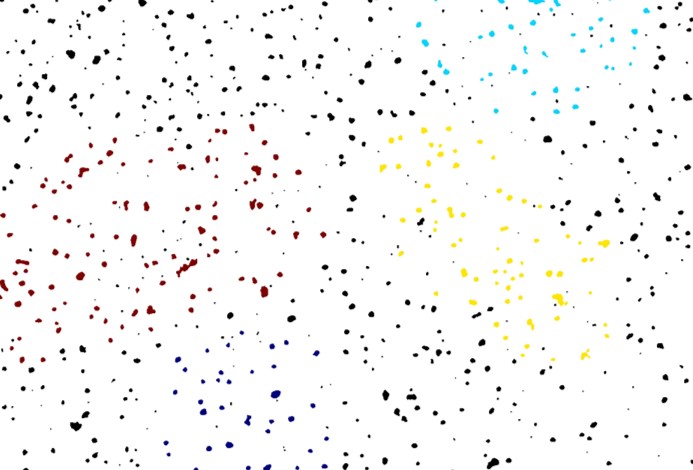}
      \caption*{Area anomaly detection}
  \end{figure}
\end{minipage}%
\caption{Example \#5.}\label{appendix:5}
\end{figure}

\newpage
\begin{figure}[H]
\centering
\begin{minipage}{.48\textwidth}
  \centering
  \begin{figure}[H]
  \centering
      \includegraphics[height=5cm, frame]{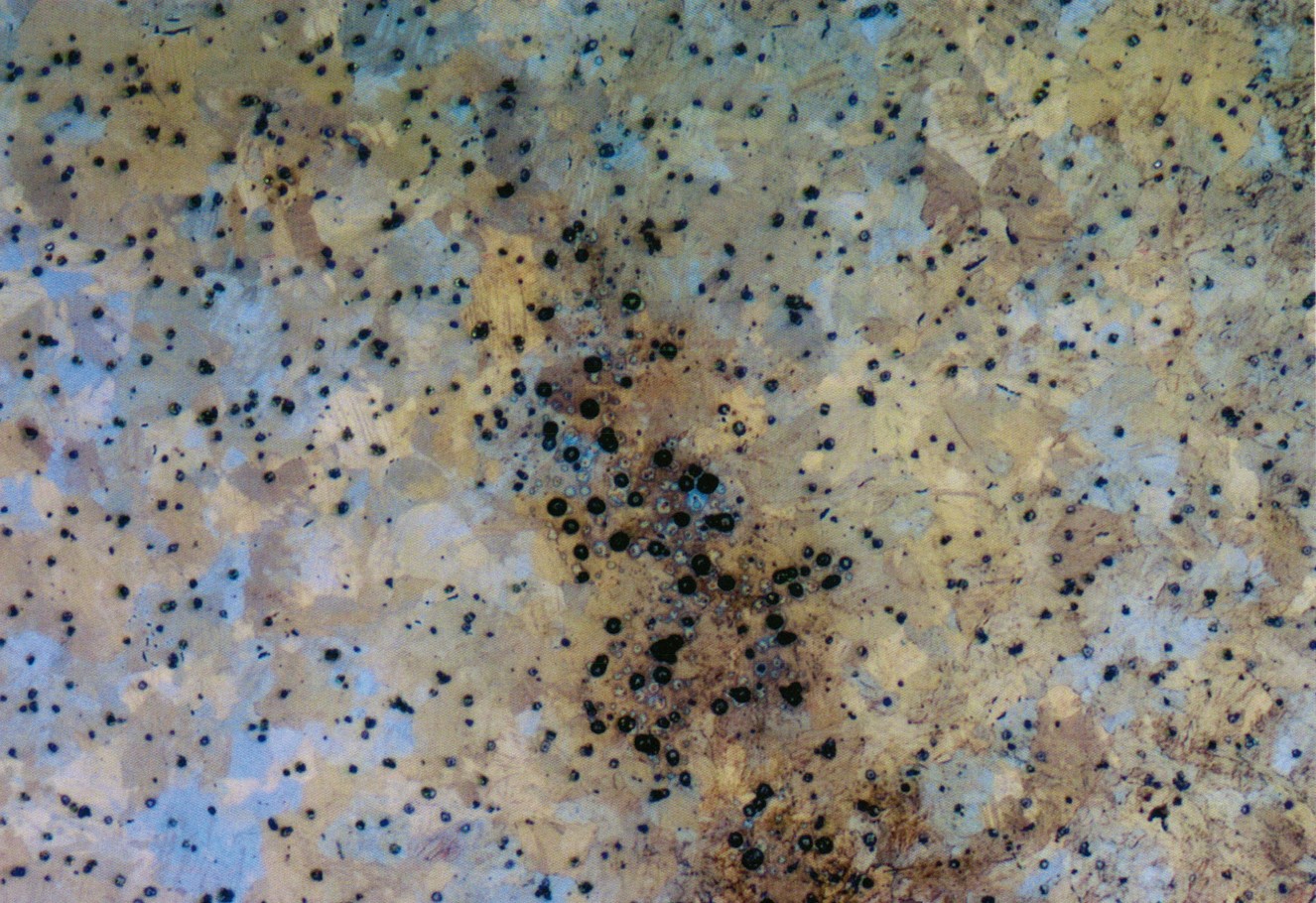}
      \caption*{Input image}
  \end{figure}
\end{minipage}%
\begin{minipage}{.48\textwidth}
  \centering
  \begin{figure}[H]
  \centering
      \includegraphics[height=5cm, frame]{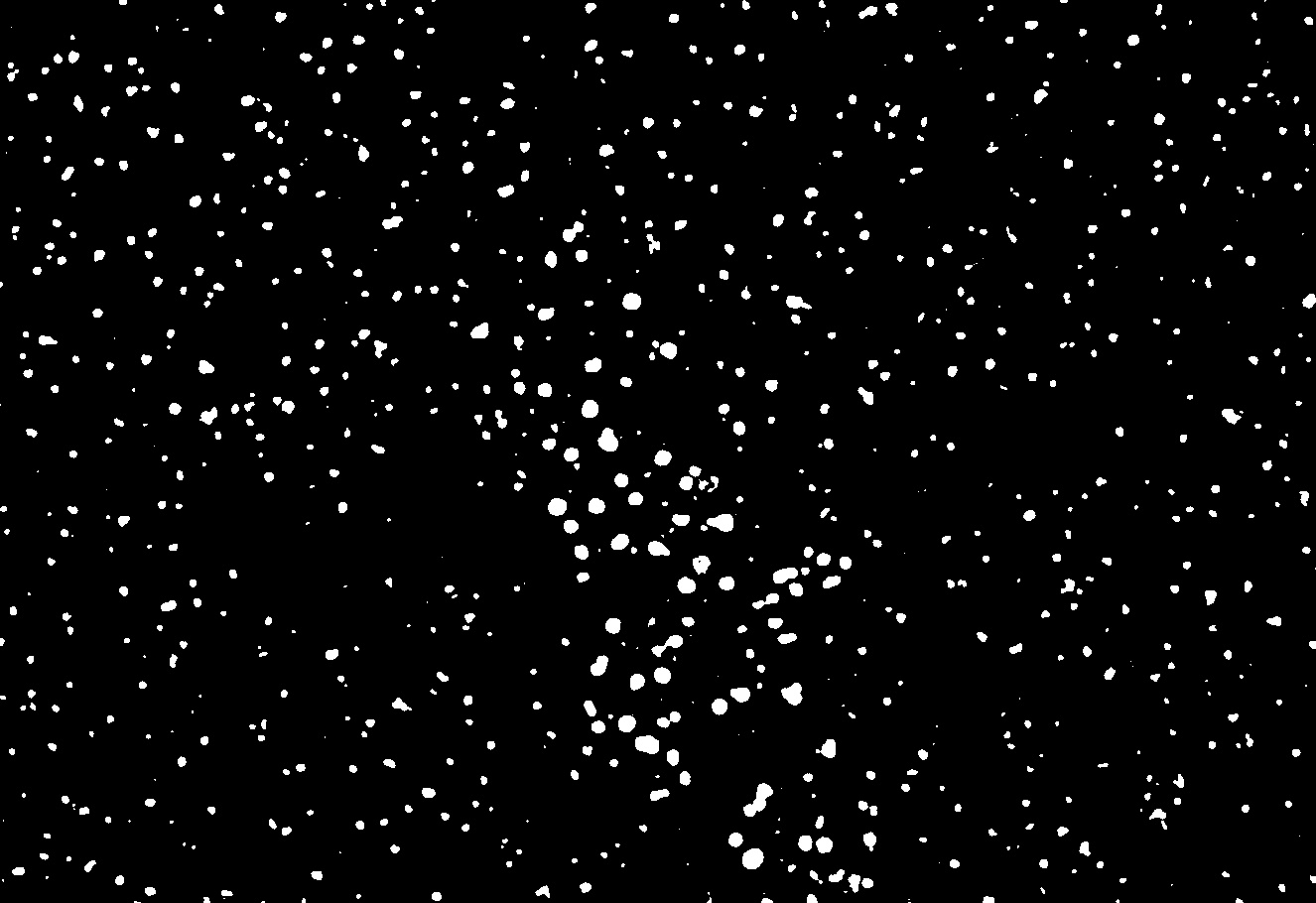}
      \caption*{Impurities segmentation}
  \end{figure}
\end{minipage}%
\end{figure}
\vspace{-0.5cm}
\begin{figure}[H]
\centering
\begin{minipage}{.48\textwidth}
  \centering
  \begin{figure}[H]
  \centering
      \includegraphics[height=5cm, frame]{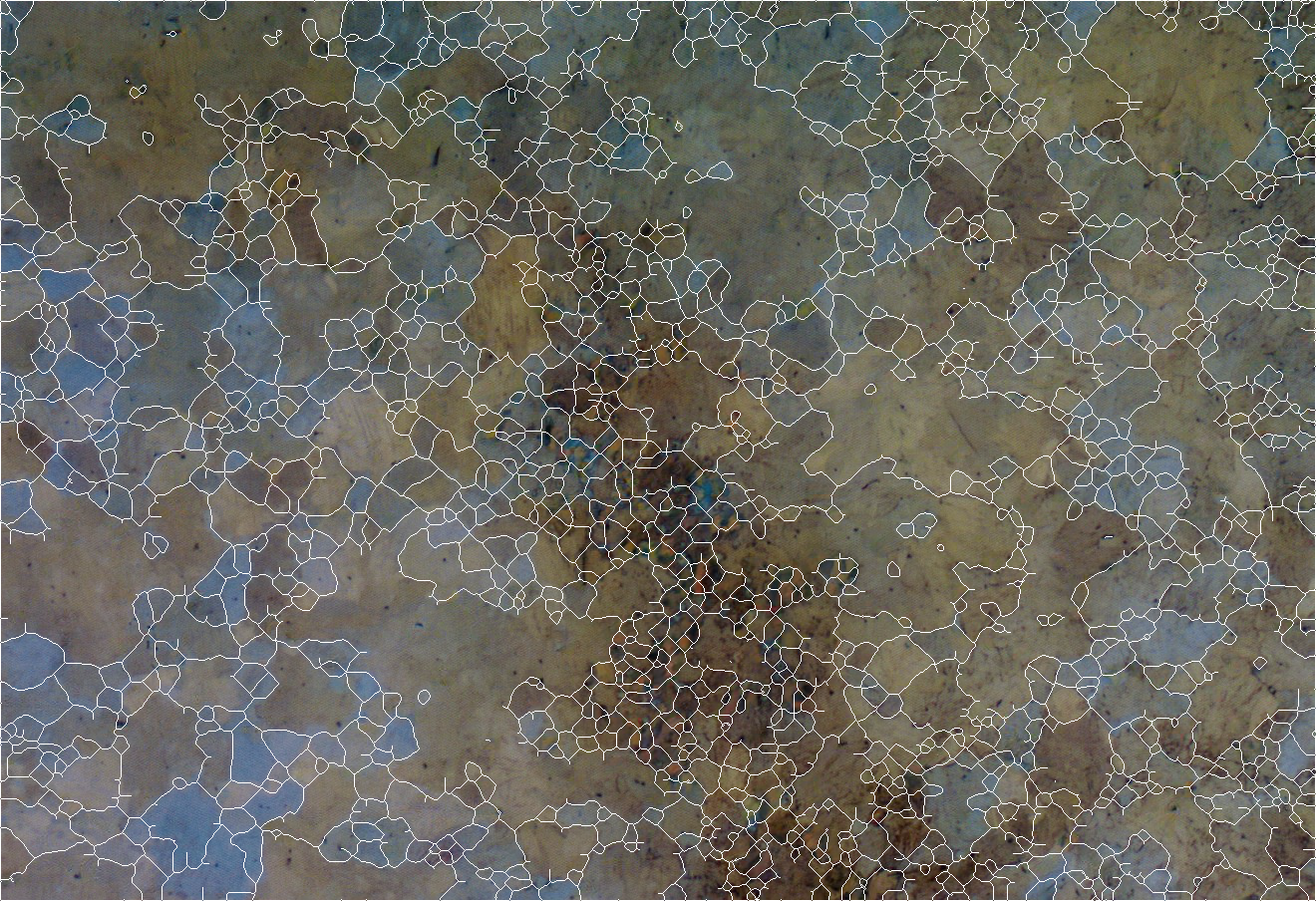}
      \caption*{Inpainting and GB segmentation}
  \end{figure}
\end{minipage}%
\begin{minipage}{.48\textwidth}
  \centering
  \begin{figure}[H]
  \centering
      \includegraphics[height=5cm, frame]{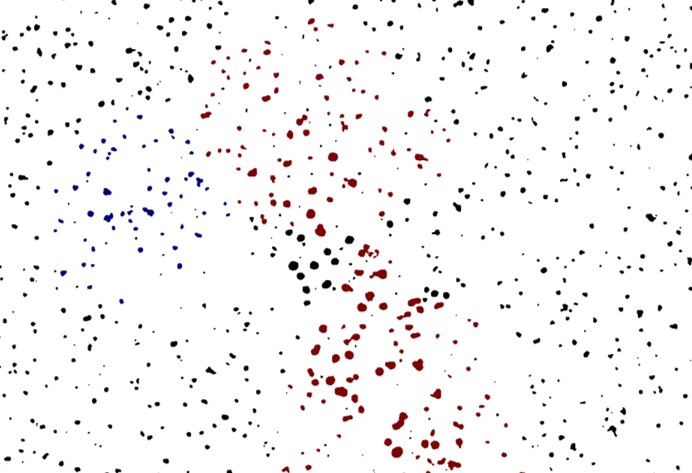}
      \caption*{Area anomaly detection}
  \end{figure}
\end{minipage}%
\caption{Example \#6.}\label{appendix:6}
\end{figure}

\end{document}